\documentstyle [12pt,eqsecnum,aps,amsfonts,axodraw] {revtex}
\tighten \draft
\input epsf
\newcommand{\beq}{\begin{equation}}
\newcommand{\eeq}{\end{equation}}
\newcommand{\beqs}{\begin{eqnarray}}
\newcommand{\eeqs}{\end{eqnarray}}

\newcommand{\lsim}{\mathrel{\raisebox{-.6ex}{$\stackrel{\textstyle<}{\sim}$}}}

\newcommand{\gsim}{\mathrel{\raisebox{-.6ex}{$\stackrel{\textstyle>}{\sim}$}}}

\begin{document}
\draft

\baselineskip 6.0mm

\title{Fermion Masses and Mixing in Extended Technicolor Models}

\vspace{6mm}

\author{
Thomas Appelquist$^{a}$ \thanks{email: thomas.appelquist@yale.edu} \and
Maurizio Piai$^{a}$ \thanks{email: maurizio.piai@yale.edu} \and Robert
Shrock$^{b}$ \thanks{email: robert.shrock@sunysb.edu}}

\vspace{6mm}

\address{(a) \ Physics Department, Sloane Laboratory \\
Yale University \\
New Haven, CT 06520}

\address{(b) \ C. N. Yang Institute for Theoretical Physics \\
State University of New York \\
Stony Brook, N. Y. 11794 }

\maketitle

\vspace{10mm}

\begin{abstract}

We study fermion masses and mixing angles, including the generation of a seesaw
mechanism for the neutrinos, in extended technicolor (ETC) theories.  We
formulate an approach to these problems that relies on assigning right-handed
$Q=-1/3$ quarks and charged leptons to ETC representations that are conjugates
of those of the corresponding left-handed fermions. This leads to a natural
suppression of these masses relative to the $Q=2/3$ quarks, as well as the
generation of quark mixing angles, both long-standing challenges for ETC
theories.  Standard-model-singlet neutrinos are assigned to ETC representations
that provide a similar suppression of neutrino Dirac masses, as well as the
possibility of a realistic seesaw mechanism with no mass scale above the
highest ETC scale of roughly $10^3$ TeV. A simple model based on the ETC group
SU(5) is constructed and analyzed. This model leads to non-trivial, but not
realistic mixing angles in the quark and lepton sectors. It can also produce
sufficiently light neutrinos, although not simultaneously with a realistic
quark spectrum.  We discuss several aspects of the phenomenology of this class
of models.

\end{abstract}

\pacs{14.60.PQ, 12.60.Nz, 14.60.St}

\vspace{16mm}

\pagestyle{empty}
\newpage

\pagestyle{plain} \pagenumbering{arabic}
\renewcommand{\thefootnote}{\arabic{footnote}}
\setcounter{footnote}{0}

\section{Introduction}

 The hypothesis that electroweak symmetry breaking (EWSB) is triggered
by a Higgs potential in which the quadratic term $\mu^2
\phi^\dagger\phi$ has a coefficient $\mu^2 < 0$, leads to the well-known
hierarchy problem, the radiative instability of Higgs sector with respect to
a much higher fundamental scale. Various approaches to this problem, such as
the incorporation of supersymmetry, have been suggested in which the Higgs
field remains elementary to energies well beyond the electroweak scale.
However, in two cases where scalar fields have been used successfully to
describe spontaneous symmetry breaking in the real world, namely the
Ginzburg-Landau free energy functional for superconductivity and the
$\sigma$ model for spontaneous chiral symmetry breaking in hadronic physics,
the scalars are clearly composite at the relevant energies, representing
bilinear fermion condensates.

   These facts have motivated an alternative approach based on dynamical EWSB
driven by a strongly coupled gauge interaction, associated with an exact gauge
symmetry, denoted generically as technicolor (TC) \cite{tc}-\cite{tcrev}.  The
EWSB is generated by the condensation of technifermion bilinears, and the
masses of quarks and charged leptons then arise via extended technicolor (ETC)
interactions \cite{etc}. It has seemed possible to understand some of the
fermion mass scales in this way, although the very light neutrino masses
present a newer and perhaps more challenging problem. Two of us have shown,
however, how light neutrino masses and lepton mixing can be obtained in
extended technicolor models \cite{nt,lrs} containing a set of standard-model
(SM) singlet neutrinos. This involves a strong suppression of both Dirac and
Majorana neutrino mass terms, along with a seesaw mechanism that does not
involve any superheavy mass scales.

In this paper, we explore in a similar way the generation of all the fermion
masses and mixing angles.  This is a very ambitious task. Grand unified
theories are less ambitious in this respect, since they do not incorporate a
dynamical theory of generations, but instead put these in as copies of the GUT
group representations. A variety of free parameters is available to accommodate
measured masses and mixing angles. Extended technicolor theories gauge the
generations and must dynamically produce all the fermion masses, the
generational hierarchy, the fermion mixing angles, etc. Our work on this
problem has met with only partial success. Nevertheless, it seems worthwhile to
report our results and discuss open problems.

We begin by reviewing the problem of fermion mass and mixing angle generation
in ETC models. The intra-generational mass splittings and CKM mixing have been
long-standing challenges for these theories.  In this paper we propose a
mechanism that succeeds in generating both.  We explore this mechanism in an
explicit ETC model based on the ETC gauge group SU(5). It incorporates a
generational U(3), and as the SU(5) breaks sequentially, it leads naturally to
a generational hierarchy and a remaining, unbroken $SU(2)_{TC}$ group. A key
feature will be the assignment of right-handed $Q=-1/3$ quarks and charged
leptons to ETC representations ${\mathcal R}$ that are conjugates of those of
the corresponding left-handed fermions. This leads to a natural strong
suppression of these masses relative to the $Q=2/3$ quark masses.

The representations of the SM singlet neutrinos are such that at all but the
highest ETC symmetry breaking scales, the left- and right-handed components
also transform according to mutually conjugate representations. Thus, the
elements of the Dirac mass matrices for neutrinos are also naturally suppressed
\cite{ssvz,at94}. This suppression, together with the dynamical generation of
Majorana mass terms for other SM-singlet neutrinos can provide a new kind of
seesaw mechanism \cite{nt,lrs} that involves only mass scales well within the
ETC range.

In Section II, we briefly review the technicolor and extended technicolor
framework, focusing on the problem of fermion masses and mixing, including the
generation of the quark and charged lepton mass hierarchies and mixing angles
as well as the typical presence of new, heavy degrees of freedom. The
corresponding discussion of neutrino masses and mixing angles is postponed to
section IV.

The models of Refs. \cite{at94,nt,lrs} are the immediate antecedents of the
model to be presented in this paper. They share in common the feature of having
two asymptotically-free gauge interactions that become strong at high scales:
${\rm SU}(5)_{ETC} \times {\rm SU}(2)_{HC}$, where the latter is a new
interaction designated as hypercolor (HC).  Ref. \cite{nt} presented a
mechanism for obtaining realistically light neutrino masses and lepton
mixing. Ref.  \cite{lrs} showed how this could be implemented in theories with
extended strong-electroweak groups. We review these models briefly in Section
III.

Section IV is the core of the paper. We begin by describing the
ingredients of our model, including the use of relatively conjugate
ETC representations.  We discuss generally the structure of the
neutrino mass matrices and the possible appearance of a seesaw
mechanism. Two possible symmetry breaking sequences are then
described, each leading to the unbroken $SU(2)_{TC}$ group. In each
case, suppressed neutrino masses emerge naturally, as do quark and
charged lepton mass splittings and mixing angles. The general
structure of the mass matrices is elucidated by a set of selection
rules that follow from residual global generational symmetries. For
both sequences, however, the success is only partial, with mixings
that are not fully realistic and an inability to achieve
simultaneously the right level of suppression for both Dirac neutrino
masses and down-type quark and charged lepton masses.

In Section V we discuss some phenomenological aspects of this class of
models. These include the constraints from precision electroweak data,
flavor-changing neutral current processes, and ETC-instanton-induced violation
of lepton number.  We also discuss global symmetries and associated
Nambu-Goldstone bosons. An interesting generic prediction of models that
incorporate the mechanism proposed in Ref. \cite{nt} for the origin of light
neutrino masses, and in particular of the class of models discussed here, is
the existence of neutrino-like mass eigenstates that are predominantly
electroweak-singlets with masses that lie in the range from a few hundred MeV
to a few hundred GeV.  We comment on experimental implications of these
particles.  We also remark on possible candidates for dark matter.

In Section VI, we summarize our work, listing the successes and shortcomings
of the mechanism employed. We suggest possible directions for future study.

\section{General Framework}

\subsection{Technicolor and Extended Technicolor}

For general discussion, we take the technicolor gauge group to be
SU($N_{TC})$. The set of technifermions includes one family, viz., $Q_L = {U
\choose D}_L$, $L_L = {N \choose E}_L$, $U_R$, $D_R$, $N_R$, $E_R$, with each
field transforming according to the fundamental or conjugate fundamental
representation of SU($N_{TC}$) and the usual representations of $G_{SM} = {\rm
SU}(3) \times {\rm SU}(2)_L \times {\rm U}(1)_Y$. The $N_R$ are SM singlets.

To generate the quark and lepton masses, this theory must be imbedded in a
larger, extended technicolor (ETC) theory, taken here to be SU($N_{ETC})$.
Constraints from flavor-changing neutral-current processes require that the ETC
vector bosons, which can mediate generation-changing transitions, must have
large masses.  We envision that they arise from self-breaking of the ETC gauge
symmetry, which can occur if ETC is a strongly coupled, chiral gauge theory.

Each fermion with standard-model interactions is embedded in a fundamental
or conjugate fundamental representation of the ETC gauge group such that the
first three components are the successive generations of this
fermion, and the remaining components are the corresponding technifermions
with the same SM quantum numbers. This entails the relation
\beq N_{ETC} = N_{gen.}+N_{TC} = 3 + N_{TC} \ . \label{nrel} \eeq
Additional standard-model-singlet fermions, some of which will play the role of
right-handed neutrinos, will also be introduced.  The three generations are
formed by the sequential breaking of the SU($N_{ETC}$) gauge symmetry to the
residual exact SU($N_{TC}$). We denote the mass scale at which each stage of
breaking takes place as $\Lambda_i$, where the $i$'th generation separates off
 from the other components of the ETC representations.

A particularly attractive choice for the technicolor group, used in the
explicit model to be studied here, is ${\rm SU}(2)_{TC}$., which minimizes the
TC contributions to the $S$ parameter \cite{precision,scalc,nutev} and can
yield walking behavior, allowing for enhanced quark and charged lepton masses.
 From Eq. (\ref{nrel}), this choice of $N_{TC}=2$ implies that our ETC group is
SU(5)$_{ETC}$.  With $N_f \simeq 16$ chiral technifermion doublets, as above,
studies suggest that the SU(2)$_{TC}$ theory could have an (approximate)
infrared fixed point (IRFP). The theory would be in the confining phase with
spontaneous chiral symmetry breaking, but near to the phase transition (as a
function of $N_f$) beyond which it would go over into a nonabelian Coulomb
phase \cite{vals,gap}. This approximate IRFP provides walking behavior,
enhancing the technifermion condensates that control the quark and charged
lepton masses.  This choice of $N_{TC}=2$ also plays a crucial role in our
mechanism \cite{nt} for getting light neutrino masses and our present approach
for explaining intra-generational mass splittings.

\subsection{Conventional ETC Mass Generation for Fermions}

The conventional ETC mechanism for the masses of quarks and charged leptons
relates these masses to the TC condensate through ETC gauge boson exchange
An estimate of the resultant masses is
\beq M^{(f)}_{ii} \simeq \left ( \frac{ g_{_{ETC}} }{\sqrt{2}} \right )^2
\frac{\eta \ \langle \bar F F  \rangle}{M_i^2} \label{mfii} \eeq
where $i$ is the generation index, $\langle \bar F F \rangle \ \equiv \
\langle \sum_{i=4,5} \bar F_i F^i \rangle$ for a technifermion $F$ (sum on
TC indices, no sum on color indices in the case where $F$ is a techniquark),
$M_i \sim g_{_{ETC}}\Lambda_i$ is the mass of the ETC gauge bosons that gain
mass at scale $\Lambda_i$, where $g_{_{ETC}}$ is the running ETC gauge
coupling at this scale.  The quantity $\eta$ is a renormalization group
factor given by
\beq \eta = \exp \left [ \int \frac{d\mu}{\mu} \gamma(\alpha_{TC}(\mu))
\right ] \label{eta} \eeq
reflecting the running of the bilinear operator $\bar f f$ between the
technicolor scale and the relevant ETC scale, where $\gamma$ is the
anomalous dimension for this operator. For a technicolor theory that
exhibits full walking behavior between $\Lambda_{TC}$ and a scale
denoted $\Lambda_{w}$, so that $\gamma \simeq 1$, it follows that
\beq \eta = \frac{\Lambda_{w}}{\Lambda_{TC}} \ . \label{etavalue} \eeq
Since only the SU(2)$_{TC}$ theory is taken to walk, $\Lambda_w$ is the lowest
ETC scale.

To evaluate Eq. (\ref{mfii}), we use the approximate relation
\beq \langle \bar F F \rangle \ = \ 4 \pi f_F^3 \left (\frac{3}{N_{TC}}
\right )^{1/2} \ , \label{condensate_F_relation} \eeq
where $f_F$ is the technipion decay constant. We note that the corresponding
QCD expression, $\langle \bar q q \rangle \simeq 4\pi f_\pi^3
({3}/{N_c})^{1/2}$, is quite accurate. Here, $\langle \bar q q \rangle$
includes a sum over color for each quark $q$. With $f_\pi = 93$ MeV, one has a
value of $(216 \ {\rm MeV})^3$ for $\langle \bar q q \rangle$. This may be
compared to the current-algebra relation $f_\pi^2 m_\pi^2 = (m_u + m_d) \langle
\bar u u + \bar d d \rangle = 2(m_u + m_d)\langle \bar q q \rangle$, where
$m_u$ and $m_d$ are the current quark masses and we have used the isospin
symmetry of QCD to equate $\langle \bar u u \rangle = \langle \bar d d \rangle
\equiv \langle \bar q q \rangle$.  The values of the $u$ and $d$ current quark
masses contain significant theoretical uncertainty; estimates range from
$m_u+m_d \simeq 15$ MeV from older current algebra methods \cite{gl} to 9 MeV
using lattice methods \cite{mlat}. Using the illustrative value $(m_u + m_d)
\simeq 9$ MeV, one finds $\langle \bar u u \rangle = \langle \bar d d \rangle =
(211 \ {\rm MeV})^3$, in good agreement with the above.

Next we recall that for a technicolor theory with one SM family of
technifermions,
\beq m_W^2 = \frac{g^2}{4}(N_c+1)f_F^2 \ , \label{mwsq} \eeq
where $g$ is the electroweak coupling and $f_F$ is the technipion decay
constant.  This yields $f_F \simeq 125$ GeV. It is convenient to express
various quantities in terms of the energy scale $\Lambda_{TC}$ at which the
technicolor interaction gets sufficiently large to cause technifermion
condensation. In QCD, $f_\pi = 93$ MeV and $\Lambda_{QCD} \simeq 180$ MeV so
that $\Lambda_{QCD} \simeq 2f_{\pi}$; using this as a guide to technicolor, one
infers $\Lambda_{TC} \simeq 250$ GeV. If one regards $N_c=3$ and $N_{TC}=2$ as
being sufficiently large so that one should include factors yielding the
correct respective large-$N_c$ and large-$N_{TC}$ behaviors, then one would
write $\Lambda_{QCD} \simeq 2 f_\pi\sqrt{3/N_c}$ and hence $\Lambda_{TC} \simeq
2 f_F \sqrt{3/N_{TC}} \simeq 300$ GeV.  We include these factors here, and
hence use $\Lambda_{TC} = 300$ GeV.

Since the ETC theory is strongly coupled, we cannot calculate precisely the
relation between $M_i$ and $\Lambda_i$.  We take
\beq M_i = \frac{a g_{_{ETC}}\Lambda_i}{4}~, \label{Mi} \eeq
where the constant $a$ is expected to be of order unity. Substituting in Eq.
(\ref{mfii}), we get
\beq M^{(u)}_{ii} \simeq \frac{\kappa \eta \Lambda_{TC}^3} { \Lambda_i^2},
\label{mfiicalc} \eeq
where $\kappa = 8\pi / 3 a^2$ ($\simeq 8\pi /3$ with $a = O(1)$). This rough
value for $\kappa$ will be used in all of our mass estimates.

 Suppose that three separate ETC scales, $\Lambda_1 > \Lambda_2 > \Lambda_3$,
emerge from the dynamical breaking. (This will be the case in the first of the
two symmetry breaking sequences in the model to be described here.)  Then, with
the SU(2)$_{TC}$ theory walking up to $\Lambda_3$,
\beq M^{(u)}_{ii} \simeq \frac{\kappa \Lambda_{TC}^2 \Lambda_3}{
\Lambda_i^2} \quad i=1,2,3 \ . \label{muii} \eeq
To compare this to the experimental values of the $Q=2/3$ quark masses, in
particular, the top quark, we may neglect off-diagonal entries in $M^{(u)}$,
at least for the higher two generations, so that the diagonal elements give
the actual quark masses. With the ETC breaking scales
\beq \Lambda_1 \simeq 10^3 \ \ {\rm TeV}, \quad \Lambda_2 \simeq 50 \ \ {\rm
TeV}, \quad \Lambda_3 \simeq 4 \ \ {\rm TeV} \label{lambda_values} \eeq
in the above formula (\ref{muii}), we get $m_t \simeq 175$ GeV, $m_c \simeq
1.3$ GeV and $m_u \simeq 3$ MeV \cite{qmass}.  Since our ETC theory appears
capable of generating the top quark mass, we shall not need to use other
approaches (e.g. \cite{tc4f}-\cite{topcolor}, \cite{tcrev}) for this purpose.
The consistency of these choices of ETC scales with precision experimental
constraints will be discussed in Section V.

Thus, while walking from $\Lambda_{TC}$ to the lowest ETC scale can
successfully generate a sufficiently heavy top quark mass, as well as charm and
up quark masses, this simple mechanism cannot be the entire story. It does not
account, for example, for quark and charged lepton mass splittings within each
generation. We next discuss this problem briefly, and introduce the mechanism
designed to address it.

\subsection{Intra-generational Mass Splittings}

A longstanding challenge for dynamical theories of fermion masses has been to
obtain the splittings $m_t >> m_b, \ m_\tau$ and $m_c >> m_s, \ m_\mu$
\cite{gen1}. The fact that $U_R$, $D_R$, $E_R$, and $N_R$ have different
hypercharges provides one source of splittings, and the fact that the
techniquarks have color interactions while the technileptons do not provides
another. However, these interactions are too weak at the scale $\Lambda_{TC}$,
to explain the observed splittings without fine tuning.

The approach here is to assign right-handed components of the $Q=-1/3$
quarks and charged leptons to representations of the extended technicolor
group that are conjugates of the representations assigned to the respective
left-handed $Q=-1/3$ quarks and charged leptons, i.e.,
\beq {\mathcal R}(f_L) = \overline{\mathcal R}(f_R) \ , \quad {\rm for} \
f=d,e \ , \label{conjugate_representation} \eeq
retaining the conventional assignment
\beq
 {\mathcal R}(u_L) = {\mathcal R}(u_R) \ .
\label{same_representation} \eeq
Here we use the notation $u_L$, $u_R$, $d_L$, $d_R$, $e_L$, $e_R$ (and,
for the SU(2)$_L$ doublets, $Q_L$, $L$) to refer to the full
SU(5)$_{ETC}$ representations with the indicated quantum numbers,
encompassing in each case the three generations and the $N_{TC}=2$
technifermion components.  Thus, for example, writing out $Q_L$ explicitly,
we have
\beqs Q_L & = &
   \left( \begin{array}{ccccc}
    u^{1,a}, & u^{2,a}, & u^{3,a}, & u^{4,a}, & u^{5,a} \\
    d^{1,a}, & d^{2,a}, & d^{3,a}, & d^{4,a}, & d^{5,a}
\end{array} \right )_L \cr\cr
& \equiv &
  \left( \begin{array}{ccccc}
    u^a, & c^a, & t^a, & U^{4,a}, & U^{5,a} \\
    d^a, & s^a, & b^a, & D^{4,a}, & D^{5,a}
\end{array} \right )_L \ .
\label{qleft} \eeqs
Note that we thus use synonymously the notation $f^i_\chi \equiv F^i_\chi$
for
$i=4,5$ and $\chi=L,R$.  Similarly, for leptons, we have, e.g.,
\beqs L_L & = &
   \left( \begin{array}{ccccc}
    n^1, &   n^2,   & n^3,   & n^4,   & n^5 \\
    e^1, &   e^2,   & e^3,   & e^4,   & e^5
\end{array} \right )_L \cr\cr
& \equiv &
  \left( \begin{array}{ccccc}
  \nu_e, &  \nu_\mu, & \nu_\tau, & N^4, & N^5 \\
   e,     &  \mu,    & \tau,     & E^4, & E^5
\end{array} \right )_L \ .
\label{nleft} \eeqs

The condition (\ref{conjugate_representation}) is similar to the device
\cite{ssvz} employed to suppress Dirac neutrino masses in Refs. \cite{nt}.
Note that this can only be done while maintaining a nonzero technicolor
condensate if the technicolor group is SU$(2)_{TC}$, corresponding here to the
ETC group SU$(5)_{ETC}$, since only in this case can one form a
TC-gauge-invariant bilinear of the form $\langle \epsilon^{ij}\bar F_{i,L}
F_{j,R}\rangle$ or $\langle \epsilon_{ij}\bar F^i_L F^j_R \rangle$, where $i,j$
are the SU(2) TC indices (and the color indices on the techniquarks are
implicit).  In this approach the diagrams giving rise to the masses of the
$Q=-1/3$ quarks and charged leptons require mixing of ETC gauge bosons, which
leads to strong suppression of these masses. In contrast, the masses of $Q=2/3$
quarks are generated in the conventional ETC manner, without any mixing of ETC
gauge bosons.  As we will show, this naturally produces large
intra-generational mass splittings with acceptable violation of custodial
$SU(2)$ symmetry. But in the models we have analyzed so far it leads to
excessive suppression of the down-type quark and charged lepton masses.

\subsection{Quark Mixing}

Another challenge for (extended) technicolor models has been to generate the
observed, inter-generational quark mixing. In general, the mass matrix 
$M^{(f)}$ ($f=u, d, e$) is diagonalized by a
bi-unitary transformation
\beq M^{(f)}_{diag.} = U^{(f)}_L M^{(f)} U^{(f) \ ^\dagger}_R
\label{mf_diagonalization} \eeq
and the Cabibbo-Kobayashi-Maskawa \cite{ckm} (CKM) quark mixing matrix $V$
defined by the charge-raising weak current
\beq J_\lambda = \sum_{i,j=1}^{N_{gen.}}\bar u_{i,L} V_{ij} \gamma_\lambda
d^{j}_L \label{j} \eeq
is given by
\beq V = U_L^{(u)}U_L^{(d) \dagger} \ . \label{v} \eeq

If both right and left-handed quarks are placed in the same representations
(the fundamental or conjugate fundamental representation of $SU(5)_{ETC}$ in
the present model), and there are no other interactions affecting the quark
mass matrices, then it is not difficult to see that no quark mixing is
produced.  As the ETC group breaks sequentially, a set of massive ETC bosons is
generated, with the $i$'th generation of quarks being connected to the
corresponding techniquarks by the ETC gauge bosons $V^i_t$, $i \in \{1,2,3\}$,
$t \in \{4,5\}$. If there is no mixing among the ETC bosons of the form $V^i_t
\leftrightarrow V^j_t$ with $i \ne j$, then the only type of mass generation is
that of Fig. 1. The mass matrix for both the up and down type quarks is
diagonal, so there is no quark mixing.

To remedy this problem, the requisite combination of ETC gauge boson mixings
must be generated. For the up-type quarks, with the conventional assignment
\ref{same_representation}, inter-generational mixing will exist, provided
that
\beq V^i_t \ \leftrightarrow \ V^j_t \ , \quad i,j \in \{1,2,3\}, \quad i
\ne j \label{fmix_possib1} \eeq
exists, where $t = 4,5$ is a TC index. For the down-type quarks, with the
unconventional assignment \ref{conjugate_representation}, inter-generational
mixing will require the presence of
\beq V^i_4 \ \leftrightarrow \ V^5_j \ , \quad V^i_5 \ \leftrightarrow \
V^4_j \ , \quad i,j \in \{1,2,3\}. \label{fmix_possib2} \eeq
In either case, the mass matrix will contain off-diagonal as well as
diagonal elements. The differences between the off-diagonal elements in
$M^{(u)}$ and $M^{(d)}$ and hence in the unitary transformations $U^{(u)}$
and $U^{(d)}$ then produce nontrivial quark mixing as specified in
(\ref{v}).

\begin{figure}[h]
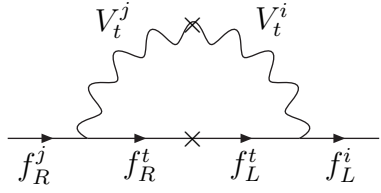

\caption{\footnotesize{Graphs generating $\bar f_{i,L} M^{(f)}_{ij} f^j_R$
where $i=1,2,3$, assuming that the indicated one-loop mixings of ETC gauge
bosons occur, for the case in which $f_L$ and $f_R$ both transform according
to the same (fundamental) representation of SU(5)$_{ETC}$.  The index $t$
takes on the values 4 and 5.  Here, $f^i$ is an up-type quark for $1 \le i
\le 3$ and techniquark for $i=4,5$.}} \label{ffgraph_samerep}
\end{figure}

In the model to be described in Section IV of this paper, the breaking of
the ETC gauge group is driven by a set of SM-singlet fermions, and this
produces both sources for off-diagonal entries in mass matrices given in
(\ref{fmix_possib1}) and (\ref{fmix_possib2}). The model takes ${\mathcal
R}(u_L) = {\mathcal R}(u_R)$ to be the fundamental representation of
SU$(5)_{ETC}$. Graphs that produce the off-diagonal mass matrix elements for
up-type quarks in this case are shown in Fig. \ref{ffgraph_samerep}.

\begin{center}
\begin{picture}(200,100)(0,0)
\ArrowLine(0,20)(30,20) \ArrowLine(30,20)(70,20) \ArrowLine(70,20)(110,20)
\ArrowLine(110,20)(140,20) \PhotonArc(70,20)(40,0,180){4}{8.5}
\Text(70,20)[]{$\times$} \Text(70,63)[]{$\times$} \Text(10,10)[]{$f^j_R$}
\Text(50,10)[]{$f^t_R$} \Text(90,10)[]{$f^t_L$} \Text(120,10)[l]{$f^i_L$}
\Text(40,65)[]{$V^j_t$} \Text(100,65)[]{$V^i_t$}
\end{picture}
\end{center}

\begin{figure}
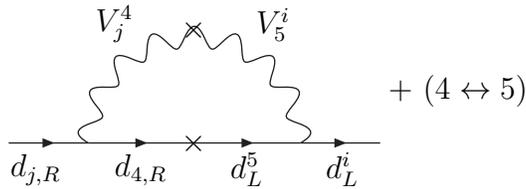

\caption{\footnotesize{Graphs generating $\bar d_{i,L} M^{(d)}_{ij} d_{j,R}$
where $i=1,2,3$, assuming that the indicated, one-loop mixings of ETC gauge
bosons occur, for the case in which $d_L$ and $d_R$ transform according to
the fundamental and conjugate fundamental representation of SU(5)$_{ETC}$.
Here $d^i$ is a down-type quark (techniquark) for $1 \le i \le 3$ ($i=4,5$).
As indicated, the graph with the indices 4 and 5 interchanged also
contributes.}} \label{ddgraph}
\end{figure}

For the down-type quarks, the model of Section IV takes ${\mathcal
  R}(d_L)$ to be the fundamental representation of SU(5)$_{ETC}$ and
${\mathcal R}(d_R)$ to be the conjugate fundamental representation.
With $N_{TC}=2$, so that the SU(2)$_{TC}$ condensates can still form,
bilinears for these quarks arise via diagrams that necessarily involve
ETC gauge boson mixing. The down-type techniquark condensate is of the
form $\langle \epsilon^{ij} \bar D_{i,a,L} D^a_{j,R} \rangle$, where
here $i,j$ and $a$ denote technicolor and color indices. This
generates off-diagonal elements in $M^{(d)}$ via Eq.
(\ref{fmix_possib2}).  The mass matrix for the down-type quarks,
generically of the form
\beq \bar d_{i,L} M^{(f)}_{ij} d_{j,R} + h.c. , \label{ddmassterm} \eeq
is generated by the graph of Fig. \ref{ddgraph}. The indicated ETC mixing
will be shown to exist in the model of Section IV.

\begin{center}
\begin{picture}(200,100)(0,0)
\ArrowLine(0,20)(30,20) \ArrowLine(30,20)(70,20) \ArrowLine(70,20)(110,20)
\ArrowLine(110,20)(140,20) \PhotonArc(70,20)(40,0,180){4}{8.5}
\Text(70,20)[]{$\times$} \Text(70,63)[]{$\times$} \Text(10,10)[]{$d_{j,R}$}
\Text(50,10)[]{$d_{4,R}$} \Text(90,10)[]{$d^{5}_L$}
\Text(120,10)[l]{$d^{i}_L$} \Text(40,65)[]{$V_j^4$} \Text(100,65)[]{$V_5^i$}
\Text(170,40)[]{$+ \ (4 \leftrightarrow 5)$}
\end{picture}
\end{center}

For the charged leptons, $e$, there are two possibilities. The $e$ fields of a
given chirality $\chi=L,R$ can transform according to the same (S) or conjugate
(C) ETC representations as the down-type quarks and techniquarks, $d$. We have
considered both possibilities, and we label the two classes of models as
DES and DEC, respectively. The model of Section IV will be of the DEC class,
with ${\mathcal R}(e_L)$ taken to be the conjugate fundamental representation
of SU(5)$_{ETC}$, and ${\mathcal R}(e_R)$ taken to be the fundamental
representation.

In this model then, the technifermion condensates take the following
forms:
\beq \langle \epsilon^{ij} \bar D_{i,a,L} D^a_{j,R} \rangle \ , \quad
\langle \epsilon_{ij} \bar E^i_L E^j_R \rangle \label{tc_condensates_dec}
\eeq
(together with their hermitian conjugates) where, as before, $i,j$ are
technicolor indices and $a$ are color indices.  Since the representations of
SU(2) are (pseudo)real, the technicolor interaction, by itself, would produce
the same magnitude for all of the technifermion condensates, both the ones in
Eq.  (\ref{tc_condensates_dec}) and the $\langle \bar U_{i,a,L}
U^{i,a}_L\rangle$ (for each color, in the case of techniquarks).  Diagrams
analogous to those in Fig. \ref{ddgraph}, with obvious changes in the DEC case,
give rise to the charged-lepton mass matrix $M^{(e)}_{ij}$.

This mechanism will lead to off-diagonal mass matrix elements for the
down-type quarks and charged leptons that are different from those of the
up-type quarks. It can therefore lead to CKM mixing among the quarks and
explain why $m_t >> m_b, m_\tau$ and $m_c >> m_s, m_\mu$.  The same
mechanism has been used to suppress the elements of Dirac mass matrices for
the neutrinos \cite{ssvz,at94}, providing an ingredient in a possible new
type of seesaw mechanism. This will be discussed in Section IV.

\section{Progenitors}

The model to be described in Section IV evolved from two earlier models,
designated here as AT94 \cite{at94} and AS02 \cite{nt}. In addition, an
embedding of AS02 in models with an extended standard model gauge group was
described in Refs. \cite{as,lrs}. Since these models use the same
representation for the left- and right-handed components of all the quarks
and charged leptons, they do not lead to nontrivial quark mixings. We review
their ingredients briefly to set the stage for the model of Section IV.

A basic condition for building any of these models is the cancellation of
the SU(5)$_{ETC}$ gauge anomaly. The SM-nonsinglet fermions and
technifermions contribute the following terms to this anomaly (written for
right-handed chiral components): $A(Q^c_R)=-2N_c=-6$, $A(L^c_R)=-2$,
$A(u_R)=A(d_R)=N_c=3$, and $A(e_R)=1$, for a total of $-1$. It follows that
the contributions of the right-handed SM-singlet fermions must be
\beq \sum_{f_R} A(f_R) =  1 \ . \label{arsum_conventional} \eeq
We describe below the ingredients of each of these models. The breaking
patterns are discussed in the original references. We then note that the
incorporation of Pati-Salam symmetry, used in two of these models, is not
compatible with the approach of the present paper.

\subsection{AT94}

In Ref. \cite{at94} an ETC model was constructed that explicitly
demonstrated the sequential self-breaking of the ETC group, producing three
generations of SM fermions.  This model was based on the gauge group
\beq G_{AT94} = {\rm SU}(5)_{ETC} \times {\rm SU}(2)_{HC} \times {\rm
SU}(4)_{PS} \times {\rm SU}(2)_L \times {\rm U}(1)_R \label{g_at94} \eeq
where SU(4)$_{PS}$ is the Pati-Salam group, which gauges baryon minus lepton
number $U(1)_{B-L}$ and combines it with color SU(3)$_c$.  This yields
quantization of electric charge and partial unification of quarks and
leptons. The U(1)$_R$ is an $R$-charge abelian group.  Thus the gauge group
involves two gauge interactions that become strong at high energies:
extended technicolor SU(5)$_{ETC}$, and one additional strong gauge
interaction called hypercolor (HC), ${\rm SU}(2)_{HC}$.  The latter is
included in order to produce the desired sequential gauge symmetry breaking.

The fermions in this theory are listed below, in a notation where the
numbers indicate the representations under ${\rm SU}(5)_{ETC} \times {\rm
SU}(2)_{HC} \times {\rm SU}(4)_{PS} \times {\rm SU}(2)_L$, and the subscript
gives the U(1)$_R$ charge:
\beqs (5,1,4,2)_{0,L} \ , \quad\quad & & (5,1,4,1)_{1,R} \ , \quad
(5,1,4,1)_{-1,R} \ , \cr\cr & & (1,1,6,1)_{0,R} \ , \quad (1,2,6,1)_{0,R}
\cr\cr & & (\bar 5,1,1,1)_{0,R} \ , \quad (\overline{10},1,1,1)_{0,R} \ ,
\quad (10,2,1,1)_{0,R} \label{fermions_at94} \eeqs
(In \cite{at94} these were written equivalently in left-handed holomorphic
form.)

A sequential breaking pattern was described in AT94, yielding some of the
features of the quark and lepton masses. The (Dirac) neutrino masses,
although much smaller than the quark and charged lepton masses, were still
much too large.

\subsection{AS02}

In Ref. \cite{nt} it was shown how one could get realistically light
neutrinos in a model with dynamical electroweak symmetry breaking. The model
used the ETC and HC gauge groups of Ref. \cite{at94}, together with only the
SM gauge group:
\beq G = {\rm SU}(5)_{ETC} \times {\rm SU}(2)_{HC} \times G_{SM} \label{g}
\eeq
where
\beq G_{SM} =  {\rm SU}(3) \times {\rm SU}(2)_L \times {\rm U}(1)_Y \ .
\label{g_sm} \eeq

The fermion content of this model is listed below, where the numbers
indicate the representations under ${\rm SU}(5)_{ETC} \times {\rm
SU}(2)_{HC} \times {\rm SU}(3)_c \times {\rm SU}(2)_L$ and the subscript
gives the weak hypercharge $Y$:
\beqs (5,1,3,2)_{1/3,L} \ , \quad\quad & & (5,1,3,1)_{4/3,R} \ , \quad\quad
(5,1,3,1)_{-2/3,R} \cr\cr (5,1,1,2)_{-1,L}  \ , \quad\quad & &
(5,1,1,1)_{-2,R} \ , \cr\cr & & (\overline{10},1,1,1)_{0,R} \ , \quad\quad
(10,2,1,1)_{0,R} \ . \label{fermions} \eeqs
Two sequential breaking patterns were studied in Ref. \cite{nt}.

\subsection{Models with Extended Strong-Electroweak Gauge Groups}

Refs. \cite{as,lrs} enlarged the model of Ref. \cite{nt} by embedding the SM
gauge theory in two models with an extended strong-electroweak gauge group.
These, too, are broken dynamically. The first is based on the group
\beq G = {\rm SU}(5)_{ETC} \times {\rm SU}(2)_{HC} \times G_{LR}
\label{gextended_lr} \eeq
where
\beq G_{LR} = {\rm SU}(3)_c \times {\rm SU}(2)_L \times {\rm SU}(2)_R \times
{\rm U}(1)_{B-L} \ . \label{glr} \eeq
The fermion content is
\beqs & & (5,1,3,2,1)_{1/3,L} \ , \quad (5,1,3,1,2)_{1/3,R} \ , \cr\cr & &
(5,1,1,2,1)_{-1,L}  \ , \quad (5,1,1,1,2)_{-1,R} \ , \cr\cr & & (\bar
5,1,1,1,1)_{0,R} \ , \quad (\overline{10},1,1,1,1)_{0,R} \ , \quad
(10,2,1,1,1)_{0,R} \ . \label{lrfermions} \eeqs

The second uses the group
\beq G = {\rm SU}(5)_{ETC} \times {\rm SU}(2)_{HC} \times G_{422}
\label{gextended_422} \eeq
where
\beq G_{422}={\rm SU}(4)_{PS} \times {\rm SU}(2)_L \times {\rm SU}(2)_R \ .
\label{g422} \eeq
and ${\rm SU}(4)_{PS}$ is the Pati-Salam group.  This achieves a higher
degree of unification, as compared with (\ref{gextended_lr}). Here the
fermion content is
\beqs & &  (5,1,4,2,1)_L \ , \quad (5,1,4,1,2)_R \ , \cr\cr & & (\bar
5,1,1,1,1)_R \ , \quad (\overline{10},1,1,1,1)_R \ , \quad (10,2,1,1,1)_R
\ . \label{422fermions} \eeqs

 These two models explain how $G_{LR}$ and $G_{422}$ break to ${\rm SU}(3)_c
\times {\rm SU}(2)_L \times {\rm U}(1)_Y$, and why this takes place at a scale
($\simeq 10^3$ TeV) large compared to the electroweak scale, but much smaller
than a GUT scale.  We note, however, that they are not compatible with the
approach of the present paper where the left- and right-handed components of
the $Q = - 1/3$ quarks and charged leptons are placed in conjugate ETC
representations (Eq. (\ref{conjugate_representation})). The ${\rm SU}(2)_L
\times {\rm SU}(2)_R$ symmetry would imply that ${\mathcal R}(u_{L,R})
={\mathcal R}(d_{L,R})$.  Since the choice ${\mathcal R}(u_{L}) ={\mathcal
R}(u_{R})$ is necessary to provide conventional mass ETC mass generation for
the $Q = 2/3$ quarks, the use of conjugate representations for the $Q = - 1/3$
quarks would not be allowed. The inclusion of SU(4)$_{PS}$ symmetry as in
$G_{422}$ would also prohibit the use of conjugate representations for the
charged leptons.

\section{A Model for Quarks and Leptons}

\subsection{Ingredients}

We employ the same gauge group G as in Ref. \cite{nt}, given in Eq.
(\ref{g}), and use relatively conjugate ETC representations for the left-
and right-handed components of $Q=-1/3$ quarks and charged leptons. All of
the nonabelian factor groups in $G$ are asymptotically free. There are no
bilinear fermion operators invariant under $G$ and hence there are no bare
fermion mass terms.  The SU(2)$_{TC}$ subsector of SU(5)$_{ETC}$, and the
SU(2)$_{HC}$ interaction, are vectorial. The representation content for the
quarks is
\beq Q_L: \ (5,1,3,2)_{1/3,L} \ , \quad\quad u_R: \ (5,1,3,1)_{4/3,R} \ ,
\quad d_R \ (\bar 5,1,3,1)_{-2/3,R} \ . \label{qtq} \eeq
The representations for the charged leptons are
\beq L_L: \ (\bar 5,1,1,2)_{-1,L} \ , \quad \quad e_R: \ (5,1,1,1)_{-2,R}
\label{dec_le} \eeq
so the model is of DEC type (Eq. (\ref{conjugate_representation})).

The model also contains SM-singlet fermions, constrained by the absence of an
SU(5)$_{ETC}$ gauge anomaly. We denote the contribution of a given chiral
fermion $f_R$, written as a right-handed field, to this anomaly, as $A(f_R)$.
The quarks and techniquarks make the contributions $A(Q^c_R) = -2N_c = -6$,
$A(u_R)=-A(d_R)=N_c=3$. In a DEC-type model, $A(L^c_R)=2$, $A(e_R)=1$, while
for a DES-type model, $A(L^c_R)=-2$, $A(e_R)=-1$.  The totals for the
contributions from these SM fermions are $-3$ and $-9$ for models of DEC and
DES type, respectively. In order that a model should have zero SU(5)$_{ETC}$
gauge anomaly, the contributions of the right-handed SM-singlet fermions to
this anomaly must be
\beq \sum_{{\rm SM-singlet} \ f_R} A(f_R) =  \cases{ 3 & for DEC-type models
\cr
                              9 & for DES-type models }
\label{arsum} \eeq

Since SU(2)$_{HC}$ is used for the hypercolor group, it is free of any local
gauge anomaly; the constraint of no global anomaly requires that there be an
even number of doublets, and this is satisfied for our model, since it has
hypercolored fermions transforming as a 10 (antisymmetric rank-2 tensor) or
SU(5)$_{ETC}$ and two doublets under SU(2)$_{HC}$.

There are a variety of solutions to the constraint Eq. (\ref{arsum}),
and we have studied a number of these. Here we focus on one
relatively simple solution.  For the SM-singlet fermions we take
\beq \psi^{ij}_R: \ (10,1,1,1)_{0,R} \ , \quad \zeta^{ij,\alpha}_R: \
(10,2,1,1)_{0,R} \ , \quad \omega^\alpha_{p,R}: \ 2(1,2,1,1)_{0,R}
\label{fermions_model1} \eeq
where $1 \le i,j \le 5$ are SU(5)$_{ETC}$ indices, $\alpha=1,2$ are
SU(2)$_{HC}$ indices, and $p=1,2$ refers to the two copies of the
$\omega^\alpha_{p,R}$ field. The SM-singlet fermions in this model thus
involve three types: (i) ETC-nonsinglet, HC-singlets, viz., the
antisymmetric rank-2 tensor $\psi^{ij}_R$; (ii) ETC-nonsinglet,
HC-nonsinglet, viz., the $\zeta^{ij,\alpha}_R$ field, which transforms as an
antisymmetric rank-2 tensor representation of SU(5)$_{ETC}$ and a
fundamental representation of SU(2)$_{HC}$, and (iii) the ETC-singlet,
HC-nonsinglet fields $\omega^\alpha_{p,R}$, which transform as fundamental
representations of SU(2)$_{HC}$ (we include an even number $p=2$ of copies
to avoid a global SU(2)$_{HC}$ anomaly).  The subsector comprised of the
$\psi^{ij}_R$ and $\zeta^{ij,\alpha}_R$ is the same as the SM-singlet sector
of the model discussed in Ref. \cite{nt}.  One can see that condition
(\ref{arsum}) is satisfied and, more generally, that this model is free of
any gauge and global anomalies.

In a theory in which lepton number is not gauged, it is a convention how one
assigns the lepton number $L$ of the SM-singlet fields.  We assign $L=1$ to
$\psi^{ij}_R$ so that the Dirac mass terms $\bar n^i_L b_{ij} \psi^{1j}_R$ that
will form conserve $L$. One could, alternatively, assign $L=0$ to
$\psi^{ij}_R$, so that the Dirac mass terms transform as $\Delta L=1$.  Indeed,
in the models of Refs. \cite{as,lrs}, where $L$ is gauged, this is the way that
the $\Delta L=2$ violation of lepton number arises. The lepton number assigned
to the $\zeta^{ij,\alpha}_R$ and $\omega^\alpha_R$ is also a convention; since
they have no Dirac terms with the electroweak-doublet neutrinos, we leave these
assignments arbitrary.

\subsection{Symmetry Breaking -- The First Stage}

Symmetry breaking down to the TC scale is driven completely by the
SM-singlet sector of the model. We identify plausible preferred condensation
channels using a generalized-most-attractive-channel (GMAC) approach that
takes account of one or more strong gauge interactions at each breaking
scale, as well as the energy cost involved in producing gauge boson masses
when gauge symmetries are broken. An approximate measure of the
attractiveness of a channel $R_1 \times R_2 \to R_{cond.}$ is taken to be
$\Delta C_2 = C_2(R_1)+C_2(R_2)-C_2(R_{cond.})$, where $R_j$ denotes the
representation under a relevant gauge interaction and $C_2(R)$ is the
quadratic Casimir invariant.

As the energy decreases from some high value, the SU(5)$_{ETC}$ and
SU(2)$_{HC}$ couplings increase. We envision that at $E \simeq \Lambda_1
\gsim 10^3$ TeV, $\alpha_{_{ETC}}$ is sufficiently strong \cite{gap} to
produce condensation in the channel
\beq (10,1,1,1)_{0,R} \times (10,1,1,1)_{0,R} \to (\bar 5,1,1,1)_0
\label{10-10channel} \eeq
with $\Delta C_2 = 24/5=4.8$, breaking ${\rm SU}(5)_{ETC} \to {\rm
SU}(4)_{ETC}$.  With no loss of generality, we take the breaking direction
in SU(5)$_{ETC}$ as $i=1$; this entails the separation of the first
generation of quarks and leptons from the components of SU(5)$_{ETC}$ fields
with indices lying in the set $\{2,3,4,5\}$. With respect to the unbroken
${\rm SU}(4)_{ETC}$, we have the decomposition $(10,1,1,1)_{0,R} =
(4,1,1,1)_{0,R} + (6,1,1,1)_{0,R}$. We denote the fundamental representation
$(4,1,1,1)_{0,R}$ and antisymmetric tensor representation $(6,1,1,1)_{0,R}$
as $\alpha^{1i}_R \equiv \psi^{1i}_R$ for $2 \le i \le 5$ and $\xi^{ij}_R
\equiv \psi^{ij}_R$ for $2 \le i,j \le 5$.  The associated
SU(5)$_{ETC}$-breaking, SU(4)$_{ETC}$-invariant condensate is then
\beq \langle \epsilon_{1 i j k \ell} \xi^{ij \ T}_R C \xi^{k \ell}_R \rangle
= 8\langle \xi^{23 \ T}_R C \xi^{45}_R - \xi^{24 \ T}_R C \xi^{35}_R +
\xi^{25 \ T}_R C \xi^{34}_R \rangle \ . \label{xixicondensate} \eeq
This condensate and the resultant dynamical Majorana mass terms of order
$\Lambda_1$ for the six $\xi^{ij}_R$ fields in Eq. (\ref{xixicondensate})
violate total lepton number as $|\Delta L|=2$.  The actual mass eigenstates are
linear combinations of these six fields involving maximal ($\pm \pi/4$) mixing
of $|\xi_{24,R}\rangle$ with $|\xi_{35,R}\rangle$, $|\xi_{34,R}\rangle$ with
$|\xi_{25,R}\rangle$, and $|\xi_{23,R}\rangle$ with $|\xi_{45,R}\rangle$,
respectively, with respective eigenvalues $\pm \Lambda_1$ (where in the case of
negative eigenvalues, one redefines fields appropriately so as to obtain
positive masses).

At energy scales below $\Lambda_1$, depending on relative strengths of
couplings, various symmetry-breaking sequences with different
condensates are plausible. As all the condensates develop, there will
naturally arise relative phases among them, leading in general to the
presence of CP violation in both the quark and neutrino sectors.  We
postpone an analysis of these phases and of CP violation from
topological terms in the gauge sector and the associated strong CP
problem for later work, concentrating in this paper on the magnitudes
of the condensates and the resultant pattern of masses and mixing
angles.

\subsection{Neutrino Mass Matrix}

Having described the SM-singlet sector of our model and the first stage of
symmetry breaking at scale $\Lambda_1$, we are now in a position to discuss
the general structure of the neutrino mass matrix and the possibility of a
seesaw mechanism. The details will then depend on which lower-energy
symmetry breaking sequence is favored.

The full neutrino mass term for the model is
\beq
  -{\cal L}_m =
 {1 \over 2}(\bar n_L \ \overline{\chi^c}_L)
             \left( \begin{array}{cc}
              M_L & M_D \\
              (M_D)^T & M_R \end{array} \right )\left( \begin{array}{c}
      n^{c}_R \\
      \chi_R \end{array} \right ) + h.c.
\label{mnugeneral} \eeq
where $n_L=(\nu_e,\nu_\mu,\nu_\tau,N_4,N_5)_L$ and $\chi_R$ is a vector of
the $n_s$ SM-singlet fields, including $\alpha^{ij}_R$, $\xi^{ij}_R$,
$\zeta^{ij,\alpha}_R$, and $\omega^\alpha_{p,R}$.  Since $(M_L)^T=M_L$ and
$(M_R)^T=M_R$, the full $(5+n_s) \times (5+n_s)$ neutrino mass matrix $M$ in
(\ref{mnugeneral}) is complex symmetric and can be diagonalized by a unitary
transformation $U_\nu$ as
\beq M_{diag.}=U_\nu^\dagger M (U_\nu^\dagger)^T \ . \label{Mnu} \eeq
This yields the neutrino masses and transformation $U_\nu$ relating the
group eigenstates $\nu_L = (\bar n, \overline{\chi^c})_L^T$ and the
corresponding mass eigenstates $\nu_{m,L}$, according to
\beq \nu_{j,L} = \sum_{k=1}^{5+n_s} (U_\nu)_{jk} \nu_{k,m,L} \ , \quad 1 \le
j \le 5+n_s \label{Unutransformation} \eeq
(the elements $(U_\nu)_{jk}$ connecting techni-singlet and technicolored
neutrinos vanish identically).  The lepton mixing matrix for the observed
neutrinos \cite{mns,lpss} $\nu_{\ell,L} = U \nu_{m,L}$ is then given by
\beq U_{ik} = \sum_{j=1}^3 (U_{\ell,L})_{ij} (U_\nu)_{jk} \ , \quad 1 \le i
\le 3, \quad 1 \le k \le 5+n_s \label{u} \eeq
where $U_{1k} \equiv U_{ek}$, etc., and where the diagonalization of the
charged lepton mass matrix is carried out by the bi-unitary transformation
in Eq. (\ref{mf_diagonalization}) for $f=e$. Thus, the charge-lowering
leptonic weak current is given by
\beq J^\lambda = \sum_{i=1}^3 \sum_{k=1}^{5+n_s} \bar e^i_L \gamma^\lambda
U_{ik} (\nu_m)_{k,L} \label{jleptonic} \eeq
where $e_i$ denotes the $i$th charged lepton mass eigenstate and $\nu_m$ is
the $(5+n_s)$-dimensional vector of neutrino mass eigenstates.

The complex symmetric mass matrix $M$ of neutrino-like (colorless and
electrically neutral) states is $39 \times 39$, with $n_s=34$. Of the 39
neutrino-like chiral components of fermion fields, $N_{gen} = 3$ are the
observed left-handed neutrinos, $N_{TC}=2$ are left-handed technineutrinos,
and the other $n_s=34$ are electroweak-singlets, comprised of the 10
$(\overline{10},1,1,1)_{0,R}$, the 20 $(10,2,1,1)_{0,R}$, and the four
fermions in the pair of $(1,2,1,1)_{0,R}$. All of its entries arise as the
high-energy physics is integrated out at each stage of condensation from
$\Lambda_1$ down to $\Lambda_{TC}$. Composite operators of various dimension
are formed, with bilinear condensation then leading to the masses. The
nonzero entries of $M$ arise either directly, as dynamical masses associated
with various condensates, or via loop diagrams involving dynamical mass
insertions on internal fermion lines with, in most cases, mixings among ETC
gauge bosons on internal lines. The different origins for the elements of
$M$ give rise to quite different magnitudes for these elements; in
particular, there is substantial suppression of the second class because the
diagrams involve ratios of small scales such as $\Lambda_{TC}$ and lower ETC
scales to larger scales such as $\Lambda_1$.

In the two symmetry breaking sequences to be considered here, either the
SU(2)$_{HC}$ symmetry or a U(1)$_{HC}$ subgroup will remain unbroken.
Therefore, the hypercolored fermions will not form bilinear condensates and
resultant mass terms with hypercolor-singlet fermions. Hence, $M$ has a
block-diagonal structure, and it is convenient to group the HC-singlet
blocks together as $M_{HCS}$ and the HC-nonsinglet blocks together as
$M_{HC}$. The matrix of primary interest, $M_{HCS}$, is defined by the
operator product
\beq -{\cal L}_{HCS} =
  {1 \over 2}(\bar n_L,  \ \overline{\alpha^c}_L, \ \overline{\xi^c}_L)
M_{HCS}
\left( \begin{array}{c}
       n^{c}_R \\
       \alpha_R \\
       \xi_R \end{array} \right ) + h.c.
\label{mhcsoperator} \eeq
so that
\beq
   M_{HCS} =
   \left( \begin{array}{ccc}
   M_L & (M_D)_{\bar n \alpha} & (M_D)_{\bar n \xi} \\
   (M_D)_{\bar n \alpha}^T & (M_R)_{\alpha \alpha} & (M_R)_{\alpha \xi} \\
   (M_D)_{\bar n \xi}^T    & (M_R)_{\alpha \xi}^T  & (M_R)_{\xi \xi}
                               \end{array} \right ) \ .
\label{mhcs} \eeq
This matrix has many vanishing entries arising from the fact that the
$SU(2)_{TC}$ symmetry is exact. It, too, could therefore be written in a
block-diagonal form by clustering the seven TC-singlet fermions,
$((n^c)^1,(n^c)^2,(n^c)^3,\alpha^{12},\alpha^{13},\xi^{23},\xi^{45})_R
\equiv
(\nu_e^c,\nu_\mu^c,\nu_\tau^c,\alpha^{12},\alpha^{13},\xi^{23},\xi^{45})_R$,
in one group, and the TC-doublet fermions is a second group. This
block-diagonal structure will be evident in many of the exact zeros in the
various matrices displayed below.

Within $M_{HCS}$, the most important Dirac submatrix is $(M_D)_{\bar n
\alpha}$, defined by the operator product
\beq
  \bar n^i_L [(M_D)_{\bar n \alpha}]_{ij} \alpha^{1j}_R
\label{alphanuterm} \eeq
with $1 \le i \le 5$, $2 \le j \le 5$.  This matrix has the general form
\beq
  (M_D)_{\bar n \alpha} =
  \left( \begin{array}{cccc}
   b_{12} & b_{13} & 0 &    0 \\
   b_{22} & b_{23} & 0 &    0 \\
   b_{32} & b_{33} & 0 &    0 \\
   0             & 0             & 0 & c_1  \\
   0             & 0             & -c_1 & 0 \end{array} \right ) \ .
\label{MDnualpha} \eeq
The vanishing entries are zero because of exact technicolor gauge
invariance. The entry $c_1$ represents a dynamical mass directly generated
by technicolor interactions corresponding to
\beq \sum_{i,j=4,5} \epsilon_{ij} \bar n^i_L \alpha^{1j}_R \ ,
\label{nbaralpha45} \eeq
so that
\beq |c_1| \simeq \Lambda_{TC} \ . \label{c1v} \eeq
In Fig. \ref{alpha-n} we show graphs that contribute to $\bar n^i_L b_{ij}
\alpha^{1j}_R$ for $i=1,2,3$ and $j=2,3$ if the indicated mixings of ETC
gauge bosons occur.  Which of these ETC gauge boson mixings do occur depends
on the symmetry-breaking sequence.

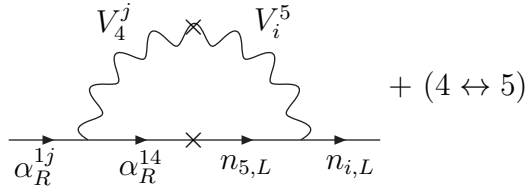
\begin{center}
\begin{picture}(200,100)(0,0)
\ArrowLine(0,20)(30,20) \ArrowLine(30,20)(70,20) \ArrowLine(70,20)(110,20)
\ArrowLine(110,20)(140,20) \PhotonArc(70,20)(40,0,180){4}{8.5}
\Text(70,20)[]{$\times$} \Text(70,63)[]{$\times$}
\Text(10,10)[]{$\alpha^{1j}_R$} \Text(50,10)[]{$\alpha^{14}_R$}
\Text(90,10)[]{$n_{5,L}$} \Text(120,10)[l]{$n_{i,L}$}
\Text(40,65)[]{$V_4^j$} \Text(100,65)[]{$V_i^5$} \Text(170,40)[]{$+ \ (4
\leftrightarrow 5)$}
\end{picture}
\end{center}

\begin{figure}
\caption{\footnotesize{Graphs generating $\bar n^i_L b_{ij} \alpha^{1j}_R$
for $i=1,2,3$ and $j=2,3$, provided that the indicated mixings of ETC gauge
bosons occur.}} \label{alpha-n}
\end{figure}

\begin{figure}[h]
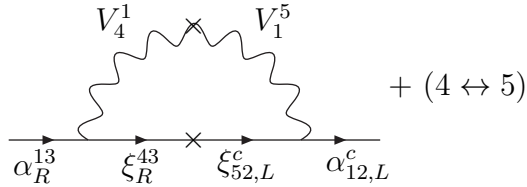

\caption{\footnotesize{Graphs for $\alpha^{12 \ T}_R C r_{23}
\alpha^{13}_R$. }} \label{alpha-alpha}
\end{figure}

The sub-matrix $(M_R)_{\alpha \alpha}$, which, depending on mass scales, can
play a key role in a seesaw mechanism \cite{nt}, is associated with the
operator product
\beq (\overline{\alpha^c})^{1i}_L r_{ij} \alpha^{1j}_R = \alpha^{1i \ T}_R C
r_{ij} \alpha^{1j}_R \ . \label{rij} \eeq
With the usual ordering of the components in the 4 of SU(4)$_{ETC}$, viz.,
$(\alpha^{12},\alpha^{13},\alpha^{14},\alpha^{15})$, this matrix is
\beq
  (M_R)_{\alpha \alpha}= \left( \begin{array}{cccc}
  r_{22} & r_{23} & 0  &  0  \\
  r_{23} & r_{33} & 0  &  0  \\
      0  &    0   & 0  &  0  \\
      0  &    0   & 0  &  0  \end{array} \right ) \ .
\label{rqmatrix} \eeq
As before, the zeros are exact due to technicolor invariance.  If the $2
\times 2$ \ $r_{ij}$ submatrix has maximal rank, it could provide the seesaw
which, in conjunction with the suppression of the Dirac entries $b_{ij}$
discussed above, could yield adequate suppression of neutrino masses. The
$r_{ij}$ submatrix plays this role because $\alpha_{12,R}$ and
$\alpha_{13,R}$ are the electroweak-singlet technisinglet neutrinos that
remain as part of the low-energy effective theory below the electroweak
scale.  In Fig. \ref{alpha-alpha} we show graphs that contribute to
$r_{23}$.

\begin{center}
\begin{picture}(200,100)(0,0)
\ArrowLine(0,20)(30,20) \ArrowLine(30,20)(70,20) \ArrowLine(70,20)(110,20)
\ArrowLine(110,20)(140,20) \PhotonArc(70,20)(40,0,180){4}{8.5}
\Text(70,20)[]{$\times$} \Text(70,63)[]{$\times$}
\Text(10,10)[]{$\alpha^{13}_R$} \Text(50,10)[]{$\xi^{43}_R$}
\Text(90,10)[]{$\xi^c_{52,L}$} \Text(120,10)[l]{$\alpha^c_{12,L}$}
\Text(40,65)[]{$V^1_4$} \Text(100,65)[]{$V_1^5$} \Text(170,40)[]{$+ \
(4\leftrightarrow 5)$}
\end{picture}
\end{center}

The other submatrices in $M$ do not have as important an effect on the
primary mass eigenstates in the observed neutrinos.  The second Dirac
submatrix in Eq. (\ref{mhcs}), $(M_D)_{\bar n \xi}$ is associated with the
operator product
\beq \bar n^i_L [(M_D)_{\bar n \xi}]_{i,kn} \xi^{kn}_R \label{nxioperator}
\eeq
with $1 \le i \le 5$ and $2 \le k < n \le 5$ and has the general form
\beq (M_D)_{\bar n \xi} =
 \left( \begin{array}{cccccc}
 d_{1,23} & d_{1,45} & 0 &  0  & 0   & 0   \\
 d_{2,23} & d_{2,45} & 0 &  0  & 0   & 0   \\
 d_{3,23} & d_{3,45} & 0 &  0  & 0   & 0   \\
 0                        & 0  & 0 & c_2 & 0   & c_3 \\
 0                        & 0  &-c_2& 0  &-c_3 & 0  \end{array} \right ) \ .
\label{MDnuxi} \eeq
Again, the zeros are exact and follow from technicolor invariance.  Because
the $\xi$ fields decouple from the theory at scales below $\Lambda_1$, they
cannot directly condense with the $n$ fields at lower scales. Thus, the
nonzero elements of $(M_D)_{\bar n \xi}$ arise only indirectly, via loop
diagrams and are highly suppressed.  These elements of $(M_D)_{\bar n \xi}$
have only a small effect on the neutrino eigenvalues because in the
characteristic polynomial for the full neutrino mass matrix $M$, they occur
as corrections to much larger terms involving $\Lambda_1$.

\begin{figure}[h]
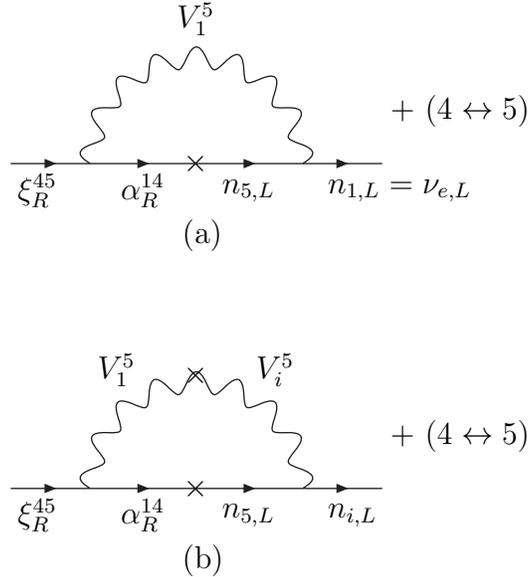

\caption{\footnotesize{Graphs that yield contributions to the Dirac bilinear
$\bar n^i_L d_{i,jk} \xi^{jk}_R$ for $j,k=4,5$ and (a) $i=1$, (b) $i=2,3$.
In the latter case, the mixing of ETC gauge bosons that is necessary is
indicated.}} \label{xi-n}
\end{figure}

In Fig. \ref{xi-n}(a,b) we show the one-loop diagrams that contribute to
$\bar n^i_L d_{i,jk} \xi^{jk}_R$ for $j,k=4,5$ and (a) $i=1$, i.e., to $\bar
\nu_{e,L} d_{1,45} \xi^{45}_R$ and (b) $i=2,3$, i.e., to $\bar \nu_{\mu,L}
d_{2,45} \xi^{45}_R$ and $\bar \nu_{\tau,L} d_{3,45} \xi^{45}_R$.  For the
$i=1$ case shown in Fig. \ref{xi-n}(a), no ETC gauge boson mixing is
necessary, but for the $i=2,3$ cases shown in Fig. \ref{xi-n}(b), one needs
the mixings $V_1^t \to V_i^t$ for $t=4,5$.

\begin{center}
\begin{picture}(200,110)(0,0)
\ArrowLine(0,30)(30,30) \ArrowLine(30,30)(70,30) \ArrowLine(70,30)(110,30)
\ArrowLine(110,30)(140,30) \PhotonArc(70,30)(40,0,180){4}{8.5}
\Text(70,30)[]{$\times$} \Text(10,20)[]{$\xi^{45}_R$}
\Text(50,20)[]{$\alpha^{14}_R$} \Text(90,20)[]{$n_{5,L}$}
\Text(120,20)[l]{$n_{1,L} = \nu_{e,L}$} \Text(70,85)[]{$V_1^5$}
\Text(170,50)[]{$+ \ (4 \leftrightarrow 5)$} \put(65,0){(a)}
\end{picture}
\end{center}

\begin{center}
\begin{picture}(200,110)(0,0)
\ArrowLine(0,30)(30,30) \ArrowLine(30,30)(70,30) \ArrowLine(70,30)(110,30)
\ArrowLine(110,30)(140,30) \PhotonArc(70,30)(40,0,180){4}{8.5}
\Text(70,30)[]{$\times$} \Text(70,73)[]{$\times$}
\Text(10,20)[]{$\xi^{45}_R$} \Text(50,20)[]{$\alpha^{14}_R$}
\Text(90,20)[]{$n_{5,L}$} \Text(120,20)[l]{$n_{i,L}$}
\Text(40,75)[]{$V^5_1$} \Text(100,75)[]{$V^5_i$} \Text(170,50)[]{$+ \ (4
\leftrightarrow 5)$} \put(65,0){(b)}
\end{picture}
\end{center}

The entries $c_2$ and $c_3$ in the matrix (\ref{MDnuxi}) correspond to the
bilinears
\beq \sum_{i,j=4,5} \epsilon^{ij} \bar n_{i,L} \xi_{2j,R} \ , \quad
\sum_{i,j=4,5} \epsilon^{ij} \bar n_{i,L} \xi_{3j,R}~, \label{nxitc} \eeq
and are allowed by technicolor invariance, but they cannot occur directly as
TC-scale dynamical masses since the $\xi$ fields, gain masses of order
$\Lambda_1$ from the condensate (\ref{xixicondensate}). They might be
induced by higher-order processes, but we will not pursue this in detail,
since they are not important for the light neutrinos.

\begin{figure}[h]
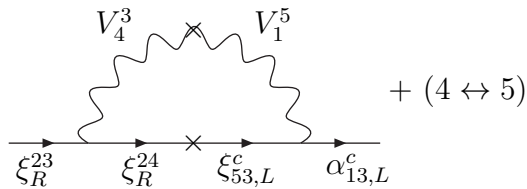

\caption{\footnotesize{Graphs that contribute to the Majorana bilinear
$\alpha^{13 \ T}_R C w_{13,23} \xi^{23}_R$ in symmetry-breaking sequence
1.}} \label{alpha-xi}
\end{figure}

The submatrix $(M_R)_{\alpha \xi}$ corresponding to the bilinear
\beq \overline{\alpha^c}^{1i}_L w_{1i,jk} \xi^{jk}_R = \alpha^{1i \ T}_R C
w_{1i,jk} \xi^{jk}_R \label{w1ijk} \eeq
has the general form
\beq (M_R)_{\alpha \xi}= \left( \begin{array}{cccccc}
 w_{12,23} & w_{12,45} & 0    &   0   &   0   &   0 \\
 w_{13,23} & w_{13,45} & 0    &   0   &   0   &   0 \\
     0     &    0      & 0    & c_4   &   0   &  c_5 \\
     0     &    0      & -c_4 &   0   & -c_5  &   0  \end{array} \right ) \
.
\label{wmatrix} \eeq
Again, all zeros in these matrices are exact and follow from SU(2)$_{TC}$
invariance.  In Eq. (\ref{wmatrix}) the $\pm c_4$ and $\pm c_5$ entries
refer to bilinear fermion operator products that are allowed by the exact
SU(2)$_{TC}$ invariance, namely
\beq \sum_{i,j=4}^5 \epsilon_{123ij} \overline{\alpha^c}^{1i}_L \xi^{kj}_R =
\sum_{i,j=4}^5 \epsilon_{123ij}\alpha^{1i \ T}_R C \xi^{kj}_R \ , \quad
k=2,3 \ . \label{alphaxitechnicolor} \eeq
Since the $\xi^{ij}_R$ fields pick up dynamical masses of order $\Lambda_1$
and then decouple, there are no technicolor-scale condensates of the form
$\langle \sum_{i,j=4}^5 \epsilon_{123ij}\alpha^{1i \ T}_R C \xi^{kj}_R
\rangle$ with $k=2,3$. However, these terms can, in general, be induced by
higher-order processes similar to those that can induce the $w_{ij,k\ell}$
shown in Eq. (\ref{wmatrix}).  In Fig. \ref{alpha-xi} we show graphs that
can contribute to the Majorana bilinear $\overline{\alpha^c}^{13}_L
w_{13,23} \xi^{23}_R = \alpha^{13 \ T}_R C w_{13,23} \xi^{23}_R$ (and do so
in the first sequence to be described below).

\begin{center}
\begin{picture}(200,110)(0,0)
\ArrowLine(0,30)(30,30) \ArrowLine(30,30)(70,30) \ArrowLine(70,30)(110,30)
\ArrowLine(110,30)(140,30) \PhotonArc(70,30)(40,0,180){4}{8.5}
\Text(70,30)[]{$\times$} \Text(70,73)[]{$\times$}
\Text(10,20)[]{$\xi^{23}_R$} \Text(50,20)[]{$\xi^{24}_R$}
\Text(90,20)[]{$\xi^c_{53,L}$} \Text(120,20)[l]{$\alpha^c_{13,L}$}
\Text(40,75)[]{$V_4^3$} \Text(100,75)[]{$V_1^5$} \Text(170,50)[]{$+ \ (4
\leftrightarrow 5)$}
\end{picture}
\end{center}

To summarize, the full $39 \times 39$-dimensional neutrino mass matrix, M,
factorizes into one diagonal block involving the HC singlet fields and
another involving the HC non-singlets. The former can be seen to
block-diagonalize further (implicit in the above discussion) into one block
involving the TC-singlet fields and another involving the TC-non-singlets.
The characteristic polynomial, $P(\lambda)$, associated with the
diagonalization of M and the determination of its eigenvalues (the masses of
the neutrino-like states), therefore factorizes into four parts, one of
which, $P_{HCS-TCS}(\lambda)$, is associated with the HC-singlet and
TC-singlet fields. The roots of the other three factors range in magnitude
 from $\Lambda_1$ down to $\Lambda_{TC}$. The factor $P_{HCS-TCS}(\lambda)$
contains two large roots of order $\Lambda_1$ and five other roots.  These
five roots will be of primary interest here since they contain three that
form the mass eigenstates in the three electroweak-doublet neutrino
interaction eigenstates, together with two that are primarily
electroweak-singlet neutrinos. The detailed forms of these polynomial
factors depend on the symmetry-breaking sequence and will be discussed
below.

This general formalism can now be applied to any symmetry breaking sequence.
We discuss two plausible sequences, depending on the relative strength of
ETC and HC couplings.

\subsection{Symmetry-Breaking Sequence 1}

This sequence, leading to two further, distinct stages of ETC-symmetry
breaking, is related to the one denoted $G_a$ in Ref. \cite{nt}, but has
further structure resulting from the additional $\omega^\alpha_{p,R}$
fermions not present there. In the effective theory below $\Lambda_1$, the
ETC interaction, which is SU(4)$_{ETC}$-symmetric, is again asymptotically
free. As the energy decreases through a scale, $\Lambda_2 \simeq 10^2$ TeV,
the SU(4)$_{ETC}$ and SU(2)$_{HC}$ couplings become sufficiently strong to
lead together to the condensation
\beq (4,2,1,1)_{0,R} \times (6,2,1,1)_{0,R} \to (\bar 4,1,1,1)
\label{4x6channel} \eeq
with $\Delta C_2 = 5/2$ for SU(4)$_{ETC}$ and $\Delta C_2 = 3/2$ for
SU(2)$_{HC}$. This breaks SU(4)$_{ETC}$ to SU(3)$_{ETC}$.  The condensate is
\beqs & & \langle \epsilon_{\alpha\beta}\epsilon_{12jk \ell}\zeta^{1j,\alpha
\ T}_R C \zeta^{k \ell,\beta}_R \rangle = \cr\cr & & 2\langle
\epsilon_{\alpha\beta}( \zeta^{13,\alpha \ T}_R C \zeta^{45,\beta}_R -
\zeta^{14,\alpha \ T}_R C \zeta^{35,\beta}_R +
         \zeta^{15,\alpha \ T}_R C \zeta^{34,\beta}_R ) \rangle \ ,
\label{4x6zetacondensate} \eeqs
and the twelve $\zeta^{ij,\alpha}_R$ fields in this condensate gain masses
$\simeq \Lambda_2$.  (To be precise, linear combinations of these fields
form eigenstates with definite masses.)

The fact that the neutrino-like fields $\alpha^{1i}_R$ transform as a 4 of
SU(4)$_{ETC}$, while the left-handed neutrinos and technineutrinos transform
as a $\bar 4$, will lead to a strong suppression of relevant entries in the
Dirac submatrix $M_D$ \cite{ssvz,at94}.  This suppression depends only on
the fact that the left- and right-handed components of the neutrinos
transform according to relatively conjugate representations.

In the effective theory below $\Lambda_2$, both the SU(3)$_{ETC}$ and the
SU(2)$_{HC}$ interactions are asymptotically free, so their couplings
continue to increase as the energy scale decreases. At a scale $\Lambda_3
\simeq 3$ TeV, these interactions together are envisioned to lead to a
condensation in the channel (where the numbers give the representations
under SU(3)$_{ETC} \times$ SU(2)$_{HC} \times G_{SM}$)
\beq (3,2,1,1)_{0,R} \times (3,2,1,1)_{0,R} \to (\bar 3,1,1,1)_0
\label{33to3barchannel} \eeq
with $\Delta C_2 = 4/3$ for SU(3)$_{ETC}$ and $\Delta C_2 = 3/2$ for
SU(2)$_{HC}$.  This breaks SU(3)$_{ETC}$ to SU(2)$_{TC}$.  Without loss of
generality, we may use the original SU(3)$_{ETC}$ gauge symmetry to orient
the condensate so that it takes the form
\beq \langle \epsilon_{123jk} \epsilon_{\alpha\beta} \zeta^{2j,\alpha \ T}_R
C \zeta^{2k,\beta}_R \rangle \ = \ 2\langle
\epsilon_{\alpha\beta}\zeta^{24,\alpha \ T}_R C \zeta^{25,\beta}_R \rangle \
. \label{33to3barcondensate} \eeq

In the low-energy effective field theory below $\Lambda_3$, the massless
SM-singlet fermions then consist of $\zeta^{ij,\alpha}_R$ with $ij=12, 23$,
and $\omega^\alpha_{p,R}$ with $p=1,2$.  In this energy interval the
SU(2)$_{HC}$ coupling continues to grow, as does the SU(2)$_{TC}$ coupling.
When the coupling $\alpha_{_{HC}}$ becomes sufficiently strong, the
hypercolor interaction can naturally produce (HC-singlet) condensates of the
various remaining HC-doublet fermions.  In each case, $\Delta C_2 =3/2$.
Since the condensate (\ref{33to3barcondensate}) was formed via a combination
of both SU(3)$_{ETC}$ and SU(2)$_{HC}$ interactions, while the present
condensates are formed only by the SU(2)$_{HC}$ interaction, and have the
same value of $\Delta C_2$, it follows that the scale at which they form,
denoted $\Lambda_s$ (where $s$ denotes SU(2)$_{TC}$-singlet) satisfies
$\Lambda_s \le \Lambda_3$.  There are six condensates of this type:
\beq \langle \epsilon_{\alpha\beta} \zeta^{12,\alpha \ T}_R C
\zeta^{23,\beta}_R \rangle \label{z12z23condensate} \eeq
\beq \langle \epsilon_{\alpha\beta} \zeta^{12,\alpha \ T}_R C
\omega^\beta_{p,R} \rangle \ , \quad p=1,2 \label{z12omegacondensate} \eeq
\beq \langle \epsilon_{\alpha\beta} \zeta^{23,\alpha \ T}_R C
\omega^\beta_{p,R} \rangle \ , \quad p=1,2 \label{z23omegacondensate} \eeq
\beq \langle \epsilon_{\alpha\beta} \omega^{\alpha \ T}_{1,R} C
\omega^\beta_{2,R} \rangle \ . \label{omega_selfcondensate} \eeq
With these condensations, the 20 fields $\zeta^{ij,\alpha}$ and the four
$\omega^\alpha_{p,R}$ fields condense out of the effective theory at energies
below $\Lambda_s$, just as the six $\xi_{ij,R}$ condense out of the theory at
energies below $\Lambda_1$. In what follows, shall take $\Lambda_s \simeq
\Lambda_3$. This is reasonable, since the condensates (\ref{z12z23condensate})
and (\ref{z23omegacondensate}) have the same symmetry behavior as the
condensate (\ref{33to3barcondensate}) that forms at $\Lambda_3$, i.e., all
three of these break SU(3)$_{ETC}$ to SU(2)$_{TC}$.

 The condensates involving the $\omega^\alpha_{p,R}$ fields are important in
the generation of the quark and lepton mass matrices. Without them, as we will
describe shortly, a residual unbroken global symmetry would prevent the
formation of some desirable matrix elements.  The fact that the generational
structure of the model is determined in this way by generation-dependent global
symmetries is a feature of dynamical symmetry breaking, in which only gauge
bosons are responsible for the communication of the breaking from the SM
singlet fermion sector to the visible sector of the theory.

Finally, at the technicolor scale $\Lambda_{TC}$, the technifermions condense,
breaking ${\rm SU}(2)_L \times {\rm U}(1)_Y$ to U(1)$_{em}$.  (Vacuum alignment
arguments motivate color and electric charge conservation by the
technicondensates \cite{vac_align}). The condition $\Lambda_{TC} < \Lambda_s$
is natural since, although both the TC and HC groups are SU(2), the leading
coefficient for the beta function in the HC group below $\Lambda_3$ is larger
than that for the TC group.  Explicitly, with $\beta(\alpha) = - b_{0} \alpha^2
/ 2\pi + ...$, we have $b_0=(11/3)N_{HC}-(2/3)(1/2)N_{hcd}=6$ for the
SU(2)$_{HC}$ theory and $b_0=(11/3)N_{TC}-(2/3)(1/2)N_{tcd}=2$ for the
SU(2)$_{TC}$ theory, since there are $N_{hcd}=4$ chiral SU(2)$_{HC}$-doublet
fermions ($\zeta^{12,\alpha}_R$, $\zeta^{23,\alpha}_R$, $\omega^\alpha_{p,R}$)
and $N_{tcd}=16$ chiral SU(2)$_{TC}$-doublet fermions active in this energy
interval.

\subsection{Fermion Mass Matrices for Sequence 1}

 The diagonal entries of the up-type quark mass matrix $M^{(u)}_{ij}$ are
the
conventional masses generated by the ETC mechanism, as given above in Eq.
(\ref{muii}). The off-diagonal entries in $M^{(u)}_{ij}$ are generated by
ETC gauge boson mixing of type (\ref{fmix_possib1}).  The general set of ETC
gauge boson mixings generated by the SM-singlet condensates listed above is
analyzed in the appendix. Here and throughout the text we use the results of
this analysis. In the present case, the only ETC gauge boson mixing of type
(\ref{fmix_possib1}) that is generated is $V^1_t \leftrightarrow V^3_t$ with
$t=4,5$.  This gives rise to nonzero off-diagonal elements $M^{(u)}_{ij}$
only for $ij=13,31$ (which, in this approach, are equal in magnitude, as a
consequence of (\ref{mfsym}).  Other off-diagonal entries will be shown to
vanish from global symmetry considerations. The above ETC gauge boson mixing
is described by a mass mixing function ${}^1_t\Pi^3_t(0)$ with $t=4,5$, as
described in the appendix.

The simplest graph contributing to the bilinear $\bar u_{a,1,L} M^{(u)}_{13}
u^{a,3}_R$ is shown in Fig.  \ref{ffgraph_samerep}, with $f=u, \ j=3, \
i=1$. The estimate is very similar to that of Eq. (\ref{muii}) with $i = 3$,
but suppressed by the ratio of the gauge boson mixing to the mass-squared of
the heavier ($\Lambda_1$-scale) ETC boson in the graph. With the TC theory
walking up to the scale $\Lambda_{3}$, the integral will be dominated by
momenta of that scale, and the gauge boson mixing estimate of Eq.
(\ref{Pi_v1up4dn_to_v3up4n_gap}) may be used.  We utilize the expression Eq.
(\ref{Mi}), with $a=1$, for the ETC gauge boson masses and thus estimate
\beq M^{(u)}_{13}  \simeq  \frac{\kappa \Lambda_{TC}^2\Lambda_3^2}{
\Lambda_1^2\,\Lambda_2}~, \label{mu13_gap} \eeq
where $\kappa \simeq 8 \pi/3$.  The result is the same order as the diagonal
element $M^{(u)}_{11}$. The off-diagonal elements $M^{(u)}_{ij}$ with $ij$
different from 13 and 31 vanish identically, as a consequence of a residual
global symmetry described below. With the ETC breaking scales of Eq.
(\ref{lambda_values}), this structure of $M^{(u)}_{ij}$ leaves the up-type
quarks masses consistent with experiment.

For the down-type quarks and charged leptons, because we employ relatively
conjugate ETC representations for the left- and right-handed components, all
the elements of their mass matrices, diagonal and non-diagonal, will be
suppressed, vanishing were it not for the ETC gauge boson mixing. The
relevant non-diagonal ETC gauge boson 2-point functions are of the form
$V_i^4 \leftrightarrow V^j_5$ and $V_i^5 \leftrightarrow V^j_4$ (see the
appendix). The graphs that yield a down-type quark mass term in this model
are given by the appropriate special cases of Fig. \ref{ddgraph}.  We find
that $M^{(f)}_{ij}$ for $f=d,e$ has nonzero entries for all $(ij)$ except
(12), (21), (23), and (32).

We estimate these entries, as above, by taking the theory to walk up to the
lowest ETC scale $\Lambda_3$. We note again that the results will be given
by the conventional ETC estimate of Eq. (\ref{muii}), but suppressed by the
relevant ratios of mixings to gauge boson squared masses. Consider first the
estimate of $M^{(d,e)}_{13}$. The appropriate gauge boson mixing term is
given by Eq. (\ref{Pi_v1up4dn_to_v3dn5up_gap}), and the full estimate is
similar to that for $M^{(u)}_{13}$:
\beq M^{(d,e)}_{13} \simeq  \frac{\kappa \Lambda_{TC}^2
\Lambda_3} { \Lambda_1^2}~. \label{Mf13_gap} \eeq

Turning next to the diagonal element $M^{(d,e)}_{22}$, we make use of the
mixing term of Eq. (\ref{Pi_v2up4dn_to_v2dn5up_gap}), which is suppressed
relative to the previous mixings by $\Lambda_{3}/\Lambda_2$. We find
\beq M^{(d,e)}_{22} \simeq \frac{\kappa \Lambda_{TC}^2 \Lambda_3^4} {
\Lambda_2^5}~. \label{Mf22_gap} \eeq
The two other diagonal entries are estimated in much the same way, making
use of the gauge boson mixing terms of Eqs. (\ref{Pi_1dn4up_to1up5dn_gap})
and (\ref{Pi_3dn4up_to3up5dn_gap}). Recall that these involve iterations of
previous mixings. The results are
\beq M^{(d,e)}_{11} \simeq \frac{\kappa \Lambda_{TC}^2 \Lambda_3^3}
{\Lambda_1^4} \label{Mf11_gap} \eeq
\beq M^{(d,e)}_{33} \simeq \frac{\kappa \Lambda_{TC}^2 \Lambda_3} {
\Lambda_1^2} \ . \label{Mf33_gap} \eeq
There is mixing only between the first and third eigenstates.  The resultant
quark masses are $m_s = M^{(d)}_{22}$ and $m_d \sim m_b \simeq
M^{(d)}_{13}$. With the $\Lambda_i$ choices of Eq. (\ref{lambda_values}),
these are $m_s \simeq 0.2$ MeV and $m_d \simeq m_b \simeq 1$ MeV. Clearly,
the down-type quark masses are suppressed far too much by this mechanism
and, in addition, do not exhibit a full three-generation structure.

The diagonalization of $M^{(u)}$ and $M^{(d)}$ yields the quark mixing
matrix $V$, according to (\ref{v}).  This mixing arises almost completely
 from the diagonalization of the down-quark mass matrix, so that $V=U^{(d) \
\dagger}_L$. Although the model does produce quark mixing, the $d$-$b$
mixing angle is close to $\pi/4$ and hence not phenomenologically
acceptable. The same structural features apply to the charged lepton mass
matrix.  To leading order, the charged lepton masses are given by the same
formulas as for the corresponding down-type quarks; however, color effects
for the quarks will modify this equality.

For the neutrinos, sequence 1 yields nonzero entries in the Dirac submatrix
$b_{ij}$ for $ij=13, \ 22, \ 33$.  A general relation in the present class
of models is
\beq b_{ij}=M^{(d,e)}_{ij}  \quad {\rm for} \ \ i=1,2,3; \quad j=2,3 \ .
\label{bij_Mdij} \eeq
(Recall that there are only two right-handed, SM-singlet states below
$\Lambda_{TC}$, and therefore there are no $j=1$ entries.) This equality is
due to the fact that both the down-quark masses and the Dirac masses for the
neutrinos are generated in the same way, employing relatively conjugate
representations for the left- and right-handed components of these fields.
Thus, Eq. \ref{bij_Mdij}) holds for any symmetry-breaking sequence in the
present model. It a defect of the present class of models, since it means
that these models have difficulty in simultaneously producing sufficiently
large masses for down-type quarks and charged leptons on the one hand, and
sufficiently small masses for neutrinos on the other hand.

Focusing on the neutrinos alone, the Dirac masses $b_{ij}$ could play a role
in a seesaw mechanism, provided that a corresponding $2\times2$ Majorana
mass matrix for the SM singlets (the $r_{ij}$'s of Eq. (\ref{rqmatrix})) is
formed with the right magnitude. The $23$ elements of this matrix {\it are}
formed in sequence 1 via Fig. \ref{alpha-alpha}, but with magnitude much
smaller than the Dirac terms. The reason for the smallness is that the
fermions and gauge bosons of Fig. \ref{alpha-alpha} have masses of order
$\Lambda_1$, and the gauge boson mixing is soft above the smaller scale
$\Lambda_{3}$. The relevant mixing, estimated in the appendix, arises from a
combination of the mixings of Figs. \ref{v1dn4up_to_v3up5dn_gap_fig} and
\ref{v1dn4up_to_v3dn4up_gap_fig}. The only scale entering this combination
is $\Lambda_3$, leading to the estimate
\beq r_{23} \simeq \frac{\kappa \Lambda_{3}^6}{ \Lambda_1^5} \ .
\label{r23_gap}
\eeq
For the above choices of $\Lambda_1 \simeq 10^3$ TeV and $\Lambda_3 \simeq
4$ TeV, $r_{23}$ is well below any of the nonzero $b_{ij}$'s.  This problem
is alleviated as $\Lambda_3/\Lambda_1$ is increased, but since $\Lambda_1$
can be no smaller than a few hundred GeV to adequately suppress
flavor-changing neutral current processes (see Section V), a realistic
seesaw is not attainable for any $\Lambda_3 \leq \Lambda_1$. This problem
will be alleviated in sequence 2, allowing a neutrino seesaw, but not at the
same time generating realistic quark masses.

It is useful to exhibit the structure of the characteristic polynomial
$P(\lambda)$ of $M$. Recall that $P(\lambda)$ factors in general into four
parts, only one of which, $P_{HCS-TCS}(\lambda)$, has some roots smaller
than $\Lambda_{TC}$. These occur as five of its seven roots (the other two
are the masses of linear combinations of $\xi^{23}_R$ and $\xi^{45}_R$, of
order $\Lambda_1$). $P_{HCS-TCS}(\lambda)$ takes the form
\beq P_{HCS-TCS}(\lambda) = \lambda P_6(\lambda) \ ,\label{p_hcs-tcs_gap}
\eeq
where the zero eigenvalue is the mass of $\nu_1$.  Since the experimental
data indicating neutrino oscillations only determines the differences of
squares of neutrino masses $|\Delta m^2_{32}|$ and $\Delta m^2_{21}$, it
allows the lightest eigenvalue to vanish.  However, the model with this
symmetry-breaking sequence is too simple to reproduce another feature from
the data, namely the fact that $\nu_1$ mixes with $\nu_2$ and $\nu_3$ to
form the interaction eigenstates. The factor $P_6(\lambda)$ is a polynomial
of the form
\beq P_6(\lambda) = \sum_{j=0}^6 p_{6,j} \lambda^j \ ,\label{p6} \eeq
where for this sequence $p_{6,5}=p_{6,3}=p_{6,1}=0$, so that
$P_6(\lambda)=P(-\lambda)$ and the roots come in opposite-sign pairs. The
nonzero coefficients in Eq. (\ref{p6}) are
\beq p_{6,4}=-(\Lambda_1^2 + b_{13}^2 + b_{22}^2 + b_{33}^2 + d_{1,45}^2 +
d_{3,45}^2 + r_{23}^2 + w_{13,23}^2) \label{p6_4_gap} \eeq
\beqs & & p_{6,2}= (b_{13}^2+b_{22}^2+b_{33}^2+r_{23}^2)\Lambda_1^2 -2
w_{13,23}( b_{13}d_{1,45} + b_{33}d_{3,45})\Lambda_1 + b_{22}^2(b_{13}^2 +
b_{33}^2)\cr\cr & & + d_{1,45}^2(b_{22}^2 + b_{33}^2 + r_{23}^2) +
    d_{3,45}^2(b_{13}^2 + b_{22}^2 + r_{23}^2) +
w_{13,23}^2(b_{22}^2 + d_{1,45}^2 + d_{3,45}^2) \cr\cr & & -2
b_{13}b_{33}d_{1,45}d_{3,45} \label{p6_2_gap} \eeqs
and
\beqs & & p_{6,0}=-b_{22}^2 \biggl [ (b_{13}^2+b_{33}^2)\Lambda_1^2 -2
w_{13,23}(b_{13}d_{1,45} + b_{33} d_{3,45}) \Lambda_1 \cr\cr & & +
b_{13}^2d_{3,45}^2 + b_{33}^2 d_{1,45}^2 + w_{13,23}^2(d_{1,45}^2 +
d_{3,45}^2) - 2b_{13}b_{33}d_{1,45}d_{3,45} \biggr ] \ . \label{p6_0_gap}
\eeqs

As noted above (see Fig. \ref{xi-n}), while $d_{1,45}$ is nonzero in
general, in order to get a nonzero $d_{i,45}$ with $i=2,3$, one needs the
ETC gauge boson mixing $V_1^t \to V_i^t$ with $i=2,3$ and $t=4,5$ to occur.
In the present sequence, we have the mixing $V_1^t \to V_3^t$ and hence
$d_{3,45}$ is also nonzero. Both $d_{1,45}$ and $d_{3,45}$ are estimated to
be of order $\kappa \Lambda_{TC}^2 \Lambda_3 / \Lambda_2$, while $d_{2,45}$
and $d_{i,23}$ for $i=1,2,3$ vanish.  The graphs that yield $w_{13,23}$ in
this sequence are shown in Fig. \ref{alpha-xi}. This entry $w_{13,23}$ is
estimated to be of order $\kappa \Lambda_3^6 / \Lambda_1^5$. The other
$w_{ij,k\ell}$ entries vanish. Because the $d_{1,45}$, $d_{3,45}$, and
$w_{13,23}$ terms enter together with much larger terms in the various
coefficients, they have a negligible effect on the eigenvalues.  Dropping
negligible terms, $P_{HCS-TCS}(\lambda)=\lambda(\lambda-\Lambda_1)
(\lambda+\Lambda_1)P_4(\lambda)$, where $P_4(\lambda)$, evident from Eqs.
(\ref{p6})-(\ref{p6_0_gap}), gives the nonzero eigenvalues of magnitude less
than $\Lambda_{TC}$.

\subsection{Residual Generational Symmetries in Sequence 1}

The presence of zeros in the fermion mass matrices for sequence 1 is
not due to the approximations made in the estimates. The zeros are
exact, as a consequence of residual global generational symmetries. To
see this, we first consider the model without the
$\omega^\alpha_{p,R}$ fermions, in which case sequence 1 reduces to
the sequence denoted $G_a$ in Ref. \cite{nt}. Then $M^{(u)}_{ij}$ is
nonzero only for the diagonal entries $ij =11,22,33$, $M^{(d)}_{ij}$
is nonzero only for $ij = 13,31,22$, the Dirac neutrino mass matrix is
nonzero only for $ij = 13,22$, and the $2\times2$ right-handed
Majorana neutrino mass matrix vanishes identically.

The reason for the zeros is that all the elements of the above matrices,
other than the diagonal elements of $M^{(u)}_{ij}$, require ETC gauge boson
mixing, and the allowed mixings are in turn determined by global symmetries.
The requirement of mixing for the Dirac matrices has already been explained.
That mixing is necessary for the $2\times2$ right-handed Majorana neutrino
mass matrix will be explained in the general discussion of the seesaw
mechanism in Section IV H. The relevant global symmetry is a certain
subgroup, U(1)$_f$, of the generational global symmetry $U(3)_f \subset
SU(5)_{ETC}$ acting on the first three ETC indices. It is generated by
$e^{i\theta T_f}$ where $T_f$ is a combination of diagonal generators of
SU(5)$_{ETC}$:
\beq T_f = \frac{1}{2}{\rm diag}(1,0,-1,0,0) \ . \label{tf} \eeq

Although this U(1)$_f$ is broken by the condensates in the $G_a$ sequence,
the ETC gauge boson masses and mixings remain invariant under it. Since only
the mixings $V^2_4 \leftrightarrow V_2^5$ and $V^1_4 \leftrightarrow V_3^5$,
$V^1_5 \leftrightarrow V_3^4$ respect this symmetry, only these are
generated. To see that the ETC gauge boson masses and mixings are invariant
under U(1)$_f$, we first consider the condensate, (\ref{xixicondensate}),
that forms at the highest scale $\Lambda_1$. It is not U(1)$_f$-invariant,
but the ETC gauge boson masses that emerge at this scale, arise via loop
diagrams with internal fermion lines carrying the insertions of the
dynamical mass terms corresponding to these condensates. Each involves only
$\xi$ fields, but every loop of these fields requires an even number of
insertions of the dynamical $\xi$ mass term. Each insert requires a
corresponding insert with conjugate fields in order to close the loop, and
this renders these diagrams invariant under the global U(1)$_f$
transformation.

At the lower scales $\Lambda_2$ and $\Lambda_3$, it is the $\zeta$ fields
that condense with each other, producing the lighter ETC gauge boson masses
as well as the ETC gauge boson mixings. But one can see by inspection that
all the condensates that form (\ref{4x6zetacondensate},
\ref{33to3barcondensate}, and \ref{z12z23condensate}) are U(1)$_f$
invariant. Hence, so too are the ETC gauge boson masses and mixings.

The mixings allowed by the U(1)$_f$ global symmetry determine, in
turn, the allowed down-quark, charged-lepton, and neutrino mass
matrices.  Only the diagonal up-type quark masses, the 22, 13 and 31
mass matrix entries for down-type quarks and charged leptons, and the
22 and 13 entries for the Dirac neutrino mass matrix can appear. The
U(1)$_f$ symmetry also forbids the appearance of any nonzero element
of the $2\times2$ right-handed Majorana neutrino mass matrix $r_{ij}$.
This symmetry can also be used to understand which $d_{i,jk}$ and
$w_{ij,k\ell}$ entries are nonzero, but, as discussed above, the
nonzero $d_{i,jk}$ and $w_{ij,k\ell}$ do not play an important role in
determining the neutrino mass eigenvalues. .

The full theory with the $\omega^\alpha_{p,R}$ fermions produces the two
condensates (\ref{z12omegacondensate} and \ref{z23omegacondensate}) which
are formed with the $\zeta$ fields. These condensates are not invariant
under $U(1)_f$.  However, the loop diagrams responsible for the ETC gauge
boson mixings require the insertion of an even number of dynamical masses
corresponding to these additional condensates, half of which are complex
conjugates of the others. Thus, the charge under $U(1)_f$ of such diagrams
is always twice the charge of the single condensate involved, and therefore
a residual discrete ${\mathbb Z}_2$ symmetry, generated by $e^{i \pi T_f}$,
remains. This symmetry forbids the 12, 21, 23, and 32 entries of quark and
charged lepton mass matrices, and the 12, 23, and 32 entries of the Dirac
neutrino mass matrix. But now the 13 and 31 entries of the up-type quarks,
the 11 and 33 entries of the down-type quarks and charged leptons, and the
33 entry of the Dirac neutrino mass matrix are nonzero. This is a
consequence of the fact that the mixings $V^1_t \leftrightarrow V^3_t$ with
$t=4,5$ and $V^1_4 \leftrightarrow V_1^5$ and $V^3_4 \leftrightarrow V_3^5$
(plus the analogous mixings with the indices 4 and 5 interchanged) are
generated, which lead, in turn, to the mixings $V_i^4 \leftrightarrow V^i_5$
and $V_i^5 \leftrightarrow V^i_4 $with $i=1$ and $i=3$.

Similarly, from the structure of the condensate at the scale $\Lambda_1$, in
conjunction with the condensates at the lower scales $\Lambda_2$ and
$\Lambda_3$, it follows that the entry $r_{23}=r_{32}$ of the (symmetric)
$2\times 2$ Majorana neutrino mass matrix is nonzero, while the diagonal
elements $r_{22}$ and $r_{33}$ vanish.  But as discussed above, these
elements are far too small to drive the seesaw mechanism. Finally, the
presence of the $\omega^\alpha_{p,R}$ fermions and their condensates allows
a few more of the (less important) entries shown in Eqs. \ref{MDnuxi} and
\ref{wmatrix} to be non-vanishing.

\subsection{Symmetry-Breaking Sequence 2}

We turn next to another symmetry breaking sequence which becomes plausible
depending on the relative strengths of the ETC and HC couplings. It is
related to the sequence denoted $G_b$ in Ref. \cite{nt}, but has further
structure resulting from the inclusion here of the ETC-singlet
$\omega^\alpha_{p,R}$ fermions. Like sequence 1, it does not lead to
completely realistic masses and mixing angles, but it has some attractive
features not present there.

At the first stage of breaking, $\Lambda_1$, the SU(5)$_{ETC}$ breaks to
SU(4)$_{ETC}$ just as in sequence 1, driven by condensation of the $\xi$
fields. As the energy decreases below $\Lambda_1$, the SU(4)$_{ETC}$ and
SU(2)$_{HC}$ couplings increase, and now, at a scale $\Lambda_{BHC} \lsim
\Lambda_1$ (BHC = broken HC), the $SU(4)_{ETC}$ interaction produces a
condensation in the channel
\beq (6,2,1,1)_{0,R} \times (6,2,1,1)_{0,R} \to (1,3,1,1)_0.
\label{661channel} \eeq
With respect to SU(4)$_{ETC}$, this channel has $\Delta C_2 = 5$ and is
hence slightly more attractive than the initial condensation
(\ref{xixicondensate}) with $\Delta C_2 = 24/5$ for the SU(5)$_{ETC}$
interaction, but it can occur at the somewhat lower scale $\Lambda_{BHC}$
because it is repulsive with respect to hypercolor ($\Delta C_2 = -1/4$).
This requires the HC coupling to be strong, but not so strong as to combine
with the ETC interaction to produce the sequence 1 condensation Eq.
(\ref{4x6channel}).

With no loss of generality, one can orient the SU(2)$_{HC}$ axes so that the
condensate is
\beq \langle \epsilon_{1 i j k \ell} \zeta^{ij,1 \ T}_R C \zeta^{k \ell,2}_R
\rangle + (1 \leftrightarrow 2) \ . \label{6x6condensate} \eeq
Since this is an adjoint representation of hypercolor, it breaks ${\rm
SU}(2)_{HC} \to {\rm U}(1)_{HC}$. We let $\alpha=1,2$ correspond to the
charges $Q_{HC}=\pm 1$ under the U(1)$_{HC}$.  This gives dynamical masses
$\sim \Lambda_{BHC}$ to the twelve $\zeta^{ij,\alpha}_R$ fields involved.

At a lower scale, $\Lambda_{23}$, a combination of the SU(4)$_{ETC}$ and
U(1)$_{HC}$ attractive interactions produces the condensation
\beq 4 \times 4 \to 6 \label{44to6channel} \eeq
with $\Delta C_2 = 5/4$ for the SU(4)$_{ETC}$ interaction and associated
condensate
\beq \langle \epsilon_{\alpha \beta} \zeta^{12,\alpha \ T}_R C
\zeta^{13,\beta}_R \rangle \ .
\label{44to6condensate} \eeq
The contraction with $\epsilon_{\alpha \beta}$ is included not to yield an
SU(2)$_{HC}$-invariant product, since this symmetry has already been broken,
but instead to yield the requisite antisymmetric product SU(4)$_{ETC}$ $(4
\times 4)_{antisym.} = 6$. The condensate (\ref{44to6condensate}) is
U(1)$_{HC}$-invariant and breaks ${\rm SU}(4)_{ETC}$ to the direct product
of ${\rm SU}(2)_{TC}$ and a local U(1)$^\prime$ symmetry generated by
$U(\theta) = e^{i\theta T_d}$, where $T_d$ is a certain linear combination
of diagonal SU(5)$_{ETC}$ generators, $T_d = (1/2){\rm diag}(0,1,-1,0,0)$.
(Here, we use the standard notation ${\rm diag}(a_1,...,a_n)$ for the matrix
with the given elements on the diagonal and other elements zero.)

The strong U(1)$_{HC}$ interaction can also produce a number of additional
condensates between fermions with opposite U(1)$_{HC}$ charge (i.e.
different values of the HC index $\alpha=1,2$).  Since the U(1)$_{HC}$
interaction is not asymptotically free, this should occur at the scale
$\Lambda_{23}$. The condensates include eight involving $\zeta$ and $\omega$
fermions,
\beq \langle \zeta^{1i,\alpha \ T}_R C \omega^\beta_{p,R} \rangle \ , \quad
i=2,3 \ ; \quad p=1,2 \ ; \quad \alpha \ne \beta
\label{z12omegacondensate_gbp} \eeq
and four that involve only $\omega$ fields,
\beq \langle \omega^{\alpha \ T}_{p,R} C \omega^\beta_{p^\prime,R} \rangle \
, \quad 1 \le p,p^\prime \le 2 \ ; \alpha \ne \beta \ .
\label{omega_selfcondensate_gbp} \eeq
Each of the condensates (\ref{z12omegacondensate_gbp}) breaks the local
U(1)$^\prime$ symmetry, leaving a residual SU(2)$_{TC} \times$ U(1)$_{HC}
\times G_{SM}$ continuous local symmetry.

Since this model involves only two ETC symmetry-breaking scales,
$\Lambda_1$, and $\Lambda_{23}$, it does not obviously lead to a full
hierarchy of three fermion generations.  It is, nevertheless, of interest
because of its different quark and lepton mixings and because it exhibits
the mechanism proposed in Ref. \cite{nt} for the origin of light neutrino
masses.

The technifermion condensation at the lower scale $\Lambda_{TC}$ is similar
to that in sequence 1, with the difference that since the technidoublet
$\zeta^{1i,\alpha}_R$, $i=4,5$ is not involved in any higher condensations,
there are 18 rather than 16 chiral SU(2)$_{TC}$ doublets present at
$\Lambda_{TC}$, viz., $u^{i,a}_\chi$, $d^{i,a}_\chi$, $e^i_\chi$ for
$\chi=L,R$, $n^i_L$, $\alpha^{1i}_R$ and $\zeta^{1i,\alpha}_R$ with $i=4,5$,
$\alpha=1,2$, and one has the additional technicondensate
\beq \langle \epsilon_{123jk} \epsilon_{\alpha\beta} \zeta^{1j,\alpha \ T}_R
C \zeta^{1k,\beta}_R \rangle = 2\langle \epsilon_{\alpha\beta}
\zeta^{14,\alpha \ T}_R C \zeta^{15,\beta}_R \rangle \ .
\label{z14z15_gbp} \eeq
Note that the presence of the $\epsilon_{\alpha\beta}$ is not due to the
SU(2)$_{HC}$ symmetry, which is broken, but rather represents the automatic
antisymmetrization of the operator product in Eq. (\ref{z14z15_gbp}) on the
indices $\alpha$ and $\beta$ due to the contraction with $\epsilon_{123jk}$
and
Fermi statistics.

The fact that the hypercolor-doublet technicolor-doublet fermions
$(\zeta^{14,\alpha}_R, \zeta^{15,\alpha}_R)$ do not form bilinear
technicondensates with hypercolor-singlet technicolor-doublet fermions
follows from the exact U(1)$_{HC}$ gauge symmetry.  The presence of these
two additional chiral technidoublets means that, when the technicolor theory
is expressed in vector-like form (as it always can be, since SU(2) has only
(pseudo)real representations), there are $N_f=9$ rather than $N_f=8$
vectorially coupled technifermions transforming according to the fundamental
representation.  Because of the strong-coupling nature of the TC theory, we
consider that this fermion content is again consistent with the assumed
existence of an approximate infrared fixed point in the confined phase, with
associated walking behavior up to the lowest ETC scale \cite{vals}.

\subsection{Quark and Charged Lepton Mass Matrices for Sequence 2}

The fermion mass matrices for sequence 2 can be analyzed in the same way as
for sequence 1.  The zero and nonzero entries are summarized in Table
\ref{matrix_elements}.  The diagonal entries of the up-quark mass matrix are
given by the general formula (\ref{muii}). But now, with only the two
ETC-breaking scales $\Lambda_{1}$ and $\Lambda_{23}$, we have
\beq M^{(u)}_{11} \simeq \frac{\kappa \Lambda_{TC}^{2}\Lambda_{23}}
 { \Lambda_{1}^2} \ , \quad  M^{(u)}_{22}= M^{(u)}_{33} \simeq
\frac{\kappa \Lambda_{TC}^2}{ \Lambda_{23}}~, \label{muii_gbp} \eeq
where we recall $\kappa \simeq 8 \pi/3$.

 Off-diagonal elements of $M^{(u)}_{ij}$ are generated for the case
$ij=23, 32$, by the ETC gauge boson mixings of the form $V^2_t
\leftrightarrow V^3_t$ with $t=4,5$ (see appendix).  The scale associated
with this mixing is $\Lambda_{23}$, and therefore the off-diagonal terms are
of the same order as the $ij=22$ and $33$ terms.  The estimate is
$M^{(u)}_{23}=M^{(u)}_{32} = \rho_u M^{(u)}_{22}$ with $\rho_u \simeq O(1)$.
Thus, $M^{(u)}$ takes the form
\beq M^{(u)} = \frac{\kappa \Lambda_{TC}^2}{\Lambda_{23}}
             \left( \begin{array}{ccc}
             \Lambda_{23}^2/\Lambda_1^2 &  0  &  0  \\
                 0  & 1     &  \rho_u    \\
                 0  & \rho_u  &  1   \end{array} \right ) \ .
\label{Mu_gap} \eeq

For $\Lambda_1 = 10^3$ TeV, this gives $m_u=3$ MeV as in sequence 1. For
$m_c$ and $m_t$, the off-diagonal elements $M_{23}$ and $M_{32}$ play an
important role. These masses are
\beq m_t, \ m_c = \frac{\kappa (1 \pm \rho_u)\Lambda_{TC}^2}{\Lambda_{23}}
\label{mt_gbp} \eeq
These are implicitly defined at the same momentum scale, say $m_t$. If
we take $\Lambda_{23} \simeq 4$ TeV, then $m_t$ can be fit as in
sequence 1. In order to fit $m_c$ one would need to have $\rho_u$
quite close to unity, namely $\rho_u = [1-(m_c/m_t)]/[1+(m_c/m_t)]
\simeq 1 - 2(m_c/m_t)$, so some fine tuning would seem to be required.
Still, it is interesting that sequence 2 can generate three mass
scales for the up-type quark masses with only two ETC breaking scales,
$\Lambda_1$ and $\Lambda_{23}$, as a result of the feature that the
lower right-hand $2 \times 2$ submatrix of $M^{(u)}$ has all entries
of comparable magnitude.

The mass matrices for the $Q=-1/3$ quarks and charged leptons arise solely
via ETC gauge boson mixing. The requisite mixings, estimated in the
appendix, are all suppressed by small mass ratios. We obtain
\beq M^{(d,e)}_{11} \simeq \frac{\kappa \Lambda_{TC}^2 \Lambda_{23}^4} {
\Lambda_1^4 \Lambda_{BHC}}, \label{md11_gbp} \eeq
\beq M^{(d,e)}_{22} = M^{(d,e)}_{33} \simeq \frac{\kappa \Lambda_{TC}^4} {
\Lambda_{23}^3}, \label{md22_gbp} \eeq
and
\beq M^{(d,e)}_{23} = M^{(d,e)}_{32} \simeq \frac{\kappa
\Lambda_{TC}^4}{\Lambda_{23}^3}, \label{md23_gbp} \eeq
where, as in sequence 1, the overall numerical factor is simply taken from
Eq. \ref{muii}. Thus, $M^{(d,e)}$ has the form
\beq M^{(d,e)} \simeq \frac{\kappa \Lambda_{TC}^4}{ \Lambda_{23}^3}
             \left( \begin{array}{ccc}
                 x  &  0  &  0  \\
                 0  &  1  &  \rho_{d}    \\
                 0  &  \rho_{d}  &  1   \end{array} \right )
\label{Mde_gbp} \eeq
where $\rho_{d} \simeq O(1)$ and
\beq x \simeq \frac{\Lambda_{23}^7}{\Lambda_{TC}^2
\Lambda_1^4\Lambda_{BHC}} \ . \label{x} \eeq

With $\Lambda_{23} \simeq 4$ TeV, $m_{\tau}$ is suppressed relative to $m_t$
by approximately $\Lambda_{TC}^{2}/\Lambda_{23}^2$, the right order of
magnitude. The fact that the lower right-hand $2 \times 2$ sub-matrix of
$M^{(d,e)}$ has all entries of comparable magnitude means that the correct
value of $m_{\mu}$ could also emerge, but possibly involving some fine
tuning. The same respective values obtain for $m_s$ and $m_b$. Recent
lattice determinations of the current quark mass $m_s$ (i.e., the running
mass evaluated at a scale well above $\Lambda_{QCD}$, say at 2 GeV) yield
values close to $m_\mu$ \cite{mlat}, the prediction of this model that $m_s
\simeq m_\mu$ is a successful one.  However, the corresponding prediction
that
$m_b \simeq m_\tau$ means that $m_b$ is too small by a factor of roughly
$2-3$. The electron mass, $m_e$, and the down-quark mass, $m_d$, are given
by Eq. (\ref{md11_gbp}), and are much too small.

For $f=u,d$, the matrices $M^{(f)}$ can be diagonalized by respective
unitary transformations as in Eq. (\ref{mf_diagonalization}) with $U^{(f)}_R
= U^{(f)}_L$. These unitary transformations have similar forms, and involve
large mixing between the generation $i=2,3$ interaction eigenstates to form
mass eigenstates in each charge subsector.  Hence, in the product
$V=U^{(u)}_L U^{(d) \ \dagger}_L$ which is the observed quark mixing matrix,
one could obtain a result that agrees with the general properties
of $V$ in the subsector of the two higher generations, namely that $V$ is
largely diagonal, with small off-diagonal entries.  These small off-diagonal
entries reflect the slightly different structures of $M^{(u)}$ and $M^{(d)}$
and hence the slightly different forms of $U^{(u)}_L$ and $U^{(d)}$.
However, the quark mixing is not realistic because the $i=1$ generation
interaction eigenstate does not mix with the corresponding $i=2,3$
eigenstates.

\subsection{Neutrino mass matrix for Sequence 2}

For sequence 2, the $b_{ij}$ Dirac neutrino matrix is identical to
$M^{(d,e)}_{ij}$ for the relevant values of indices $i=1,2,3$ and $j=2,3$.
Thus,
\beq b_{12} = b_{13}=0 \label{b1j0} \eeq
\beq b_{23} = b_{32} \simeq \frac{\kappa \Lambda_{TC}^4}{ \Lambda_{23}^3}
\label{b23_gbp} \eeq
\beq b_{22}=b_{33} \simeq \frac{\kappa \Lambda_{TC}^4}{ \Lambda_{23}^3} \ .
\label{b22_gbp} \eeq
For the $2\times 2$ Majorana matrix, $r_{ij}$, we obtain $r_{22}=r_{33}=0$
and, from the evaluation of the graphs in Fig. \ref{alpha-alpha},
\beq r_{23} \simeq \frac{\kappa \Lambda_{BHC}^3\Lambda_{23}^3}{\Lambda_1^5}
\ .
\label{r23_gbp} \eeq
As in the case of sequence 1, the many inverse powers of $\Lambda_1$ emerge
because the fermions and gauge bosons of Fig. \ref{alpha-alpha} have masses
of order $\Lambda_1$, and because the gauge boson mixing (described in the
appendix) is soft above the scales $\Lambda_{BHC}$ and $\Lambda_{23}$.

The Majorana mass $r_{23}$ can be much larger with this sequence than in
sequence 1.  The reason is the presence here of the scale $\Lambda_{BHC}$
which can be as large as $O(\Lambda_{1})$. The ratio of the Dirac matrix
elements and the right-handed Majorana elements is
\beq \frac{b_{23}}{r_{23}} \simeq \frac{\Lambda_{TC}^4 \Lambda_1^5}
{\Lambda_{23}^6 \Lambda_{BHC}^3} \ .  \label{br_ratio} \eeq
This symmetry-breaking sequence can produce a successful seesaw mechanism of
the type discussed in Ref. \cite{nt}, yielding light neutrinos if one uses
$\Lambda_1 \simeq 10^{3}$ TeV as above, $\Lambda_{BHC} \simeq 0.5\Lambda_1$,
and $\Lambda_{23} \simeq 10^2$ TeV.  With these values, the ratio in Eq.
(\ref{br_ratio}) is $<< 1$, the necessary condition for the seesaw of Ref.
\cite{nt}. It can be seen that this is not possible in sequence 1 while
maintaining a hierarchy among the ETC scales. However, even here the
requisite value of $\Lambda_{23}$ is too large to produce a sufficiently
heavy top quark, so that the model does not appear capable of explaining
both light neutrino masses and the top quark mass.

We discuss the neutrinos further, supposing that $\Lambda_{23}$ is allowed
to be much larger than 4 TeV as above. For this purpose, it is sufficient
\cite{nt} to consider the $5 \times 5$ submatrix $M_0$ of $M_{HCS}$ that
involves only $(\nu_e^c,\nu_\mu^c,\nu_\tau^c,\alpha^{12},\alpha^{13})_R$.
Other fermions have gained larger masses. Diagonalizing $M_0$, we find that
the model yields the following neutrino masses.  Here the eigenvalue
corresponding to $m(\nu_3)$ is negative, so we absorb this minus sign in an
appropriate redefinition of the neutrino fields.
\beq m(\nu_3) \simeq \frac{(b_{23}+b_{22})^2}{r_{23}} + O \left (
\frac{b_{ij}^4}{r_{23}^3} \right )  \simeq \frac{\kappa (1+y)^2
\Lambda_{TC}^8 \Lambda_1^5}
     { \Lambda_{23}^9 \Lambda_{BHC}^3}
\label{mnu3_gbp} \eeq
\beq m(\nu_2) \simeq \frac{(b_{23}-b_{22})^2}{r_{23}} + O \left (
\frac{b_{ij}^4}{r_{23}^3} \right ) \simeq \frac{\kappa (1-y)^2
\Lambda_{TC}^8 \Lambda_1^5}
     { \Lambda_{23}^9 \Lambda_{BHC}^3}
\label{mnu2_gbp} \eeq
where $y=b_{22}/b_{23} \simeq O(1)$, and
\beq m(\nu_e)=m(\nu_1)=0 \ . \label{mnue} \eeq

Since $y$ is positive and $O(1)$, sequence 2 gives the normal hierarchy
$m(\nu_3) > m(\nu_2)$. Thus, one can extract the value of $m(\nu_3)$ from
the measured value of $|\Delta m^2_{32}|$ \cite{atm,k2k}, namely $m(\nu_3)
\simeq \sqrt{|\Delta m^2_{32}|} \simeq 0.05$ eV, and $m(\nu_2)$ from the
measured value of $\Delta m^2_{21}$ \cite{sol,kamland}, namely $m(\nu_2)
\simeq \sqrt{\Delta m^2_{21}} \simeq 0.008$ eV. With $\Lambda_1 = 10^3$ TeV,
$\Lambda_{BHC} = 0.5\Lambda_1$, and $\Lambda_{23} = 10^2$ TeV, these
experimental values are fit with the choice $y \simeq 0.4$.

The non-vanishing Dirac masses, $b_{ij}$, are then $O(10-100)$ KeV and the
right-handed Majorana masses, $r_{23}$, are or order $O(10^2)$ GeV. The two
large eigenvalues of $M_0$ are
\beq \lambda_{h_1} = r_{23} + \frac{(b_{23}+b_{22})^2}{r_{23}} + O \left (
\frac{b_{ij}^4}{r_{23}^3} \right ) \label{mnuh1} \eeq
and
\beq \lambda_{h_2} = -r_{23} - \frac{(b_{23}-b_{22})^2}{r_{23}} + O \left (
\frac{b_{ij}^4}{r_{23}^3} \right )~. \label{mnuh2} \eeq
To leading order, the resultant mass eigenstates involve maximal mixing (i.e.,
with an angle of $\pm \pi/4$) of $|\alpha^{12}_R\rangle$ and $|\alpha^{13}_R
\rangle$.  We label the corresponding neutrino-like states as $|\nu_{h_j}
\rangle$ with masses $m(\nu_{h_1})=\lambda_{h_1}$ and, after appropriate
rephasing, $m(\nu_{h_2})=-\lambda_{h_2}$.

Experimentally, there is evidence for large lepton mixing
\cite{atm}-\cite{theta13}.  Because of the dominant off-diagonal structure of
$r_{ij}$ and the fact that $b_{ij}$ has a lower $2 \times 2$ submatrix with
comparable-size entries for $ij=22,23,32,33$, there is naturally large mixing
between the second and third generations of neutrino mass eigenstates to form
the interaction eigenstates $\nu_\mu$ and $\nu_\tau$. Indeed, the $2 \times 2$
submatrix in $U_\nu$ (the matrix that diagonalizes the full neutrino mass
matrix), acting on $(\nu_\mu,\nu_\tau)$ is the rotation $R(\theta_{\nu,23})$
with $\theta_{\nu,23}=\pi/4$, up to small corrections. The observed lepton
mixing matrix defined by Eq. (\ref{u}) involves the product of the relevant
terms from $U_\nu$ with those from the matrix $U^{(e)}_L$ that is involved in
the diagonalization of the charged lepton mass matrix.  Since the analogous $2
\times 2$ submatrix of $U^{(e)}$ acting on the $(\mu,\tau)$ subsector is also
of the form $R(\theta_{e,23})$ with $\theta_{e,23}=-\pi/4$ up to small
corrections (owing to the relation $M^{(e)}_{22} = M^{(e)}_{33}$), these
maximal mixings cancel, and the resultant observed lepton mixing matrix $U$
with this sequence does not exhibit maximal mixing.

In summary, the model with the symmetry-breaking sequence 2 can exhibit the
mechanism proposed in Ref. \cite{nt} for explaining light neutrino masses in
theories with dynamical electroweak symmetry breaking.    The sequence has
the defect that $m_e$ and $m_d$ are too small, and it appears difficult to
obtain satisfactory simultaneous fits to $m_t$ (and $m_{\tau}$) and to the
neutrino masses.

We note again that the characteristic polynomial, $P(\lambda)$, of the full
$39 \times 39$ mass matrix contains 34 roots of magnitude $\Lambda_{TC}$ or
larger. The remaining five roots, of primary interest, are those for
$m(\nu_i)$, $i=1,2,3$ and for $m(\nu_{h_j})$ discussed above. These,
together with two roots approximately equal to $\pm \Lambda_1$, occur as the
roots of the factor $P_{HCS-TCS}(\lambda)$ for the seven HC-singlet,
TC-singlet fermions. For the present sequence,
\beq P_{HCS-TCS}(\lambda) =
\lambda(\lambda^2-\Lambda_1^2-d_{1,45}^2)P_4(\lambda) \label{p_hcs-tcs_gbp}
\eeq
where
\beqs & & P_4(\lambda) = \lambda^4 -
(b_{22}^2+b_{23}^2+b_{32}^2+b_{33}^2+r_{23}^2)\lambda^2 \cr\cr & &
-2r_{23}(b_{22}b_{23} + b_{32}b_{33})\lambda + (b_{22}b_{33}-b_{23}b_{32})^2
\ .
\label{p4_gbp} \eeqs
As was true for sequence 1, one sees that $d_{1,45}$ has a negligible effect on
the eigenvalues, since it enters in the combination $\Lambda_1^2+d_{1,45}^2$
with the much larger quantity $\Lambda_1^2$.  Other $d_{i,jk}$'s and
$w_{ij,k\ell}$ terms are zero.

\subsection{Residual Generational Symmetries in Sequence 2}

As with sequence 1, we can understand the zeros in the various mass
matrices from the point of view of residual global generational
symmetries and selection rules governing the breaking of these
symmetries by the condensates. Again, we first consider the truncated
model without the $\omega^\alpha_{p,R}$ fields. In this case, the
condensates involving the $\zeta$ fields are invariant under a local
U(1)$_f$ symmetry generated by $e^{i\theta(x) T_f^\prime}$ where
$T_f^\prime$ is a different combination of diagonal generators of
SU(5)$_{ETC}$:
\beq T_f^\prime = \frac{1}{2}{\rm diag}(0,1,-1,0,0) \ . \label{tfprime} \eeq
Proceeding to the full model with the $\omega^\alpha_{p,R}$ fields, one can
apply the discussion for sequence 1 with obvious changes to conclude that
$M^{(u)}=0$ for $ij=12, 21, 13, 31$; $M^{(d,e)}=0$ for $ij=12, 21, 13, 31$;
$b_{ij}=0$ for $ij = 12,13$; and $r_{22}=r_{33}=0$.  Similar reasoning can
be applied to the $d_{i,jk}$ and $w_{ij,k\ell}$ to conclude that among these
only $d_{1,45}$ is nonzero.

There is a general relation connecting sequences 1 and 2 in insofar as they
involve condensates of the $\zeta$ fields and resultant ETC gauge boson
mixings, namely that these condensates and mixings in sequence 2 are related
to those for sequence 1 by the interchange of the ETC indices 1 and 2
(holding other ETC indices fixed) with appropriate changes in the
condensation scale. This is because, before the inclusion of the
$\omega^\alpha_{p,R}$ fields, the generators of the U(1)$_f$ symmetries in
each case are related by precisely this interchange of the ETC indices. This
interchange also relates the respective $\zeta$ condensates in sequences 1
and 2. Thus, the condensate (\ref{4x6zetacondensate}) at scale $\Lambda_2$
in sequence 1 goes to the condensate (\ref{6x6condensate}) at scale
$\Lambda_{BHC}$ in sequence 2; the condensate (\ref{33to3barcondensate}) at
$\Lambda_3$ in sequence 1 goes to the condensate (\ref{z14z15_gbp}) at
$\Lambda_{TC}$ in sequence 2;  the condensate (\ref{z12z23condensate}) at
$\Lambda_3$ in sequence 1 goes to (\ref{44to6condensate}) at $\Lambda_{23}$
in sequence 2; the condensate (\ref{z12omegacondensate}) at $\Lambda_3$ in
sequence 1 goes to the same condensate, now at $\Lambda_{23}$, in sequence
2; and the condensate (\ref{z23omegacondensate}) at $\Lambda_3$ in sequence
1 goes to the $i=3$ case of (\ref{z12omegacondensate_gbp}) at $\Lambda_{23}$
in sequence 2. This relation between sequences 1 and 2 is evident in the
summary tables \ref{etc_mixings} and \ref{matrix_elements}.

\section{Phenomenology}

While the explicit model presented here is not fully realistic, it
does have several realistic features such as natural intra-family mass
splittings and non-trivial mixing angles. These features suggest that
some ingredients in the general class of models we have described
could appear in a successful theory of fermion masses and mixing. In
this section we discuss some generic experimental constraints and
phenomenological properties of this class of models.

\subsection{Precision Electroweak Constraints}

A natural first step in theories of dynamical electroweak symmetry breaking is
to focus on corrections to the $W$ and $Z$ propagators, in particular the $S$
and $T$ parameters \cite{pt,pdg}.  Since the class of models considered here
produces large intra-generational mass splittings, it is important to evaluate
the contributions to $\Delta \rho = \alpha_{em} T$ from ETC interactions.
These contributions arise because these ETC interactions (unlike TC) do not
respect a custodial, SU(2)$_R$ symmetry and lead, for example, to $m_t >> m_b$,
as a consequence of the representation assignments
(\ref{conjugate_representation}) and (\ref{same_representation}).  The dominant
contribution to $\Delta \rho$ may be estimated by concentrating on the lowest
ETC scale, where $m_t$ and $m_b$ are generated. There will be, generically,
contributions to $\Delta \rho$ arising from corrections to the basic
technicolor mass generation mechanism for the $W$ and $Z$, due to the emission
and reabsorption of single ETC gauge bosons of this lightest mass scale.

These contributions to $\Delta \rho$ can be roughly estimated by recalling
that the momentum scale of the technicolor mass generation mechanism
is set by $\Lambda_{TC} \simeq 300$ GeV, and the emission and reabsorption
of an ETC gauge boson will lead to a denominator factor of $1/ \Lambda_l^2$,
where $\Lambda_l$ is the lightest ETC scale ($= \Lambda_3$ in sequence 1 and
$\Lambda_{23}$ in sequence 2). Note that since the momentum integrals here
are rapidly damped at scales above $\Lambda_{TC}$, there can be no walking
enhancement as in the case of the fermions mass estimates (\ref{muii}).
Drawing on the discussion leading to Eq. (\ref{Mi}) (with $a \simeq O(1)$)
and noting that the factors of $g_{ETC}$ cancel in $\Delta \rho$, we
estimate
\beq \Delta \rho_{ETC} \simeq \frac{2 b \Lambda_{TC}^2}{3 \Lambda_{l}^2} \ ,
\label{delta_rho} \eeq
where $b$ is a factor of order unity reflecting the relative strength of
custodial symmetry breaking at the lowest ETC scale.  With $\Lambda_{TC} \simeq
300$ Gev and $\Lambda_l \simeq 4$ Tev, this gives $\Delta \rho \simeq (3.8
\times 10^{-3})b$ or equivalently, $T \simeq 0.48b$.  In comparing this
estimate with experiment, we first caution that with the inclusion of the
high-statistics, high-precision data from the NuTeV experiment at Fermilab,
studies have found that attempts at global fits to precision electroweak data
using only the corrections represented by the parameters $S$ and $T$ give poor
values of $\chi^2$ per degree of freedom \cite{nutev}.  In the absence of a
satisfactory fit with the $S$ and $T$ parameters, it is difficult to draw a
firm conclusion regarding the comparison of the estimate (\ref{delta_rho}) with
this data.  Constraints obtained from global fits using pre-NuTeV are
summarized in (Fig. 10.3) of Ref. \cite{pdg}. The allowed elliptical regions in
the $S$, $T$ plane are plotted for three illustrative values of SM Higgs mass,
115, 300, and 1000 GeV.  Choosing the $10^3$ GeV value to correspond most
closely to a technicolor theory, one observes that the ellipse has a central
value of about $(S,T) \simeq (-0.1,0.3)$ and a 1$\sigma$ upper boundary that
extends up to about 0.6 in $T$ and, in a correlated manner, about 0.1 in $S$.
Our estimate above of the (E)TC contribution to $T$ of about 0.5 is consistent
with this bound if $b \sim 1$, and would lie closer to the central value for $b
\lsim 1$, which is possible, given the theoretical uncertainties in the
strong-coupling estimate.

Roughly speaking, the $S$ parameter is sensitive to all new physics at the
electroweak/technicolor scale. The smallness of $S$ is an indication, subject
to the uncertainties associated with the strong technicolor interactions, that
the number of degrees of freedom there is small. That was one of the reasons
that we chose to use the technicolor group SU(2)$_{TC}$.  The evaluation of $S$
is difficult in a strongly coupled theory such as (extended) technicolor.
Having chosen the effective Higgs mass of 1000 GeV in order to compare with the
fit to data in Ref. \cite{pdg}, we have already, in effect, included some
strong TC contributions.  In addition to these strong TC contributions, one
must also include the effect of SM-nonsinglet (pseudo)-Nambu-Goldstone
bosons.  It was noted in Ref. \cite{scalc}, that for a theory with walking,
that is, with an approximate IR conformal fixed point, the TC contribution to
$S$ is naturally reduced. This observation is based on an application of the
Weinberg spectral function sum rules and includes the effects of both
technifermions and (pseudo)-Nambu-Goldstone bosons.  It is possible that this
would lead to an $S$ parameter in agreement with current experimental limits.
We assume here that this is the case.

Another quantity to check is the ETC correction to the $Zb \bar b$ vertex and
hence to the comparison of the measured value $R_b = \Gamma(Z \to b \bar b)/
\Gamma(Z \to {\rm hadrons}) = 0.21664 \pm 0.00068$ with the SM prediction $R_b
= 0.21569 \pm 0.00016$ (from the global fit given in the current PDG listings
\cite{pdg}).  There are two main contributions to the corrections to this
vertex, from graphs in which the $Z$ produces (i) a virtual techniquark $D \bar
D$ pair which exchange a $V^t_3$, $t=4,5$, becoming a $b \bar b$ pair, and (ii)
a virtual $b \bar b$ pair which exchange the ETC gauge boson corresponding to
the lightest diagonal SU(5)$_{ETC}$ generator coupling to the third generation,
viz., $T_{15} = {\rm diag}(0,0,-2,1,1)$ in our notation.  The contributions (i)
and (ii) were studied for a conventional ETC theory in Ref. \cite{rbbar} and
were found to enter with opposite sign and hence to tend to cancel each other.
We find that the same is true in our ETC model with conjugate representations.
Because of the opposite-sign nature of these ETC contributions and in view of
the theoretical uncertainty in the calculation due to the strong ETC coupling,
we conclude that the ETC correction to $R_b$ can be consistent with the
experimentally measured value of $R_b$ and its comparison with the SM
prediction.

\subsection{Flavor-Changing Neutral Processes}

In early studies of extended technicolor, it was sensibly assumed that the ETC
theory would generate flavor-changing neutral currents (FCNC), and it was
argued that these led to severe constraints on the models.  The measured rates
for processes such as $K^0 - \bar K^0$ mixing, $K_L \to \mu^+ \mu^-$, and the
upper limits on the branching ratios for decays such as $K^+ \to \pi^+ \mu^\pm
e^\mp$, $K_L \to \mu^\pm e^\mp$, and $\mu \to e \gamma$ led to the conclusion
that the ETC scales had to be very high.  However, since nothing approaching a
realistic ETC theory was written down, the assumed mixing was put in by hand,
typically into a set of four-fermion couplings. Since in the present paper we
have constructed a model in which quark mixing is actually generated by the
dynamics, we can re-examine the question of FCNC operators.  Although the model
is not fully realistic, it can perhaps provide some new insight into this
issue.

\subsubsection{$K^+ \to \pi^+ \mu^\pm e^\mp$}

We first review the situation in a conventional ETC model in which the quarks
and charged leptons of both chiralities transform according to the same
representation of $G_{ETC}$.  We focus on $K^+ \to \pi^+ \mu^\pm e^\mp$ decays
because it is these for which recent experimental limits have been obtained
(the charge-conjugate decay modes $K^- \to \pi^- \mu^\mp e^\pm$ are, up to
small CP-violating effects, equivalent).  At the quark level, the decay $K^+
\to \pi^+ \mu^\pm e^\mp$ corresponds to the process $\bar s \to \bar d \mu^\pm
e^\mp$ with a spectator $u$ quark.  At tree level, only one of these
transitions occurs, namely $\bar s \to \bar d \mu^+ e^-$; this arises via a
diagram in which an $\bar s$ antiquark emits a virtual $V^1_2$ ETC gauge boson,
thereby transforming into a $\bar d$ antiquark, and the $V^1_2$ ETC gauge boson
produces the pair $\mu^+ e^-$.  (For the $\bar s \to \bar d \mu^- e^+$ process
one would need the ETC gauge boson mixing $V^1_2 \leftrightarrow V^2_1$.)  From
the lowest-order graph, one obtains the quark-level amplitude
\beq
Amp(\bar s \to \bar d \mu^+ e^-) \simeq
 \left (\frac{g_{_{ETC}}}{\sqrt{2}}\right )^2
[\bar s \gamma_\lambda d] \frac{1}{M_1^2} [\bar e \gamma^\lambda \mu] \ .
\label{ampkpme} \eeq
Since $g_{ETC} \simeq O(1)$ at the scale $M_1$, this is only an approximate
estimate of the amplitude.  Normalizing to $K^+_{\mu3}$ decay and taking
into
account that $ \langle \pi^+ | U_+ | K^+ \rangle = \sqrt{2} \langle \pi^0 |
V_-
| K^+ \rangle$, where $T_\pm$, $U_\pm$, and $V_\pm$ denote the usual
flavor-SU(3) shift operators, we have
\beq \frac{\Gamma(K^+ \to \pi^+ \mu^+ e^-)}{\Gamma(K^+ \to \pi^0 \mu^+
\nu_\mu)} = \frac{16}{|V_{us}|^2} \left (\frac{g_{_{ETC}}}{g_2} \right )^4
\left ( \frac{m_W}{M_1} \right )^4
\label{kpme_ratio} \eeq
where $g_{_{ETC}}$ denotes the ETC gauge coupling at the scale $\Lambda_1$, and
$g_2$ denotes the weak SU(2)$_L$ gauge coupling (at the scale $m_W$, given by
$g_2 = e/\sin\theta_W = 0.65$). With $|V_{us}| = 0.22$, this yields the lower
limit
\beq M_1 \gsim (1.6 \times 10^3 \ {\rm TeV})\left (\frac{g_{_{ETC}}}{2\pi}
\right ) \biggl [ \frac{10^{-12}}{B(K^+ \to \pi^+ \mu^+ e^- )} \biggr
]^{1/4} \ . \label{kpmelimit} \eeq
The current upper limit is $BR(K^+ \to \pi^+ \mu^+ e^-) < 2.9 \times
10^{-11}$
\cite{e865} from the Yale-BNL experiment E865.  Using the published limit
and
the relation $M_i = (g_{_{ETC}}/4)a\Lambda_i$, we get the rough lower bound
\beq \Lambda_1
 \gsim  \frac{1}{a}  O(450) \quad {\rm TeV},
\label{lam1_limit} \eeq
where $a$ is expected to be $O(1)$.

The calculation is different in the class of models considered in this paper,
because the left- and right-handed down-type quarks transform according to
conjugate representations of $G_{ETC}$.  In turn, the structure of the
amplitudes is different for models of DEC and DES type.  We focus here on the
DEC-type model discussed in the text.  At the quark level the process $\bar s
\to \bar d \mu^+ e^-$ arises from two contributions (at the leading, tree
level): (i) a $\bar s_L$ makes a transition to the $\bar d_L$ and emits a
virtual $V^1_2$, which can only couple (without ETC gauge boson mixing) to the
current $\bar e_R \gamma^\lambda \mu_R$, and (ii) a $\bar s_R$ makes a
transition to $\bar d_R$, emitting a $V^2_1$, which couples to the current
$\bar e_L \gamma^\lambda \mu_L$.  Hence, the leading contributions to the
amplitude at the quark level is
\beq
Amp(\bar s \to \bar d \mu^+ e^-) = \frac{g_{_{ETC}}^2}{2 M_1^2} \biggl [
  [\bar s_L \gamma_\lambda d_L] [\bar e_R \gamma^\lambda \mu_R]
+ [\bar s_R \gamma_\lambda d_R] [\bar e_L \gamma^\lambda \mu_L] \biggr ]
\label{amp_sbar_to_dbar_mupemin_dec}
\eeq
This should be contrasted with Eq. (\ref{ampkpme}) for a conventional ETC
model.  Another difference is the fact that, whereas $\bar s$ does not go to
$\bar d \mu^-e^+$ at tree level in a conventional ETC model, it does in the
present class of models, in particular, a DEC-type model.  One finds
\beq
Amp(\bar s \to \bar d \mu^- e^+) = \frac{g_{_{ETC}}^2}{2 M_1^2} \biggl [
  [\bar s_L \gamma_\lambda d_L] [\bar \mu_L \gamma^\lambda e_L]
+ [\bar s_R \gamma_\lambda d_R] [\bar \mu_R \gamma^\lambda e_R] \biggr ] \ .
\label{amp_sbar_to_dbar_mminep_dec}
\eeq
One obtains a lower limit on $M_1$ and hence $\Lambda_1$ that is
comparable with the limits given above for conventional ETC models.
Our choice $\Lambda_1 = 10^3$ TeV is consistent with these lower
bounds.

\subsubsection{$K^0_L \to \mu^\pm e^\mp$}

In conventional ETC models, the tree-level contribution to the decay $K^0_L \to
\mu^\pm e^\mp$ vanishes because the amplitude picks out the axial-vector part
of the hadronic current, $[\bar s \gamma^\lambda(c_V - c_A\gamma_5)d]$, but
this current is vectorial in these conventional models. However, this is no
longer the case in the type of model considered here, so one does get a
tree-level contribution to this amplitude, mediated by the $V^1_2$ and $V^2_1$
ETC gauge bosons.  The amplitudes for $K_L \to \mu^\pm e^\mp$ can be obtained
in a straightforward manner from the quark amplitudes given above in
Eq. (\ref{amp_sbar_to_dbar_mupemin_dec}) and
(\ref{amp_sbar_to_dbar_mminep_dec}).  The current upper bound on the branching
ratio for this decay, from an experiment at BNL, is \cite{e871_lfv} $BR(K_L \to
\mu^+ e^-) + BR(K_L \to \mu^- e^+) < 4.7 \times 10^{-12}$.  Our choice of
$\Lambda_1 = 10^3$ TeV is consistent with this limit.

\subsubsection{Neutral Meson Systems}

 Because of the transitions $M^0 \leftrightarrow \bar M^0$, where $M=K$, $B_d$,
$B_s$, or $D$, the mass eigenstates of these neutral non-self-conjugate mesons
involve linear combinations of $|M^0\rangle$ and $|\bar M^0\rangle$ with mass
differences between the respective heavier ($h$) and lighter ($\ell$)
eigenstates $m_{M^0_h}-m_{M^0_\ell} \equiv \Delta m_M$ given by $2 Re(\langle
\bar M^0 | {\cal H}_{eff} | M^0 \rangle)$ for the kaon system, where $K_L$ and
$K_S$ have quite different lifetimes, and by $2|\langle \bar M^0 | {\cal
H}_{eff} | M^0 \rangle|$ for the $B$ and $D$ systems, where the heavier and
lighter states have essentially the same lifetimes. The smallness of the
$K_L-K_S$ mass difference provided early evidence that the weak neutral current
should be diagonal.  Experimentally~\cite{pdg}:
\beqs \Delta m_K & = & (0.530 \pm 0.001) \times 10^{10} \ {\rm s}^{-1} = 
(3.49 \pm 0.006) \times 10^{-12} \ {\rm MeV}\\ 
\Delta m_{B_d} & = & (0.489 \pm 0.008) \times 10^{12} \ {\rm s}^{-1} = 
(3.27 \pm 0.05) \times 10^{-10} \ {\rm MeV}\\ 
\Delta m_{B_s} & > & 13 \times 10^{12} \ {\rm s}^{-1} = 0.89 \times
10^{-8} \ {\rm MeV} \quad (95 \ \% \ {\rm CL}) \\ 
\Delta m_{D} & < & 7 \times 10^{10} \ {\rm s}^{-1} =
0.5 \times 10^{-10} \ {\rm MeV}  \quad (95 \ \% \ {\rm CL})  .  
\label{dm_values}
\eeqs
The standard model accounts for the two measured mass differences and
agrees with the limits on the other two \cite{pdg,parodi}.  This
thereby places constraints on non-SM contributions such as those from
ETC gauge boson exchanges.

Early studies of constraints on conventional ETC theories,
in which
the SM fermions of both chiralities transform according to the same ETC
representations, from $K^0 - \bar K^0$ mixing amplitude assumed an ETC contribution of strength $g_{_{ETC}}^2/(8
M_{ETC}^2) \simeq 1/\Lambda_{ETC}^2$, as would arise if an $d \bar s$ pair
would annihilate to form a virtual ETC gauge boson in the $s$-channel, which
would then produce a $s \bar d$ pair, and similarly for $t$-channel ETC gauge
boson exchange.  Using this assumption, these studies obtained lower bounds of
order $\Lambda_{ETC} \gsim 10^3$ TeV.  However, the virtual $V^1_2$ ETC gauge
boson produced by the $d \bar s$ pair cannot produce a $s \bar d$ pair in these
theories; rather, one must have the mixing $V^1_2 \leftrightarrow V^2_1$ to do
this.  (This type of ETC gauge boson mixing is also necessary for box diagram
contributions involving exchange of two ETC gauge bosons.)  This ETC gauge
boson mixing suppresses the transition substantially, by a factor of order
${}^2_1 \Pi^1_2(0)/\Lambda_1^2 << 1$ (see appendix).  Hence, the lower bound on
$\Lambda_1$ from $K^0 - \bar K^0$ mixing in conventional ETC theories  is less stringent than $10^3$ TeV.  (As discussed above, a
bound of this order does hold in these conventional ETC theories, but it is due
to the experimental upper bound on $BR(K^+ \to \pi^+ \mu^+ e^-)$.) The same
comment about the necessity of ETC gauge boson mixing applies for the other
$M^0 - \bar M^0$ transitions.

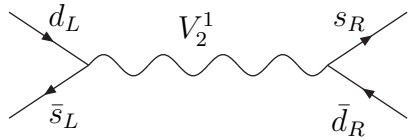
\begin{center}
\begin{picture}(160,100)(0,0)
\ArrowLine(40,30)(10,10) \ArrowLine(10,50)(40,30)
\Photon(40,30)(130,30){4}{4} \ArrowLine(130,30)(160,50)
\ArrowLine(160,10)(130,30) \Text(81,50)[t]{$V^1_2$} \Text(25,45)[lb]{$d_L$}
\Text(25,15)[lt]{$\bar s_L$} \Text(145,45)[rb]{$s_R$}
\Text(145,15)[rt]{$\bar d_R$}
\end{picture}
\end{center}

\begin{figure}
\caption{\footnotesize{Graph that contributes in the $s$-channel to $K^0 - \bar
K^0$ transition amplitude in the present type of ETC theory, where the left-
and right-handed components of the $Q=-1/3$ quarks transform according to
conjugate representations of $G_{ETC}$.  The analogous graph with chiralities
$L$ and $R$ interchanged and $V^1_2$ replaced by $V^2_1$ also contributes to
this transition. These do not require any $V^1_2 \leftrightarrow V^2_1$ ETC
gauge boson mixing.  The corresponding $t$-channel graphs also contribute.}}
\label{kkbar_here}
\end{figure}

 The situation concerning these transitions is different in our class of ETC
models.  Here there are diagrams that can contribute at tree level, without any
ETC gauge boson mixing.  These are the same for both DEC and DES-type
models. An example for $K^0 - \bar K^0$ mixing is shown in Fig.
\ref{kkbar_here}.  One finds
\beq
\Delta m_K \simeq 2 \frac{(g_{_{ETC}}/\sqrt{2})^2}{M_1^2}
Re(\langle \bar K^0 | [\bar s_L \gamma_\lambda d_L]
[\bar s_R \gamma^\lambda d_R] | K^0 \rangle) \ . 
\label{kkb_amp_here}
\eeq
A similar formula applies for $B^0_d - \bar B^0_d$ mixing, with the change
noted above and the replacement $s \to b$, and similarly for $B^0_s - \bar
B^0_s$ mixing, with the replacements $s \to b$, $d \to s$, and $M_1 \to M_2$.
These estimates involve theoretical uncertainty owing to the strong ETC
coupling and the fact that the hadronic matrix elements are not measured and
must be estimated, e.g. by lattice gauge theory simulations; in addition, there
is a question of the degree of short-distance dominance for the $K^0 - \bar
K^0$ amplitude. We find that with the values of $\Lambda_1 \sim 1000$ TeV used
above, our model predicts values of $\Delta m_K$ and $\Delta m_{B_d}$ larger
than the measured values.  For example, using values of $f_B$ and hadronic
matrix elements from lattice measurements \cite{lat2002}, we estimate $\Delta
m_{B_d} \simeq 8 \times 10^{-10}$ MeV. This is a potential problem for the
class of ETC models considered in this paper.  It can be ameliorated if the
model has somewhat larger values of $\Lambda_1$.  In addition, with our choice
of $\Lambda_2 \sim 50$ TeV, we find $\Delta m_{B_s} \sim 3 \times 10^{-7}$ MeV,
a value well above the SM prediction.  The measurement of this quantity will
clearly be very important for the class of models considered here.

Our predictions could be reduced also in models that have the
property of walking not only from $\Lambda_{TC}$ to the lowest ETC
scale, but to higher energy scales such as was assumed in
Ref. \cite{nt}.  This could make possible higher ETC breaking scales
and thereby suppress these flavor-changing neutral current effects
while maintaining reasonable agreement with experimental values of SM
fermion masses.

By contrast, the contribution of our model to $D^0 - \bar D^0$
mixing is quite small, because of the necessity of ETC gauge boson mixing
(since the up-type quarks $u_L$ and $u_R$ are in the same ETC representations).
The main graphs contributing to this transition include an $s$-channel diagram
in which the $c \bar u$ pair annihilates to form a $V^2_1$ gauge boson, which
must go through a ${}^1_2 \Pi^2_1(0)$ to yield a $V^1_2$ which then produces
the final-state $u \bar c$ pair, and the corresponding $t$-channel graph.  This
yields a resulting value of $\Delta m_D$ which is negligibly small compared
with the current upper limit on this mass difference. 

 We remind the reader that we are not considering CP-violating phases in this
paper. Furthermore, in the specific model examined in this paper, with either
symmetry breaking sequence, there is not sufficient generatonal mixing to
generate CKM CP violation.  Assuming that the class of models being considered
can lead to realistic mixing and CP violation (both dynamical in origin), there
is good reason to expect that the CP violation associated with
non-standard-model physics will be suppressed by mixing effects among both the
fermions and the ETC gauge bosons. If this is the case, then the ETC scales we
have used will not lead to unacceptable levels of CP violation in the neutral
kaon system.

Other processes can be analyzed similarly.  For example, the contribution to
$b \to s \gamma$ is safely smaller than the value of this process inferred from
experiment because of the necessity of ETC gauge boson mixing, which results
because the relevant operator $\propto [\bar s \sigma_{\mu\nu} b] F^{\mu\nu}$
flips chirality (where $F^{\mu\nu}$ denotes the electromagnetic field strength
tensor).

\subsubsection{Non-diagonal Neutrino Neutral Currents}

If the leptons of a given charge and chirality have different weak $T$ and
$T_3$, then the neutral leptonic current is non-diagonal in terms of mass
eigenstates \cite{ls}. The class of models considered here includes SM-singlet
neutrino-like states that form Dirac mass terms with the left-handed neutrinos.
It is a convention whether one writes these as right- or left-handed; if one
writes them as left-handed and applies the criteria of Ref. \cite{ls}, it
follows that the neutrino part of the weak neutral current is non-diagonal in
terms of mass eigenstates.  This non-diagonality is a generic feature of this
class of models. The coefficients of terms that are non-diagonal in mass
eigenstates are $\simeq b_{ij}/r_{23}$ and hence are extremely small in a
realistic model that incorporates the seesaw mechanism of Ref. \cite{nt} to
yield light neutrinos.

\subsection{Intermediate-Mass Dominantly Electroweak-Singlet Neutrinos}

\subsubsection{General}

Our mechanism \cite{nt} for getting light neutrinos generically leads to
neutral leptons with masses that are intermediate between the higher ETC scales
and the mass scales characterizing the primary mass eigenstates in the three
observed electroweak-doublet neutrinos.  As linear combinations of interaction
eigenstates, they are composed almost completely of SM-singlet fields, with
only a small admixture of electroweak-doublet neutrinos.  They have masses of
order $O(10^{-1}-10^2)$ GeV, depending on the type of symmetry-breaking
sequence.  The lower end of the mass range for these intermediate-mass
neutrinos is illustrated by the model of Ref. \cite{nt}, while the rest of the
range is illustrated by sequence 2.  (The absence of light neutrino mass
eigenstates that are primarily electroweak-singlets, often called (light)
``sterile'' neutrinos, is in good agreement with indications from neutrino
oscillation experiments \cite{atm,k2k,sol,kamland} ).

The mass eigenstates $(\nu_1,\nu_2,\nu_3,\nu_{h_1},\nu_{h_2})$ mix to form the
interaction eigenstates $\nu_{\ell}$ with $\ell=e,\mu,\tau$ and the SM-singlets
$\alpha^{12}$ and $\alpha^{13}$.  Writing out Eq. (\ref{Unutransformation})
explicitly (dropping the $m$ subscript on the mass eigenstates), we thus have
(in a basis where the charged lepton mixing matrix is diagonal)
\beq
 |\nu_{\ell}\rangle = \sum_{i=1}^3 U_{\ell i}|\nu_i\rangle +
                                   U_{\ell h_1}|\nu_{h_1}\rangle +
                                   U_{\ell h_2}|\nu_{h_2}\rangle
\label{u5}
\eeq
where $U_{e j} \equiv U_{1j}$, $U_{\mu j} \equiv U_{2j}$, etc.  The small
admixture coefficients $U_{\ell h_1}$ and $U_{\ell h_2}$ are of order the ratio
of the Dirac to right-handed Majorana mass entries $b_{ij}/r_{23}$, for the
relevant (dominant) coefficients.  In our model, the diagonalization of the
full neutrino mass matrix leads to the eigenstates $|\nu_{h_j} \rangle$ with
$j=1,2$; these mass eigenstates contain small admixtures of electroweak-doublet
neutrinos, with coefficients of order $b_{ij}/r_{23}$.  As discussed in
\cite{nt}, and exhibited also in our present model with symmetry-breaking
sequence 2, a dominant $b_{ij}$ element has a typical size given by $b_{23}
\simeq M^{(d)}_{23}$, and the relevant entry in $r_{ij}$ is displayed in
Eq. (\ref{r23_gbp}). The ratio $b_{23}/r_{23}$ is given by Eq.
(\ref{br_ratio}) and has a value of order $10^{-7}$.  Hence, generically, one
expects that the coefficients $|U_{\ell j}|$ would be of this size.

\subsubsection{Massive Neutrino Emission Via Mixing in Weak Decays}

At the lower end of the mass range for the intermediate-mass neutrinos, there
are some interesting and testable experimental consequences: the emission, via
lepton mixing, of heavy neutrinos in weak decays, as constrained by the
available phase space. Tests for such heavy neutrino emission via lepton mixing
have been proposed, and data analyzed to set bounds \cite{rs80}.  These tests
were carried out in a number of experiments on $\pi^+ \to \mu^+ \nu_\mu$,
$\pi^+ \to e^+ \nu_e$, $K^+ \to \mu^+ \nu_\mu$ and $K^+ \to e^+ \nu_e$
decays. A decay of this type would consist of one decay of a given charged
pseudoscalar meson $\pi^+$ or $K^+$ into the set of light neutrinos, which
propagate effectively coherently and recoil against an outgoing charged lepton,
and another decay yielding the heavy neutrino mass eigenstate(s) $|\nu_{h_j}
\rangle$ with $j=1,2$, also recoiling against the outgoing charged lepton.  The
signature of the heavy neutrino emission would thus be a peak in the charged
lepton momentum spectrum at an anomalously low value.  Since one can calculate
the branching ratio for the emission of the massive neutrino as a function of
the mixing parameter and neutrino mass, an upper limit on the branching ratio
can be converted into an upper limit on $|U_{\ell h_j}|$ for this assumed
neutrino mass.  Experimental searches in $\pi^+_{e2}$, $K^+_{\mu 2}$, and
$K^+_{e 2}$ decays have yielded upper limits of order $|U_{\ell h_j}|^2 \lsim
10^{-7}$ for $m(\nu_{h_j})$ in the interval 100-300 MeV \cite{pdg}, for $\ell =
e, \mu$.  For the type of model considered here, with heavy neutrinos having
masses above the phase space limit, this constraint is clearly satisfied.  Even
if the heavy neutrinos were as light as a few hundred MeV, as in the type of
model of Ref.  \cite{nt}, one would expect that this squared mixing matrix
element would be of order $10^{-(10 \pm 2)}$, with the theoretical uncertainty
in the exponent indicated.  Hence, the model of Ref. \cite{nt} is in accord
with these limits.

These limits could be improved drawing on more recent experiments. Data on the
decay $K^+ \to \mu^+ \nu_\mu$ that was recorded as an auxiliary part of the
very high-statistics BNL experiments E787 and E949 that observed $K^+ \to \pi^+
\nu \bar\nu$ decay \cite{e787} could be used to search for the emission of a
heavy neutrino down to a branching ratio sensitivity that is estimated to be of
order $10^{-9}$ and perhaps better \cite{llrs}.

At the extreme low end of the range of expected masses of the
intermediate-scale neutrino, it could also be emitted in muon decay, e.g., as
$\mu^+ \to \bar\nu_{h_j} e^+ \nu_e$.  These decays into heavy neutrinos would
be suppressed, relative to the usual decay into light neutrinos, $\mu^+ \to
\bar\nu_\mu e^+ \nu_e$, by the respective leptonic mixing matrix factors
$|U_{\mu h_j}|^2$ and $|U_{e h_j}|^2$, in addition to reduced phase space.  The
observed $e^+$ spectrum would be due to all of the kinematically allowed
decays.  In turn, this would lead to modifications of the observed $\mu$ decay
spectral parameters $\rho$, $\eta$, $\xi$, and $\delta$.  One can use the
agreement of these parameters with the standard model values to set upper
bounds on heavy neutrino emission \cite{rs81b}.  Again, these bounds allow the
current type of intermediate-mass neutrino.

\subsubsection{Non-orthogonality of Observed $|\nu_e\rangle$ and
$|\nu_\mu \rangle$}

Here we mention an effect that is expected to be quite small but is of
conceptual interest. The abstract Hilbert space states $|\nu_e\rangle$ and
$|\nu_\mu \rangle$ are ortho-normal.  However, the experimental definition of
the states $|\nu_e\rangle$ and $|\nu_\mu \rangle$ is as the neutrinos
accompanying the emission of $e^+$ and $\mu^+$ in the pseudoscalar decays $Ps^+
\to e^+ \nu_e$ and $Ps^+ \to \mu^+ \nu_\mu$ and their conjugates, where $Ps^+$
denotes $K^+$ or $\pi^+$. It is via these sources that standard accelerator
neutrino beams are formed \cite{nuc}.  In a theory with heavy neutrinos, these
neutrino states, as experimentally defined, are not, in general, orthogonal
\cite{lpss,ls}.  This follows from the fact that in the linear combinations
(\ref{u5}), the states with sufficiently large masses are kinematically
forbidden from occurring in the above particle or nuclear decays.  For example,
if $\nu_{h_1}$ and $\nu_{h_j}$ are sufficiently heavy that they do not occur in
$\pi^+ \to \ell^+ \nu_\ell$ or $K^+ \to \ell^+ \nu_\ell$ decays, then
\beq \langle \nu_e | \nu_\mu \rangle_{exp.} = U_{1 h_1}^*U_{2 h_1} +
U_{1 h_2}^*U_{2 h_2} \label{nu_nonorthog} \eeq
and similarly, $\langle \nu_e | \nu_\tau \rangle_{exp.}$ and
$\langle \nu_\mu | \nu_\tau \rangle_{exp.}$ would, in general, be nonzero
\cite{imp}.

\subsubsection{ $\Delta L=2$ Decays}

Because the physical intermediate-mass neutrinos are Majorana states, there are
contributions to $|\Delta L| = 2$ decays such as neutrinoless double beta decay
of nuclei and the particle decays $K^+ \to \pi^- \ell^+ \ell^{\prime +}$, where
$\ell,\ell^\prime = ee, e\mu, \mu\mu$.  Searches for neutrinoless double beta
decay of nuclei yield an upper limit, the form of which depends on whether the
mass of the virtual neutrino in the propagator is large or small compared with
the Fermi momentum, $\sim 200$ MeV, of the nucleons in the nucleus.  The
primary mass eigenstates $\nu_i$, $i=1,2,3$, entering into $\nu_e$ produce an
amplitude involving $\langle m_{\nu} \rangle = |\sum_{i=1}^3 U_{e i}^2
m(\nu_i)|$.  In the present model, the intermediate-mass neutrinos $\nu_{h_j}$,
$j=1,2$, will, in general, also contribute to this amplitude, through the
couplings $U_{e h_j}$, $j=1,2$.  If the masses $m(\nu_{h_j})$ are near the
lower end of the expected range, then the resultant $\langle m_{\nu} \rangle =
|\sum_{i=1}^3 U_{e i}^2 m(\nu_i) + \sum_{j=1,2} U_{e h_j}^2 m(\nu_{h_j})|$.
Model-dependent calculations of the relevant nuclear matrix elements are
employed in order to extract an upper limit on $\langle m_\nu \rangle$.
Current searches yield limits of $\langle m_\nu \rangle \lsim O(1)$ eV
\cite{ev}.  At present there are a number of efforts to perform new experiments
to search for neutrinoless double beta decay with sensitivities considerably
below this limit in $\langle m_\nu \rangle$ (e.g., \cite{ev}).  The mechanism
for light neutrinos in Ref.  \cite{nt} provides a further motivation for these
searches.  This can be seen as follows: if $m(\nu_{h_j})$ for $j=1,2$ are near
the lower end of the expected range, say of order 100 MeV and $|U_{e h_j}|$ is
as large as $10^{-5}$, we estimate the resultant $\langle m_{\nu} \rangle
\simeq 10^{-2}$ eV.  While this is in accord with current limits, it might
yield a signal in future searches.  However, with walking only up to the lowest
ETC scale, $\Lambda_3$ or $\Lambda_{23}$ as in the present paper, $m(\nu_{h_j}$
are much heavier.  This alters the form of the neutrino propagator and leads to
a negligible contribution by these intermediate-mass neutrinos to the
amplitude.  Moreover, since our model predicts a normal hierarchy $m(\nu_3) >
m(\nu_2) > m(\nu_1)$ and since $m(\nu_3)$ enters in the amplitude for
neutrinoless double beta decay suppressed by the very small mixing matrix
factor $u_{e3}^2$, one does not expect the primary mass eigenstates $\nu_i$,
$i=1,2,3$ to produce this type of decay at a significant level.

Another possible manifestation of the violation of total lepton number
is in $|\Delta L|=2$ particle decays such as $K^+ \to \pi^- \mu^+\mu^+$;
searches for these were discussed in Refs. \cite{littenshrock,e865}.  The
current limit is \cite{e865} $BR(K^+ \to \pi^- \mu^+\mu^+) < 3.0 \times
10^{-9}$.  The intermediate-mass neutrinos $\nu_{h_j}$, $j=1,2$ would
contribute to this decay.  However, in the present model this contribution is
expected to be negligible compared to the current limit.

\subsubsection{Astrophysical and Cosmological Constraints}

There are also astrophysical and cosmological constraints on a heavy unstable
neutrino with a mass in the range of order 0.1 GeV to 100 GeV.  To discuss
these, one has to calculate the lifetime of the neutrino.  The decay amplitude
involves two types of contributions.  First, there are charged-current
contributions arising from graphs in which the $\nu_{h_j}$, via its admixture
in $\nu_\ell$, for $\ell=e$ and $\mu$, emits a virtual $W^+$, producing an
outgoing $\ell$, with the $W^+$ then producing an $e^+\nu_e$.  In the case of
$\ell=\mu$, this decay would be suppressed by small phase space if
$m(\nu_{h_j})$ were only slightly greater than $m_\mu$ and would be forbidden
if $m(\nu_{h_j}) < m_\mu$. Second, there are contributions arising from
neutral-current processes since, in this type of model with electroweak-singlet
neutrinos the neutral current is non-diagonal \cite{ls}.  These decays are
dominantly of the form $\nu_{h_j} \to \nu_i \bar \nu_\ell \nu_\ell$, where the
$\nu_i$ and $\nu_\ell$ refer to the light neutrino mass eigenstates.  The
contribution of each decay involves a small coefficient reflecting
Eq. (\ref{u5}).  From the charged-current contributions we have
\beq \Gamma_{\nu_{h_j}} = \tau_{\nu_{h_j}}^{-1} =
\left (\frac{m(\nu_{h_j})}{m_\mu} \right )^5 \left (
\sum_\ell |U_{\ell h_j}|^2 \right )\Gamma_\mu \label{gamma_alpha} \eeq
where $\tau_\mu = \Gamma_\mu^{-1} = 2 \times 10^6$ sec. is the muon
lifetime. There are also contributions from the non-diagonal neutral-current
couplings.  Combining these, we estimate, as an example, that for $m(\nu_{h_j})
\simeq 100$ GeV, $\tau_{\nu_{h_1}} \simeq \tau_{\nu_{h_2}} \simeq 10^{-7}$ sec.
Owing to the fifth power dependence, $\Gamma_{\nu_{h_j}}$ varies rapidly as a
function of $m(\nu_{h_j})$, and approaches existing limits when $m(\nu_{h_j})$
decreases to O(100) MeV.

Since the intermediate-mass neutrino decays on a short time scale, it is, of
course, not subject to the mass limits for neutrinos that are stable on the
time scale of the age of the universe and thus affect the mean mass density of
the universe.  The relevant astrophysical and cosmological constraints include
those from (i) primordial (big-bang) nucleosynthesis and (ii) large-scale
structure formation.  A recent analysis of these constraints is given in
\cite{bbn}.  A perusal of these constraints suggests that a neutrino with the
mass of order hundreds of MeV to hundreds of GeV, and the corresponding
lifetime given by (\ref{gamma_alpha}) is allowed.

\subsection{Dark Matter Candidates}

Models incorporating the mechanism of Ref. \cite{nt} for light neutrinos can
yield interesting candidates for dark matter.  Although the model studied in
this paper is not fully realistic, we believe that it does yield some generic
predictions for possible dark matter candidates that are of interest.

A first important point concerns a state which might appear to be an
appealing dark matter candidate in technicolor theories, namely the
technibaryon composed of technineutrinos.  In our model, this would be the
particle state created by the action of the adjoint of the operator
\beq
\frac{1}{\sqrt{2}} \epsilon_{123jk} \bar n^j_L \alpha^{1k}_R
\label{dm_seq2na}
\eeq
on the vacuum.  This has a mass $\simeq 2\Lambda_{TC} \simeq O(600)$ GeV.  If
one neglects ETC gauge boson mixing, one might infer that this state is stable.
However, in our model with either symmetry-breaking pattern, the particle
(\ref{dm_seq2na}) decays via a process in which the $\alpha^{15}$, say, makes a
transition to $\alpha^{12}$, emitting a virtual ETC gauge boson $V^5_2$ which,
via mixing, goes to $V^2_4$; the $V^2_4$ is absorbed by the $\bar n^4$, going
to $\bar n^2 = \bar\nu_\mu$ (where mass eigenstates are understood for the
actual final-state particles).  This decay is kinematically allowed since the
final state has a mass $m(\nu_{h_j})$, which is less than the mass of the
initial state.  (The $\alpha^{12}$, or more precisely, the lighter of the
$\nu_{h_j}$, decays, e.g. to $e^-e^+\nu_e$.)

We next display dark matter candidates for symmetry-breaking sequences 1 and 2.
For sequence 1, we recall that at a scale $\Lambda_s$, which we take to be
$\simeq \Lambda_3$, the hypercolor-invariant condensates
(\ref{z12z23condensate})-(\ref{omega_selfcondensate}) form, due to the strong
hypercolor interaction .  Consequently, linear combinations of the states
$\zeta^{12,\alpha}_R$, $\zeta^{23,\alpha}_R$, and $\omega^\alpha_{p,R}$ with
$p=1,2$ form mass eigenstates, which may be denoted $s^\alpha_{q,R}$ with $1
\le q \le 4$, in order of increasing mass.  The SU(2)$_{HC}$-singlet bound
state of the lightest two of these, formed by the action of the adjoint of the
operator
\beq
\frac{1}{\sqrt{2}} \epsilon_{\alpha\beta} s_{1R}^{\alpha \ T} C s^\beta_{2R}
\label{state12_seq1}
\eeq
on the vacuum, is stable.  The stability can be understood as a consequence of
the fact that this particle is the lightest hypercolor baryon (a boson, since
$N_{HC}$ is even) and SU(2)$_{HC}$ is an exact symmetry.  Hence, the
corresponding particle, with a mass $\simeq 2\Lambda_3 \simeq 8$ TeV, might
provide a dark matter candidate.  Since this technibaryon is composed of
SM-singlet fermions, its interactions with regular matter would be highly
suppressed, since they would proceed dominantly via exchange of ETC gauge
bosons, whose masses are considerably larger than the electroweak scale.  It
would thus be a very weakly coupled WIMP.

With symmetry-breaking sequence 2, the candidate(s) for dark matter arise(s)
differently.  We recall that the fermions $\zeta^{12,\alpha}_R$,
$\zeta^{13,\alpha}_R$, and $\omega^\alpha_{p,R}$ form condensates at the scale
$\Lambda_{23}$, as given by Eqs. (\ref{44to6condensate}),
(\ref{z12omegacondensate_gbp}), and (\ref{omega_selfcondensate_gbp}), with the
difference that these are only invariant under U(1)$_{HC}$, since SU(2)$_{HC}$
has been broken at the higher scale $\Lambda_{BHC}$.  Linear combinations of
these fermions thus gain dynamical masses $\sim \Lambda_{23}$.  However, the
ETC-nonsinglets among these fermions can decay.  For example, a
$\zeta^{12,\alpha}_R$ can make a transition to the $\zeta^{14,\alpha}$, a
member of a lighter mass eigenstate, emitting a virtual $V^2_4$ ETC gauge boson
which then produces fermion-antifermion pairs such as $u^{5,a} \bar u_{2,a} =
U^{5,a} \bar c_a$ ($a=$ color index), etc.  This decay is kinematically allowed
since the initial fermion is part of a mass eigenstate with mass $\sim
\Lambda_{23}$, while the illustrative final state given above has mass $\sim 2
\Lambda_{TC}$, and similarly for other final states.  The same comment applies
to $\zeta^{13,\alpha}_R$.  The $\omega^\alpha_{p,R}$'s with $p=1,2$, which are
part of mass eigenstates with masses of order $2\Lambda_{23} \simeq 8$ TeV, do
not decay in this manner, but can annihilate as $\omega^{+}_{p,R} +
\omega^{-}_{q,R} \to \ge n \gamma_{HC}$ where $n \ge 2$, we have used our
labelling of the U(1)$_{HC}$ charge $+$ and $-$ for $\alpha=1,2$, respectively,
and $\gamma_{HC}$ denotes the gauge boson corresponding to the U(1)$_{HC}$
gauge symmetry.  If there is an initial asymmetry in the U(1)$_{HC}$ charge in
the universe, this is preserved, and the resultant lighter linear combination
of $\omega^{\pm}_{p,R}$, $p=1,2$ could serve as a dark matter candidate.

The model with sequence 2 has another dark matter candidate. To show this, we
recall that in this sequence the fermions $\zeta^{1j,\alpha}_R$, where $j=4,5$,
transforming as TC doublets remain light down to the $\Lambda_{TC}$ scale (in
contrast to sequence 1, where they gain dynamical masses $\sim \Lambda_2$).
Because of the exact SU(2)$_{TC}$ symmetry, these fermions do not mix with
technisinglets, and because of the residual exact U(1)$_{HC}$ gauge symmetry,
they also do not mix with HC-singlets.  The condensate (\ref{z14z15_gbp}) has
the consequence that the mass eigenstates are $t^\alpha_{1,2,R} =
(\zeta^{14,\alpha}_R \pm \zeta^{15,\alpha}_R)/\sqrt{2}$ with equal and opposite
mass eigenvalues of magnitude $\simeq \Lambda_{TC}$. We label these as
$t^{1j,\alpha}_R$.  Let us consider the particle state created by the action of
the adjoint of the operator
\beq
\frac{1}{\sqrt{2}} \epsilon_{123jk} \epsilon_{\alpha\beta} t^{1j,\alpha \ T}_R
C t^{1k,\beta}_R
\label{dm_seq2}
\eeq
on the vacuum.  (As before, the $\epsilon_{\alpha\beta}$ is included not to
represent an SU(2)$_{HC}$-singlet contraction, since SU(2)$_{HC}$ has been
broken to U(1)$_{HC}$ at the higher scale $\Lambda_{BHC}$; instead, it
represents the antisymmetrization on $\alpha$ and $\beta$ produced
automatically by the contraction with $\epsilon_{123jk}$ and Fermi statistics.)
This particle has a mass $\simeq 2\Lambda_{TC} \simeq O(600)$ GeV, but, because
of the extra binding energy due to the attractive U(1)$_{HC}$ interaction, is
lighter than the particle corresponding to (\ref{dm_seq2na}).  Note that the
particle (\ref{dm_seq2}) is a technibaryon (and is bosonic, since $N_{TC}$ is
even).  It is expected to be lighter than other technibaryons and hence to be
stable and can provide a dark matter candidate. Since this technibaryon is
composed of SM-singlet fermions, its interactions with regular matter would be
highly suppressed, since they would proceed only via exchange of ETC gauge
bosons.  The interactions between two of the particles (\ref{dm_seq2}) would
also only proceed via higher-order exchanges and, in particular for the case of
U(1)$_{HC}$, van der Waals-type interactions, since it is a technisinglet and
U(1)$_{HC}$-singlet.

\subsection{Global Symmetries and (Pseudo)-Nambu-Goldstone Bosons}

The ETC model of this paper is based on the gauge group (\ref{g}), and contains
a total of six, $U(1)$ global symmetries, associated with the representations
(\ref{qtq}), (\ref{dec_le}), and (\ref{fermions_model1}) of the ETC group. 
(The Nambu-Goldstone bosons associated with global symmetries carried by $\zeta$ and  $\omega$ fields will be weakly coupled to SM fields, and therefore we disregard them here.)  
One can label these $U(1)$'s according to the representation they
act on as:
\beq
 U(1)_{Q_L}\,\times\,U(1)_{u_R}\,\times\,U(1)_{d_R}\,
\times\,U(1)_{L_L}\,\times\,U(1)_{e_R}\,\times\,U(1)_{\psi_R}\,.
\eeq
Four linear combinations are spontaneously broken by the TC condensates, and
one by the ETC condensate at scale $\Lambda_1$, leaving one exact global
symmetry (to be identified with baryon number conservation U(1)$_B$, generated
by the sum of the generators of U(1)$_{Q_L}$, U(1)$_{u_R}$, and U(1)$_{d_R}$)
and five Nambu-Goldstone bosons. Of these five, one linear combination acquires
a mass via ETC instantons, which explicitly break the global symmetries of the
theory, and another can be identified with the Nambu-Goldstone boson associated
with spontaneous breaking of $U(1)_Y$, so that it gets eaten to become the
longitudinal component of the $Z^0$ vector boson. We are left with three
exactly massless Nambu-Goldstone bosons, each of which, being a composite of SM
non-singlets, has potentially dangerous couplings to SM fields. A similar
situation arises also in conventional TC theories, where it is
known~\cite{pngb} that two such massless states are present. We will comment on
the possible solutions to this phenomenological problem at the end of this
subsection.

While baryon number is not spontaneously broken, instanton effects
could in principle induce in the low-energy effective theory operators leading
to phenomenologically unacceptable $\Delta B=1$ (proton decay) and $\Delta B=2$
(neutron-antineutron oscillations) transitions. It is easy to see that this is
not the case: the triangle anomaly $[U(1)_B][SU(5)_{ETC}]^2$ sums up to zero,
and so $U(1)_B$, being free from $SU(5)_{ETC}$ anomalies, is an exact symmetry
(we neglect in this discussion $SU(2)_{L}$ electroweak instanton effects, which
are negligible in the zero temperature limit). By contrast, lepton number
(defined as the sum of the generators of U(1)$_L$, U(1)$_{e_R}$, and
U(1)$_{\psi_R}$) is spontaneously broken at the scale $\Lambda_1$. But, due to
the assignment of $\psi$ fields to the antisymmetric representation of
$SU(5)_{ETC}$, this symmetry is anomalous, and hence is explicitly broken by
instanton effects. In order to clarify this issue, it is useful to exhibit the
operator encoding in the low energy theory the information about ETC instanton effects.

Following the analysis by 't Hooft \cite{thooft}, one characterizes
instanton-induced violation of global symmetries by constructing appropriate
product(s) ${\cal O}$ involving all of the fields that are nonsinglets under
the ETC gauge group with a coefficient involving a dimensionless factor
$A(\Lambda)$ and a factor of the form $1/\Lambda^{d_{\cal O}-4}$, where
$d_{\cal O}$ denotes the dimension of ${\cal O}$ and $\Lambda$ denotes the
relevant ETC scale.  The quantity $A(\Lambda)$ consists of a power of
$\pi/g_{_{ETC}}(\Lambda)$ and an exponential factor $A(\Lambda) = {\rm exp}(-8
\pi^2/g_{_{ETC}}^2(\Lambda))$. For a weak-coupling situation such as
electroweak SU(2)$_L$, there is severe suppression due to the exponential
factor; however, for the strong-coupling ETC theory under consideration here
the power of $\pi/g_{_{ETC}}$ can produce an enhancement comparable to the
suppression from the exponential, and there is significant uncertainty in this
dimensionless factor.  Accordingly, for our estimates we take the
dimensionless factor to be of order unity.  In the present case we need to
generalize the 't Hooft analysis (i) from a simple nonabelian group to a group
which is a direct product of nonabelian factors; (ii) from a theory with
fermions transforming only according to the fundamental representation to a
theory with fermions transforming according to several different
representations (fundamental, conjugate fundamental, and antisymmetric
second-rank tensor representations for SU(5)$_{ETC}$); and (iii) from a theory
with only one gauge symmetry-breaking scale to a theory with a sequence of
different breaking scales.

At the scale $\Lambda_1$, hypercolor interactions are still weak,
and we neglect them together with the associated SU(2)$_{HC}$
instantons. Since SM gauge interactions
are even weaker at this scale, we also consider the limit in which SM gauge interactions are turned off.
The ETC Lagrangian at the SU(5)$_{ETC}$ symmetric level for energies
$E > \Lambda_1$ is then invariant under the global symmetry
\beq
G_{glob.} = {\rm U}(9) \times {\rm U}(6) \times {\rm U}(1) \ .
\label{g_global_su5_only}
\eeq
To see this, we note that in this limit, the Lagrangian is invariant under
transformations among the nine components of $Q_L$ and $d^c_L$, each of which
is a 5 representation of SU(5)$_{ETC}$, and among the six components of
$L_L$, $e^c_L$, and $u^c_L$, each of which is a $\bar 5$ representation.
The additional U(1) is for $\psi_R$.

We construct an operator encompassing effects of ETC instantons by
starting in the SU(5)$_{ETC}$-symmetric theory, combining
SU(5)$_{ETC}$ fermion representations in such a manner as to respect
the global symmetries, and incorporating the physics of the breaking
of this group via the coefficient of the operator, which involves the
breaking scale, $\Lambda_1$.  Just as in the SU(2)$_L$ case
\cite{thooft}, the existence of this breaking scale cuts off
integrations over instanton size and makes it possible to work with a
local instanton operator.

For energies $E < \Lambda_1$, we obtain the local operator given by
\beq
\frac{A(\Lambda_1)}{\Lambda_1^{32}} {\cal O}
\label{Leff_instanton_1}
\eeq
where $A(\Lambda_1) \simeq O(1)$ and, in a compact notation,
\beq
{\cal O} = [\Pi u_L d_L d^c_L][n_L e_L e^c_L u^c_L] [\Pi \psi^c_L]
[\Pi \zeta^c_L] \ .
\label{O1_instanton}
\eeq
In Eq. (\ref{O1_instanton}) the meaning of the notation is as follows: (i)
$[\Pi u_L d_L d^c_L]$ is an antisymmetrized 9-fold product of fermion fields
that including all $N_c=3$ colors and all flavors so as to yield an invariant
under global SU(9) (ii) $[\Pi n_L e_L e^c_L u^c_L]$ is an antisymmetrized
6-fold product including a product over all colors, so as to yield an invariant
under global SU(6); (iii) $[\Pi \psi^c_L]$ denotes a product over three
$\psi^c_L$ fields, reflecting the fact that the contribution to the anomaly of
a U(1) number transformation, and hence to the effective instanton operator, of
a fermion field transforming according to the anti-symmetric rank-2
representation of SU(5) is 3 times that of a fermion transforming according to
the fundamental representation; (iv) $[\Pi \zeta^c_{\alpha,L}]$ denotes a
similar cubic product of the $\zeta$ fields for each value of the HC index,
antisymmetrized on this index so as to yield an SU(2)$_{HC}$-invariant.  There
are thus $N_F=24$ fermion fields in the operator ${\cal O}$, so that $d_{\cal
O}=36$, and the dimensional coefficient in Eq.  (\ref{Leff_instanton_1}) is
$1/\Lambda_1^{32}$.  Since $\Lambda_1$ is much larger than the electroweak
symmetry-breaking scale, we also require that ${\cal O}$ be invariant under
$G_{SM}$.

As anticipated, the operator ${\cal O}$ is invariant under baryon number
$U(1)_{B}$ and transforms as $|\Delta L| =2$.  We have already observed and
made central use of the fact that lepton number is broken spontaneously by the
bilinear ETC condensate (\ref{xixicondensate}) at scale $\Lambda_1$.  Now, one
could imagine generating a similar bilinear operator by contracting 22 of the
24 fields in (\ref{O1_instanton}) (all but two of the $\psi$ fields).  The
attendant integrations would be cut off at a scale no larger than $\Lambda_1$,
and therefore the coefficient of the bilinear would be no larger than
$\Lambda_1$.  This operator would necessarily be an SU(5)$_{ETC}$-non-singlet,
and would consistently align in the same direction as (\ref{xixicondensate})
(taken to be the $i=1$ direction).

Finally, we return to the issue of the three exactly massless, electrically
neutral, Nambu-Goldstone bosons enumerated above. Since the operator ${\cal O}$
breaks U(1)$_{\psi_R}$, each of these necessarily includes SM non-singlet
constituents. In our model, where these particles are formed at the electroweak
(technicolor) scale, this is phenomenologically unacceptable~\cite{Raff}.  The
model of this paper must therefore be regarded as incomplete.  Some new, higher
energy interactions must be invoked to break explicitly the three corresponding
global symmetries.  An example of such interactions is discussed in
Ref.~\cite{pngb}, in the context of ETC theories which employ vector-like
representations for SM fermions: a four-fermion operator that one could obtain
in a Pati-Salam extension of the SM gauge group is shown to give masses to
massless states.  It is not possible to construct a conventional Pati-Salam
extension of the gauge symmetry of models in the class considered here, in
which the left- and right-handed chiral components of the down-type quarks and
charged leptons transform according to conjugate representations of the ETC
group.  However, one can envision a different type of extended gauge symmetry
with interactions that would break these three global U(1) symmetries, if some
of the ETC representations come from unified representations of a larger gauge
symmetry group.  Assuming such interactions to be present at scales not too far
above $\Lambda_1$, and using Dashen's formula as in Ref.~\cite{pngb}, masses of
the order of 10 GeV can be anticipated for these particles. They would then
satisfy astrophysical and cosmological constraints~\cite{Raff}, as well as
constraints coming from the decay $\Upsilon(1S) \to \gamma + $ pseudoscalar
\cite{pdg}.  It will be important to refine these mass estimates in the context
of an explicit extension of our model and to explore other phenomenological
signatures of these particles.

We note that all other pseudo-Nambu-Goldstone bosons in our model acquire mass
via QCD or ETC interactions, and are then heavy enough not to cause
phenomenological problems. This is to be contrasted with ETC models which
employ vector-like representations for the SM fermions. There, some
(electrically charged, color neutral) pseudo-Nambu-Goldstone bosons acquire
mass only through electroweak interactions, and thus remain relatively light.
The assignment of up-type and down-type right-handed quarks to different ETC
representations (and a similar assignment for the leptons) explicitly breaks
(via ETC interactions) the symmetries giving rise to these additional light
states.

\section{Discussion and Conclusions}

In this paper we have addressed the problem of quark and lepton masses and
mixing angles in the ETC framework. The production of intra-generational mass
splittings and CKM mixing has been a long-standing challenge for these
theories.  We have proposed a framework to achieve these goals and have
illustrated it with a specific model.  The model is based on the relation
(\ref{nrel}), which, with the choice $N_{TC}=2$, implies that the ETC group is
SU(5)$_{ETC}$. There are three reasons for using $N_{TC}=2$: (i) it minimizes
TC contributions to precision electroweak observables, in particular the $T$
and $S$ parameters, both of which are proportional to $N_{TC}$; (ii) with $N_f
\simeq 8$ vectorially coupled technifermions (as is true of our model with both
sequences 1 and 2), perturbative beta function calculations suggest that the
technicolor theory will exhibit walking (approximately conformal) behavior
which enhances both standard-model fermion masses and pseudo-Nambu-Goldstone
boson masses; and (iii) the choice $N_{TC}= 2$ allows our mechanism \cite{nt}
to operate and lead to small but nonzero neutrino masses and intra-generational
mass splittings.

We have formulated and analyzed an approach based on the assignment of the
right-handed components of the $Q=-1/3$ quarks and charged leptons to a
representation of the ETC gauge group that is conjugate to the representation
of the corresponding left-handed components. This leads to a natural
suppression of these masses relative to those of the $Q=2/3$ quarks, as well as
to nontrivial quark mixing angles. For the neutrinos, a set of (right-handed)
standard-model singlets was included, and assigned to representations that also
lead to a suppression of their masses, as well as to nontrivial mixing angles
and a potential seesaw mechanism. We note that these assignments are ad hoc,
made to accommodate the observed features of the fermion mass matrices. Within
our framework, they do not offer a direct explanation of why the $Q=2/3$ quarks
are heavier than the corresponding $Q=-1/3$ quarks in the higher two
generations or why the colored fermions (quarks) are heavier than the
corresponding color-singlet fermions (leptons). A natural goal for extensions
of the present work would be to explain these correlations.

We have provided a simple illustrative model involving one additional strong
gauge interaction denoted as hypercolor (HC), and an economical representation
content. We identified two plausible sequences of ETC symmetry breaking in this
model depending on the relative strengths of the ETC and HC
interactions. Although this simple model is not fully realistic, it has many
features of a successful theory.

With respect to the quarks and charged leptons, the model can generate a
sufficiently heavy top quark and satisfactory inter-generational mass
splittings in the up-quark sector.  It naturally achieves the observed
intra-generational mass splittings for the higher two generations: $m_t >> m_b,
\ m_\tau$ and $m_c >> m_s$ and $m_\mu$; and some quark mixing. It achieves
these splittings without prohibitively large contributions to $\Delta \rho =
\alpha_{em} T$. Of the two sequences considered, sequence 1 has three ETC
breaking scales and can account for all the up-type quark masses. The down-type
quark masses and charged lepton masses are suppressed, as desired, compared to
these, because of the use of relatively conjugate ETC representations for the
left- and right- handed components. But with the up-type quark masses fit,
there is excessive suppression of down-quark and charged lepton masses,
especially for those of the first generation. In sequence 2, there are only two
ETC breaking scales.  Hence, a hierarchy between $m_c$ and $m_t$ is more
difficult to achieve, although the non-diagonal structure of $M^{(u)}$ offers a
possibility, as we discussed. There is the same excessive suppression of
first-generation down-quark and charged lepton masses.  Although the model does
not give fully realistic CKM mixing, it does offer the prospect of spontaneous
(dynamical) generation of a CP-violating phase, and this will be an
interesting avenue for further study, in conjunction with implications for the
strong CP problem. Leptonic CP violation in this class of models is also worthy
of study.

For the neutrinos, sequence 1 has little success. It leads to Dirac masses
suppressed relative to the $Q=2/3$ quark masses, to right-handed Majorana
masses, and to some neutrino mixing. But the Dirac masses remain too large, and
the Majorana masses are too small for seesaw mechanism. Sequence 2, which can
occur if the HC coupling is somewhat weaker at the relevant scales, has
advantages and disadvantages. A major advantage is that purely within the
neutrino sector, it can incorporate the seesaw mechanism of Ref. \cite{nt},
yielding experimentally acceptable light neutrino masses. However, the value of
the lower ETC breaking scale $\Lambda_{23}$ that produces a successful fit to
neutrino masses does not yield sufficiently large second or third-generation
quark and lepton masses.  

The are also some potential phenomenological problems. The class of models
considered leads to flavor-changing neutral currents at a level that could
exceed the measured values for $\Delta m_K$ and $\Delta m_{B_{d}}$, and to a
value for $\Delta m_{B_s}$ well in excess of the current range of SM
predictions.  This could be ameliorated if the model exhibited walking beyond
the lowest ETC scale. This class of models also has a potential problem with
massless (electrically neutral) Nambu-Goldstone bosons.  One can envision
additional interactions that will raise their masses to the level of 10 GeV or
higher, satisfying current experimental bounds.

The shortcomings of the simple model and the symmetry breaking patterns
considered here, suggest a variety of directions for further exploration.  An
evident problem is the limited number of gauge boson mixings which yield many
zero entries in the quark and lepton mass matrices. These zeros are traceable
to selection rules related to residual global symmetries. But since these
selection rules seem specific to the breaking patterns considered, one can
imagine that other breaking patterns, if they are favored dynamically, could
lead to more nonzero mass entries and hence more realistic quark and lepton
mixing.

One possibility along these lines can arise if the interaction strengths are
such that the condensates of both sequences 1 and 2 form.  This involves the
occurrence of misaligned condensates, i.e., condensates occurring at comparable
energy scales that respect different symmetries. This could improve the
predictions of fermion mixing, but would have the drawback that there would be
only two ETC scales, $\Lambda_1$ and $\Lambda_{23}$. This contrasts with
sequence 1, which has three such scales, in one-to-one correspondence with the
generations, and sequence 2, which has two ETC scales $\Lambda_1$ and
$\Lambda_{23}$, but also an additional scale $\Lambda_{BHC}$ associated with
the breaking of hypercolor. Consequently, while the misaligned-condensate
possibility may produce more complete and realistic quark and lepton mixing, it
seems unable to produce realistic scales for quark, lepton, or neutrino masses.

Another direction is to retain the group structure, (\ref{g}), of the model,
but to consider different fermion representations. An example is a model of DES
type (see Section II.D for a definition) in which the SM-nonsinglet fermions
transform according to (\ref{qtq}) and
\beq \quad\quad  L_L: \ (5,1,1,2)_{-1,L} \ ,
 \quad\quad e_R: \ (\bar 5,1,1,1)_{-2,R},
\label{des_le} \eeq
and the SM-singlet fermions transform as
\beq \psi^{ij}_R: \ (10,1,1,1)_{0,R} \ , \quad \zeta^{ij,\alpha}_R: \
(10,2,1,1)_{0,R} \ , \quad A^i_{jk,R}: \ (\overline{45},1,1,1)_{0,R}.
\label{fermions_model45} \eeq
This model includes a mixed tensor representation of SU(5)$_{ETC}$, the
$(\overline{45},1,1,1)_{0,R} = A^i_{jk,R}$. The mixed tensor representation
$A^i_{jk}$ is antisymmetric in the indices $j,k$ and satisfies the
tracelessness condition $\sum_{i=1}^N A^i_{ik}=0$, which amounts to $N$
equations, so that its dimension is $N(N-2)(N+1)/2$. The contributions of the
SM-singlet fermions to the SU(5)$_{ETC}$ gauge anomaly satisfy the DES subcase
of condition (\ref{arsum}): $A(\psi_R)=1$, $A(\zeta_R)=2$,
$A(A)=-A(45_{SU(5)})=6$. It is therefore anomaly-free.  Although the
particular model of this type that we studied did not lead to an overall
improvement in predictions, we believe that the analysis of models with
different types of fermion representations is worthwhile.  

Some of the defects of the class of models considered here, including the
explicit model analyzed in section IV, derive from the fact that the Dirac
submatrix that exerts dominant control over the generation of masses for the
primary mass eigenstates in the observed electroweak-doublet neutrinos
satisfies the relation (\ref{bij_Mdij}). This, in turn, reflects the fact that
there is only one, common mass suppression mechanism for the down-type quarks,
charged leptons, and Dirac neutrino masses relative to the up-type quarks (the
use of conjugate ETC representations for the left-and
right-handed down-type quarks and charged leptons, and a similar arrangement
for the left- and right-handed neutrinos.)  The relation
(\ref{bij_Mdij}) prevents a simultaneous fit to all of these masses. 

One salient property of the present class of models is that the $b_{ij}$ Dirac
neutrino mass terms relevant to the seesaw connect the left-handed
electroweak-doublet neutrinos to right-handed SM-singlet fermions whose masses
arise via loop effects rather than directly, via participation in
condensates. The models do contain a different type of Dirac mass term (the
$d_{i,jk}$) which connect the left-handed electroweak-doublet neutrinos with
right-handed SM-singlet fermions that participate directly in condensates and
hence gain larger dynamical masses.  In our models, the $d_{i,jk}$ terms do not
play an important role in driving a seesaw mechanism.  It would be valuable to
develop models in which the $d_{i,jk}$ terms do play a central role, thus
avoiding the constraint~(\ref{bij_Mdij}).

Two interesting directions for future research suggest themselves.  First, it
is important to construct extensions of the class of models considered here
which contain interactions yielding sufficiently large masses for the light,
SM-singlet Nambu-Goldstone bosons.  Second, in the present work, and its
progenitors~\cite{at94,nt,lrs}, the extended technicolor symmetry breaking at
levels below $\Lambda_1$ involves a combination of the ETC interaction and
another strongly coupled gauge interaction, namely hypercolor.  It would be
worthwhile to try to construct models in which all of the ETC symmetry breaking
could be achieved without recourse to another gauge interaction that is
strongly coupled at a similar energy scale.  Ideally, one would be able to
achieve both of these goals in a single new model.

  This research was partially supported by the grants DE-FG02-92ER-4074
(T.A. and M.P.) and NSF-PHY-00-98527 (R.S.).  We thank Sekhar Chivukula, Alex
Kagan, and Ken Lane for helpful comments and acknowledge the Aspen Center for
Physics, where some of this research was done.

\newpage

\appendix

\section{ETC Gauge Boson Mixing}

  The estimates in Section IV of the fermion mass matrices for our model
depend critically on mixing among the ETC gauge bosons. Here we list for
each symmetry-breaking sequence those mixings that are nonzero, and estimate
their magnitudes. The fact that some mixings vanish can be understood from
global symmetry considerations. This discussion is provided in Section IV
for each symmetry-breaking sequence.

\subsection{Definitions and Identities}

 For a fermion $f_L$, transforming, say, according to the fundamental
representation of SU($N_{ETC}$), the basic coupling is given by
\beq {\cal L} = g_{_{ETC}} \bar f_{i,L} T_a (V_a^\lambda)^i_j \gamma_\lambda
f^j_L \label{Lcal} \eeq
where the $T_a$, with $1 \le a \le N_{ETC}^2-1$, are the generators of
SU($N_{ETC})$, and the $V_a$ are the corresponding ETC gauge bosons.  Just
as in weak interactions one switches from the Cartesian basis $A^\lambda_a$,
with $a=1,2,3$, to a basis comprised of the linear combinations of gauge
bosons $(W^\pm)^\lambda \equiv (A^\lambda_1 \mp i A^\lambda_2)/\sqrt{2}$
corresponding to the SU(2) shift operators $T_\pm = T_1 \pm i T_2$, so also
it is convenient to use the analogous basis for the ETC gauge bosons
corresponding to non-diagonal generators here.  In the SU(2) case, one may
label $W^+ = V^1_2$ and $W^- = V^2_1 = (V^1_2)^\dagger$ (suppressing the
Lorentz index), meaning that for a fermion in the fundamental
representation, say $f_L = {f^1 \choose f^2}_L$, one has a vertex in which
$f^1_L$ absorbs a $V^2_1$ or emits a $V^1_2$, making a transition to $f^2_L$
with coupling $g_{wk}/\sqrt{2}$. Similarly, for SU($N_{ETC}$) we use a basis
in which the gauge bosons corresponding to non-diagonal, shift operators are
labelled as $V^i_j = (V^j_i)^\dagger$, where $1 \le i, j \le N_{ETC}$, $i
\ne j$.  The associated vertex is $(g_{_{ETC}}/\sqrt{2}) \bar f_{i,L}
(V^i_j)^\lambda \gamma_\lambda f^j_L$.  Analogous comments apply to the
couplings of the ETC gauge bosons to the right-handed components of $f$, and
also, with appropriate changes to theories with the left- or right-handed
component of $f$ transforms according to a conjugate fundamental
representation.  We retain the Cartesian basis for the ETC gauge bosons
corresponding to diagonal Cartan generators of SU($N_{ETC}$) and hence write
them as $V_a$, where $a=n^2-1$ for $2 \le n \le N_{ETC}$.

To discuss the ETC gauge boson mixings that will be important for generating
various contributions to fermion mass matrices, we consider the
one-particle-irreducible vacuum polarization tensor ${}^k_n
\Pi^i_j(q)_{\mu\lambda}$ producing the transition $V_j^i \rightarrow V_n^k$.
We assume that the diagonal masses of these fields have already been
included, and therefore exclude the case $i=k,j=n$ from consideration here.
Because the momenta in the loop in Fig. \ref{ddgraph} and its analogue for
the charged leptons are cut off by the technicolor dynamical mass insertion
on the internal technifermion line at a scale $\Lambda_{TC}$ which is small
compared to the lowest ETC scale, we shall need only the expressions for ETC
gauge boson mixing evaluated effectively at zero momentum, i.e., ${}^k_n
\Pi^i_j(0)_{\mu\lambda}$. While ${}^k_n \Pi^i_j(q)_{\mu\lambda}$ must be
transverse because the underlying currents are conserved, the appearance of
a Nambu-Goldstone pole associated with the symmetry breaking means that
${}^k_n\Pi^i_j(0)_{\mu\lambda}$ can be nonzero. We focus on the $g_{\mu\nu}$
piece (it is simplest to imagine working in Landau gauge for this purpose)
and define the zero-momentum coefficient of $g_{\mu\nu}$ to be
${}^k_n\Pi^i_j(0)$.

Mixing among group eigenstates $(V^i_j)_\mu$, is in general to be expected,
setting in as the SU(5)$_{ETC}$ breaks sequentially to SU(2)$_{TC}$.  Because
of the mixings among the ETC gauge bosons $V^i_j$, the corresponding mass
eigenstates are comprised of linear combinations of these group
eigenstates. Since these masses are hierarchical, it will be useful to continue
to refer loosely to the ``masses'' of various ETC gauge bosons.  When a gauge
group SU($N$) breaks to SU($N-1$), the $2N-1$ gauge bosons in the coset ${\rm
SU}(N)/{\rm SU}(N-1)$ pick up mass.  For example, as the energy decreases
through the scale $\Lambda_1$, SU(5)$_{ETC}$ breaks to SU(4)$_{ETC}$ and nine
ETC gauge bosons gain masses. These consist of the four gauge bosons
$(V_1^i)_\mu$, $i=2,3,4,5$, together with their adjoints $(V^1_i)_\mu$, and the
the gauge boson corresponding to one diagonal generator. Similar comments apply
for subsequent breakings. The existence of mixing among these group eigenstates
will be seen to occur in each of the symmetry breaking sequences considered
here.  In the presence of this mixing among ETC gauge bosons, one must
rediagonalize the associated 2-point functions to obtain the actual mass
eigenstates.  Our procedure is equivalent to this diagonalization, to the order
that we work.  We note here that the relation $(V^i_j)^\dagger_\mu =
(V^j_i)_\mu$, leads to some basic identities for ${}^k_n
\Pi^i_j(q)_{\mu\lambda}$:
\beq |{}^k_n \Pi^i_j(q)_{\mu\lambda}| = |{}^i_j \Pi^k_n(q)_{\mu\lambda}| =
|{}^j_i \Pi^n_k(q)_{\mu\lambda}| \ . \label{identity1} \eeq

\subsection{ETC Gauge Boson Mixing for Symmetry-Breaking Sequence 1}

For the symmetry-breaking sequence 1, it turns out that the following ETC
gauge boson mixings occur:
\beq V_1^4 \leftrightarrow V^3_5 \ , \quad V_1^5 \leftrightarrow V^3_4
\quad + \quad (4 \leftrightarrow 5) \label{v1up4dn_to_v5up3dn_gap} \eeq
\beq V_2^4 \leftrightarrow V^2_5 \ , \quad V_2^5 \leftrightarrow V^2_4
\label{v2up4dn_to_v2dn5up_gap} \eeq
\beq V^t_1 \leftrightarrow V^t_3 \ , \quad t=4,5
\label{v1dn4up_to_v3dn4up_gap} \eeq
It is useful to classify these mixings according to their selection rules
under the maximal subgroup U(3)$\times$ SU(2)$_{TC}$ of SU(5)$_{ETC}$, where
the U(3) is the group of transformations operating on the three ETC
generation indices $i=1,2,3$, while the SU(2)$_{TC}$ subgroup operates on
the remaining $i=4,5$ indices.  Thus, the mixings of Eq.
(\ref{v1up4dn_to_v5up3dn_gap}) and (\ref{v2up4dn_to_v2dn5up_gap}) are of the
form
\beq
 (\bar 3,2) \leftrightarrow (3,\bar 2) \approx (3,2)
\label{3bar2_to_32bar} \eeq
where the second equality $\bar 2 \approx 2$ follows from the fact that
SU(2) has only (pseudo)real representations.  The mixings of Eq.
(\ref{v1dn4up_to_v3dn4up_gap}) are of the form
\beq
 (3,2) \leftrightarrow (3,2)
\label{32_to_32} \eeq
The mixings (\ref{v1up4dn_to_v5up3dn_gap}) and
(\ref{v2up4dn_to_v2dn5up_gap}) were present in the $G_a$ sequence of Ref.
\cite{nt}. Diagrams for the mixings
(\ref{v1up4dn_to_v5up3dn_gap})-(\ref{v1dn4up_to_v3dn4up_gap}) are given in
Figs.  \ref{v1dn4up_to_v3up5dn_gap_fig}, \ref{v2dn4up_to_v2up5dn_gap_fig},
and \ref{v1dn4up_to_v3dn4up_gap_fig}.  The fact that only these mixings are
nonzero is explained from global symmetry considerations in Section IV.  One
can see how the transformation property (\ref{3bar2_to_32bar}) arises by
inspection of the gauge-fermion vertices and dynamical mass insertions.
Consider, for example, the graph in Fig. \ref{v1up4dn_to_v5up3dn_gap}).  The
incoming $V_1^4$ transforms as $(\bar 3,2)$ under the maximal subgroup
U(3)$_{gen.} \times$ SU(2)$_{TC}$ of SU(5)$_{ETC}$; at the leftmost vertex,
it produces a virtual $\zeta \bar \zeta$ pair; here the $\bar \zeta$ is
written as the $\zeta^{12,\alpha}_R$ moving backward on the loop and
transforming as $(\bar 3,2,2)$ under ${\rm U}(3)_{gen.} \times {\rm
SU}(2)_{TC} \times {\rm SU}(2)_{HC}$, while the $\zeta$ in the pair is the
$\zeta^{24,\alpha}_R$ moving forward on the loop and transforming as a
$(3,2,2)$ under the above group.  Thus, the U(3)$_{gen.}$ flow at the
leftmost vertex involves three $\bar 3$'s flowing into the vertex, for a
U(3)$_{gen.}$ singlet.  The dynamical mass insertion on the upper fermion
line of the loop, in which the $\zeta^{24,\alpha}_R$ goes to a
$\zeta^c_{25,\beta,L}$ is such that two 2's of SU(2)$_{TC}$ flow into the
vertex, and two 2's of SU(2)$_{HC}$ flow into the vertex, so that it is
invariant under SU(2)$_{TC}$ and SU(2)$_{HC}$; however, this vertex
transforms a $\bar 3$ of U(3)$_{gen.}$ into a 3 of U(3)$_{gen.}$.  On the
lower fermion line in the loop, reading in the direction of the arrow, from
right to left, the $\zeta^c_{23,\beta,L}$ is transformed into a
$\zeta^{12,\alpha}_R$, i.e., a 3 into a $\bar 3$ of U(3)$_{gen.}$. Combining
these with the rightmost gauge-fermion vertex then yields the overall
transformation property (\ref{3bar2_to_32bar}) of this ETC gauge boson
mixing.  Similar analyses can be made for each of the other such diagrams.

Each nonzero mixing will be estimated from the corresponding graph(s) by
assigning a dynamical mass to the fermions in each loop. Given the
uncertainties associated with strong coupling, this can at most provide an
order of magnitude estimate of the mixing. As an illustration, consider the
mixing ${}^3_5 \Pi_1^4(0)$ arising from the graph in Fig.
\ref{v1dn4up_to_v3up5dn_gap_fig}.  (Recall that we will need this quantity
only at zero momentum.) We estimate its size in terms of the integral (after
Wick rotation)
\beqs |{}^3_5 \Pi_1^4(0)| & = & |{}^3_4 \Pi_1^5(0)| \simeq \frac{1}{2} \left
( \frac{g_{_{ETC}}}{\sqrt{2}}\right )^2 \frac{N_{HC}}{4\pi^2} \int
\frac{(k^2 dk^2) \ k^4 \ \Sigma_3(k)^2}{(k^2 + \Sigma_3^2)^4} \cr\cr &
\simeq & \frac{g_{_{ETC}}^2 \Lambda_3^2}{24}.
\label{Pi_v1up4dn_to_v3dn5up_gap} \eeqs
Here, the first factor of 1/2 reflects the chiral nature of the
gauge-fermion couplings, $N_{HC}=2$ arises from the sum over hypercolors in
the loop, the $(k^2 dk^2)$ is the measure after angular integration, the
factor of $k^4$ arises from the numerators of the four fermion propagators,
and $\Sigma_i$ is the dynamical mass of the fermions that condense at the
scale $\Lambda_i$ (an ETC or TC scale).

The crude estimate, $g_{_{ETC}}^2 \Lambda_3^{2}/24$, is made as follows. If
the condensation occurs in an SU($N$) gauge theory, then for Euclidean $k
\lsim \Lambda_i$, $\Sigma_{i}(k) \simeq \Sigma_{i,0} = 2\pi f_n/\sqrt{N}$,
where $f_n$ is the associated pseudoscalar decay constant.  For example, in
QCD, one can define the dynamical (constituent) quark mass as $m_\rho/2$ or
$m_{p,n}/N_c$; these definitions yield the values 385 MeV and 313 MeV,
respectively, which also indicate the size of the theoretical uncertainty in
this quantity. Taking the average, 350 MeV, and comparing this with the
value of 337 MeV obtained from the above relation, one finds agreement to
about 5 \%. In general, using the relation $f_i \simeq
(\Lambda_{i}/2)\sqrt{3/N}$ yields the relation $\Sigma_{i,0} \simeq
(\pi/\sqrt{3})\Lambda_i$. This, too, works well in the case of QCD, where,
with $\Lambda_{QCD} \simeq 180$ MeV, the right-hand side is $\simeq 330$
MeV, in agreement, to within the uncertainty, with the 350 MeV estimate
given above for the constituent quark mass. For high momentum $k >>
\Lambda_i$, $\Sigma_{i}(k)$ has a behavior that depends on the type of
theory: like $\Sigma_{i,0}^3/k^2$ up to logarithms for a QCD-like theory,
and, in contrast, roughly like $\Sigma_{i,0}^2/k$ for a walking theory. In
Eq. (\ref{Pi_v1up4dn_to_v3dn5up_gap}) the condensation scale for the
fermions in the loop is $\Lambda_3$, whence the appearance of $\Sigma_3(k)$.
We make the conservative assumption that the TC theory walks only up to the
lowest ETC scale, $\Lambda_3$, and hence use $\Sigma_3(k) \simeq
\Sigma_{3,0}^3/k^2$ for the large-$k$ behavior of this dynamical mass.

Consider next the loop diagram for the mixing $V_1^t \leftrightarrow V_3^t$
with $t=4,5$, shown in Fig. \ref{v1dn4up_to_v3dn4up_gap_fig}, and involving
fermion masses that are all of order $\Lambda_3$. Using the same approach as
above (and taking into account the contributions of both
$\omega^c_{\beta,p,R}$ fields with $p=1,2$), this yields
\beq |{}_1^t \Pi_3^t(0)| \simeq \frac{g_{_{ETC}}^2 \Lambda_3^2}{12}  \quad
{\rm for} \ \ t=4,5~. \label{Pi_v1up4dn_to_v3up4n_gap} \eeq

The loop diagram for the mixing $V_2^4 \leftrightarrow V^2_5$ shown in Fig.
\ref{v2dn4up_to_v2up5dn_gap_fig}, involves fermion masses with two different
scales, $\Lambda_2$ for $\zeta^{34,\alpha}_R$ and $\zeta^c_{15,\beta,L}$ (as
well as $\zeta^{14,\alpha}_R$ and $\zeta^c_{35,\beta,L}$), and $\Lambda_3$
for $\zeta^{23,\alpha}_R$ and $\zeta^c_{12,\beta,L}$. In this case, the mass
mixing contribution for each of the two set of $\zeta$'s that contribute,
summed over hypercolor, is
\beqs & & \frac{1}{2} \left ( \frac{g_{_{ETC}}}{\sqrt{2}}\right )^2
\frac{N_{HC}}{4\pi^2} \int \frac{(k^2 dk^2) \ k^4 \ \Sigma_2(k) \
\Sigma_3(k)}{(k^2 + \Sigma_2^2)^2 \ (k^2 + \Sigma_3^2)^2 } \cr\cr & \simeq &
\left ( \frac{g_{_{ETC}}^2 N_{HC}}{4 \pi^2} \right ) \left (
\frac{\Sigma_{3,0}^3 \Sigma_{2,0}}{\Sigma_{2,0}^2} \right ) \cr\cr & \simeq
& \frac{g_{_{ETC}}^2}{12} \left ( \frac{\Lambda_3}{\Lambda_2} \right )
\Lambda_3^2 \ . \label{Pi_unequal_Sigmas} \eeqs
Here since the loop momentum $k$ extends above the smaller condensation
scale $\Lambda_3$, we have used the form $\Sigma_3(k) \simeq
\Sigma_{3,0}^3/k^2$, and since the important contributions to the integral
do not extend much beyond the higher scale $\Lambda_2$, we have used
$\Sigma_2(k) \simeq \Sigma_{2,0}$.  The $\Sigma_{2,0}^2$ in the denominator
is from a propagator factor that can be pulled out of the integral, given
its degree of convergence. Thus, more generally, a loop integral of this
type has a quadratic dependence on the smaller of the two condensation
scales, further suppressed by the ratio of the smaller to the larger scale.
Multiplying by another factor of 2 to take account of the fact that, for
each set of hypercolors, there are two sets of $\zeta$'s contributing to the
loop, we obtain the estimate
\beq |{}^2_5 \Pi_2^4(0)| = |{}^2_4 \Pi_2^5(0)| \simeq \frac{g_{_{ETC}}^2
\Lambda_3^3}{12\Lambda_2} \ . \label{Pi_v2up4dn_to_v2dn5up_gap} \eeq

\begin{center}
\begin{picture}(240,200)(0,0)
\Photon(0,100)(25,100){3}{4.5} \ArrowArcn(60,100)(35,180,90)
\Text(60,135)[]{$\times$} \ArrowArcn(60,100)(35,90,0)
\ArrowArcn(60,100)(35,0,270) \Text(60,65)[]{$\times$}
\ArrowArcn(60,100)(35,270,180) \Photon(95,100)(120,100){3}{4}
\Text(0,115)[]{$(V_1^4)_\mu$} \Text(122,115)[]{$(V^3_5)_\nu$}
\Text(25,135)[]{$\zeta^{24,\alpha}_R$} \Text(20,70)[]{$\zeta^{12,\alpha}_R$}
\Text(105,135)[c]{$\zeta^c_{25,\beta,L}$}
\Text(105,70)[c]{$\zeta^c_{23,\beta,L}$}
\end{picture}
\end{center}

\begin{figure}
\caption{\footnotesize{A one-loop graph contributing to the ETC gauge boson
mixing $V_1^4 \leftrightarrow V^3_5$ in symmetry-breaking sequence 1.}}
\label{v1dn4up_to_v3up5dn_gap_fig}
\end{figure}

\begin{center}
\begin{picture}(240,200)(0,0)
\Photon(0,100)(25,100){3}{4.5} \ArrowArcn(60,100)(35,180,90)
\Text(60,135)[]{$\times$} \ArrowArcn(60,100)(35,90,0)
\ArrowArcn(60,100)(35,0,270) \Text(60,65)[]{$\times$}
\ArrowArcn(60,100)(35,270,180) \Photon(95,100)(120,100){3}{4}
\Text(0,115)[]{$(V_2^4)_\mu$} \Text(122,115)[]{$(V^2_5)_\nu$}
\Text(25,135)[]{$\zeta^{34,\alpha}_R$} \Text(20,70)[]{$\zeta^{23,\alpha}_R$}
\Text(105,135)[c]{$\zeta^c_{15,\beta,L}$}
\Text(105,70)[c]{$\zeta^c_{12,\beta,L}$}
\end{picture}
\end{center}

\begin{figure}
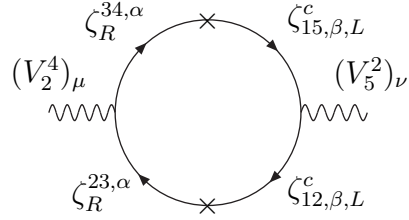

\caption{\footnotesize{A one-loop graph contributing to the ETC gauge boson
mixing $V_2^4 \leftrightarrow V^2_5$ in symmetry-breaking sequence 1. The
analogous graph with the ETC indices 1 and 3 interchanged on internal
fermion lines also contributes to this mixing.}}
\label{v2dn4up_to_v2up5dn_gap_fig}
\end{figure}

\begin{center}
\begin{picture}(240,200)(0,0)
\Photon(0,100)(25,100){3}{4.5} \ArrowArcn(60,100)(35,180,90)
\ArrowArcn(60,100)(35,90,0) \ArrowArcn(60,100)(35,0,270)
\ArrowArcn(60,100)(35,270,180) \Photon(95,100)(120,100){3}{4}
\Text(45,69)[]{$\times$} \Text(75,69)[]{$\times$}
\Text(0,115)[]{$(V_1^t)_\mu$} \Text(122,115)[]{$(V_3^t)_\nu$}
\Text(62,148)[]{$\zeta^{2t,\alpha}_R$} \Text(18,72)[]{$\zeta^{12,\alpha}_R$}
\Text(60,51)[]{$\omega^c_{\beta,p,L}$}
\Text(107,72)[c]{$\zeta^{23,\alpha}_R$}
\end{picture}
\end{center}

\begin{figure}
\caption{\footnotesize{A one-loop graph contributing to the ETC gauge boson
mixing $V_1^t \leftrightarrow V_3^t$, where $t=4,5$, in symmetry-breaking
sequence 1.}} \label{v1dn4up_to_v3dn4up_gap_fig}
\end{figure}

Insertions of these non-diagonal mixings on ETC gauge boson lines give rise
to further mixings.  Thus,
\beq V_1^4 \leftrightarrow V^3_5 \leftrightarrow V^1_5 \ \Longrightarrow \
V_1^4 \leftrightarrow V^1_5 \label{v1dn4up_to_v1up5dn_gap} \eeq
and
\beq V_3^4 \leftrightarrow V^1_5 \leftrightarrow V^3_5 \  \Longrightarrow \
V_3^4 \leftrightarrow V^3_5 \label{v3dn4up_to_v3up5dn_gap} \eeq
so that all of the three mixings,
\beq V_i^4 \leftrightarrow V^i_5 \ , \quad V_i^5 \leftrightarrow V^i_4 \ ,
\quad i=1,2,3 \label{vidn4up_to_viup5dn_gap} \eeq
are present with this sequence.  We estimate
\beq {}_1^5 \Pi^1_4(0) = {}^1_5 \Pi^3_5(0) \ \frac{1}{M_3^2} \ {}^3_5
\Pi^4_1(0) \simeq \left ( \frac{g_{_{ETC}}^2 \Lambda_3^2}{12} \right )
\frac{16}{( a g_{_{ETC}} \Lambda_3)^2} \left ( \frac{g_{_{ETC}}^2
\Lambda_3^2}{24} \right )
 \simeq \frac{g_{_{ETC}}^2 \Lambda_3^2}{18 a^2 }
\label{Pi_1dn4up_to1up5dn_gap} \eeq
and
\beq {}^3_5 \Pi^4_3(0) = {}^3_5 \Pi^1_5(0) \ \frac{1}{M_1^2} \ {}^1_5
\Pi^4_3(0) \simeq \left ( \frac{g_{_{ETC}}^2 \Lambda_3^2}{24} \right )
\frac{16}{(a g_{_{ETC}} \Lambda_1)^2} \left ( \frac{g_{_{ETC}}^2
\Lambda_3^2}{12} \right )
 \simeq \frac{g_{_{ETC}}^2 \Lambda_3^4}{18 a^2 \Lambda_1^2} \ .
\label{Pi_3dn4up_to3up5dn_gap} \eeq

For the neutrino masses and mixing angles, we need an estimate of $r_{23}$
in the mass matrix $(M_R)_{\alpha \alpha}$ of Eq. (\ref{rqmatrix}), which
plays the role as the large mass in the seesaw mechanism. This is
generated as in Fig. \ref{alpha-alpha}.  Since the dynamical mass insertion
on the internal fermion line in this graph is of order the largest ETC
scale, $\Lambda_1$, the loop momenta extend up to this scale, so that one
needs to evaluate the ETC gauge boson mixing $V^1_4 \leftrightarrow V_1^5$
(which occurs via the combination (\ref{v1dn4up_to_v1up5dn_gap})) for
momenta $q$ up to $\Lambda_1$. Thus, in the contribution to this transition
involving two successive fermion loops on the ETC gauge boson line, and
hence four $\Lambda_3$-scale dynamical fermion mass insertions, each of
these will be of the form $\Sigma_3 \propto \Lambda_3^3/Q^2$, where $Q$
denotes the maximum of the incoming ETC gauge boson momentum $q$ and the
fermion loop momentum $k$. This leads to a strong suppression of the
resultant $r_{23}$.

ETC gauge boson loop diagrams with various insertions yield still further
mixings.  An example is shown in Fig. \ref{v1up2dn_to_v2up1dn_gap_fig},
which involves $V_2^4 \to V^2_5$ and $V^1_4 \to V^3_4 \to V_1^5$ transition
on virtual ETC gauge boson lines in the loop, yielding the overall
transition $V^1_2 \to V^2_1$.

\begin{center}
\begin{picture}(240,200)(0,0)
\Photon(0,100)(25,100){4}{3} \PhotonArc(60,100)(35,0,180){-4}{8.5}
\PhotonArc(60,100)(35,180,0){-4}{8.5} \Text(60,133)[]{$\times$}
\Text(45,69)[]{$\times$} \Text(75,69)[]{$\times$}
\Photon(95,100)(120,100){4}{3} \Text(0,115)[]{$(V^1_2)_\mu$}
\Text(122,115)[]{$(V^2_1)_\nu$} \Text(25,135)[]{$V_2^4$}
\Text(97,135)[c]{$V^2_5$} \Text(18,72)[]{$V^1_4$} \Text(60,51)[]{$V^3_4$}
\Text(105,72)[c]{$V^5_1$}
\end{picture}
\end{center}

\begin{figure}
\caption{\footnotesize{One-loop graph contributing to the ETC gauge boson
mixing $V^1_2 \leftrightarrow V^2_1$ in sequence 1.  The indices on the ETC
gauge bosons in the loop are written with the convention that both upper and
lower lines go from left to right.  The graph with indices 4 and 5
interchanged on the internal gauge boson lines also contributes.}}
\label{v1up2dn_to_v2up1dn_gap_fig}
\end{figure}

The ETC gauge boson mixing $V^3_2 \rightarrow V^2_1$ is generated, e.g., by
a diagram analogous to Fig. \ref{v1up2dn_to_v2up1dn_gap_fig}, in which the
upper and lower lines (reading from left to right) contain the respective
transitions $V_2^4 \to V^2_5$ and $V^3_4 \to V^1_4 \to V^3_5 \to V^5_1$.
The mixing $V^1_3 \to V^3_1$ is generated, e.g., by an analogous diagram in
which the upper and lower lines contain the respective transitions $V_3^4
\to V^4_1 \to V^3_5$ and $V^1_4 \to V^3_4 \to V^5_1$.  The combination
$V^3_2 \leftrightarrow V^2_1 \leftrightarrow V^1_2 \leftrightarrow V^2_3$
yields the mixing $V^3_2 \leftrightarrow V^2_3$.

 From these mixings and the identity (\ref{identity1}), it follows that, for
this model with any symmetry-breaking sequence,
\beq |M^{(f)}_{ij}|=|M^{(f)}_{ji}| \ . \label{mfsym} \eeq

\subsection{ETC Gauge Boson Mixing for Symmetry-Breaking Sequence 2}

For symmetry-breaking sequence 2, we obtain, to begin with, the ETC gauge
boson mixings
\beq V_2^4 \leftrightarrow V^3_5 \ , \quad V_2^5 \leftrightarrow V^3_4
\label{v2up4dn_to_v5up3dn_gbp} \eeq
\beq V_1^t \leftrightarrow V_3^t \ , \quad t=4,5
\label{v1up4dn_to_v3up4dn_gbp} \eeq
and
\beq V^i_4 \leftrightarrow V^5_i \ , \quad V^i_5 \leftrightarrow V^4_i \ ,
\quad i = 1,2,3 \label{viup4dn_to_v5upidn_gbp} \eeq

The mixings (\ref{v2up4dn_to_v5up3dn_gbp}) and
(\ref{v1up4dn_to_v3up4dn_gbp}) were present in the $G_b$ sequence of Ref.
\cite{nt}.  We show in Figs. \ref{v3dn4up_to_v2up5dn_gbp_fig} -
\ref{v2dn4up_to_v3dn4up_gbp_fig} some diagrams that give rise to the mixings
(\ref{v2up4dn_to_v5up3dn_gbp})- (\ref{viup4dn_to_v5upidn_gbp}).

It is useful to note a general relation connecting the sequences $G_a$ and
$G_b$ of Ref. \cite{nt} insofar as they involve condensates of the $\zeta$
fields and resultant ETC gauge boson mixings, namely that these condensates
and mixings in sequence $G_b$ are related to those for sequence $G_a$ by the
interchange of the ETC indices 1 and 2 (holding other ETC indices fixed)
with appropriate changes in the condensation scale.  Thus, (i) the
condensate (\ref{4x6zetacondensate}) occurring at scale $\Lambda_2$ in
sequence 1 goes over, under this interchange of indices, to the condensate
(\ref{6x6condensate}) occurring at scale $\Lambda_{BHC}$ in sequence 2; (ii)
the condensate (\ref{33to3barcondensate}) at scale $\Lambda_3$ in sequence 1
goes over to the condensate (\ref{z14z15_gbp}) at $\Lambda_{TC}$ in sequence
2; (iii) the condensate (\ref{z12z23condensate}) at $\Lambda_3$ in sequence
1 goes over to (\ref{44to6condensate}) at $\Lambda_{23}$ in sequence 2; and
(iv) the condensate (\ref{z12omegacondensate}) at $\Lambda_3$ in sequence 1
goes to the same condensate, now at $\Lambda_{23}$, in sequence 2, and the
condensate (\ref{z23omegacondensate}) at $\Lambda_3$ in sequence 1 goes to
the $i=3$ case of (\ref{z12omegacondensate_gbp}) at $\Lambda_{23}$ in
sequence 2.  We shall denote this interchange symmetry with corresponding
changes in condensation scales as $S12$.

\begin{center}
\begin{picture}(240,200)(0,0)
\Photon(0,100)(25,100){3}{4.5} \ArrowArcn(60,100)(35,180,90)
\Text(60,135)[]{$\times$} \ArrowArcn(60,100)(35,90,0)
\ArrowArcn(60,100)(35,0,270) \Text(60,65)[]{$\times$}
\ArrowArcn(60,100)(35,270,180) \Photon(95,100)(120,100){3}{4}
\Text(0,115)[]{$(V_3^4)_\mu$} \Text(122,115)[]{$(V^2_5)_\nu$}
\Text(25,135)[]{$\zeta^{14,\alpha}_R$} \Text(20,70)[]{$\zeta^{13,\alpha}_R$}
\Text(105,135)[c]{$\zeta^c_{15,\beta,L}$}
\Text(105,70)[c]{$\zeta^c_{12,\beta,L}$}
\end{picture}
\end{center}

\begin{figure}
\caption{\footnotesize{A one-loop graph contributing to the ETC gauge boson
mixing $V_3^4 \leftrightarrow V^2_5$ for the symmetry-breaking sequence 2.}}
\label{v3dn4up_to_v2up5dn_gbp_fig}
\end{figure}

\begin{center}
\begin{picture}(240,200)(0,0)
\Photon(0,100)(25,100){3}{4.5} \ArrowArcn(60,100)(35,180,90)
\Text(60,135)[]{$\times$} \ArrowArcn(60,100)(35,90,0)
\ArrowArcn(60,100)(35,0,270) \Text(60,65)[]{$\times$}
\ArrowArcn(60,100)(35,270,180) \Photon(95,100)(120,100){3}{4}
\Text(0,115)[]{$(V_1^4)_\mu$} \Text(122,115)[]{$(V^1_5)_\nu$}
\Text(25,135)[]{$\zeta^{24,\alpha}_R$} \Text(20,70)[]{$\zeta^{12,\alpha}_R$}
\Text(105,135)[c]{$\zeta^c_{35,\beta,L}$}
\Text(105,70)[c]{$\zeta^c_{13,\beta,L}$}
\end{picture}
\end{center}

\begin{figure}
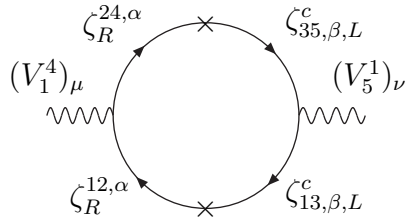

\caption{\footnotesize{A one-loop graph contributing to the ETC gauge boson
mixing $V_1^4 \leftrightarrow V^1_5$ for the symmetry-breaking sequence 2.
The graph with the index interchange $2 \leftrightarrow 3$ on the internal
fermion lines also contributes.}} \label{v1dn4up_to_v1up5dn_gbp_fig}
\end{figure}

\begin{center}
\begin{picture}(240,200)(0,0)
\Photon(0,100)(25,100){3}{4.5} \ArrowArcn(60,100)(35,180,90)
\ArrowArcn(60,100)(35,90,0) \ArrowArcn(60,100)(35,0,270)
\ArrowArcn(60,100)(35,270,180) \Photon(95,100)(120,100){3}{4}
\Text(45,69)[]{$\times$} \Text(75,69)[]{$\times$}
\Text(0,115)[]{$(V_2^t)_\mu$} \Text(122,115)[]{$(V_3^t)_\nu$}
\Text(62,148)[]{$\zeta^{1t,\alpha}_R$} \Text(18,72)[]{$\zeta^{12,\alpha}_R$}
\Text(60,51)[]{$\omega^c_{\beta,p,L}$}
\Text(107,72)[c]{$\zeta^{13,\alpha}_R$}
\end{picture}
\end{center}

\begin{figure}
\caption{\footnotesize{A one-loop graph contributing to the ETC gauge boson
mixing $V_2^t \leftrightarrow V_3^t$, where $t=4,5$, in symmetry-breaking
sequence 2.}} \label{v2dn4up_to_v3dn4up_gbp_fig}
\end{figure}

By methods similar to those above, we estimate
\beq |{}^3_5 \Pi_2^4(0)| = |{}^3_4 \Pi_2^5(0)| \simeq \frac{g_{_{ETC}}^2
\Lambda_{TC}^3}{24\Lambda_{23}} \label{Pi_v2up4dn_to_v3dn5up_gbp} \eeq
\beq |{}^1_5 \Pi_1^4(0)| = |{}^1_4 \Pi_1^5(0)| \simeq \frac{g_{_{ETC}}^2
\Lambda_{23}^3}{12\Lambda_{BHC}} \label{Pi_v1up4dn_to_v1dn5up_gbp} \eeq
\beq |{}_2^t \Pi_3^t(0)| \simeq \frac{g_{_{ETC}}^2 \Lambda_{23}^2}{12}
\quad {\rm for} \ \ t=4,5 \label{Pi_v2up4dn_to_v3up4n_gbp} \eeq
Note that for the mixing in Eq. (\ref{Pi_v1up4dn_to_v1dn5up_gbp}), for a
given set of hypercolors, two pairs of $\zeta$'s contribute, while in the
case of (\ref{Pi_v2up4dn_to_v3up4n_gbp}) we sum over $p=1,2$ for the
contributions of the $\omega^c_{\beta,p,L}$.

As was the case for sequence 1, insertions of these non-diagonal mixings on
ETC gauge boson lines give rise to further mixings.  Thus,
\beq V_2^4 \leftrightarrow V^3_5 \leftrightarrow V^2_5 \ \Longrightarrow \
V_2^4 \leftrightarrow V^2_5 \label{v2dn4up_to_v2up5dn_gbp} \eeq
and
\beq V_3^4 \leftrightarrow V^2_5 \leftrightarrow V^3_5 \  \Longrightarrow \
V_3^4 \leftrightarrow V^3_5 \label{v3dn4up_to_v3up5dn_gbp} \eeq
so that all of the three mixings
\beq V_i^4 \leftrightarrow V^i_5 \ , \quad V_i^5 \leftrightarrow V^i_4 \ ,
\quad i=1,2,3 \label{vidn4up_to_viup5dn_gbp} \eeq
occur.  We estimate
\beq {}_2^5 \Pi^2_4(0) = {}^2_5 \Pi^3_5(0) \ \frac{1}{M_{23}^2} \ {}^3_5
\Pi^4_2(0) \simeq \frac{g_{_{ETC}}^2 \Lambda_{TC}^2}{18 a^2 }
\label{Pi_2dn4up_to2up5dn_gbp} \eeq
\beq {}_3^5 \Pi^3_4(0) = {}^5_3 \Pi^5_2(0) \ \frac{1}{M_{23}^2} \ {}^5_2
\Pi^3_4(0) \simeq \frac{g_{_{ETC}}^2 \Lambda_{TC}^2}{18 a^2 }
\label{Pi_3dn4up_to3up5dn_gbp} \eeq
where again we note that there is significant theoretical uncertainty in
the coefficients because of the strong ETC coupling.
ETC gauge boson loop diagrams with various insertions yield still further
mixings in a manner similar to that discussed above.

We summarize the ETC gauge boson mixings for the two symmetry-breaking
sequences that we consider in this paper and, for comparison, the two
simpler ones denoted $G_a$ and $G_b$ in Ref. \cite{nt}, in Table
\ref{etc_mixings}.  The symmetry S12 connecting sequences 1 and 2 is evident
here.

\begin{table}
\caption{\footnotesize{Some ETC gauge boson mixings in the model with the
two symmetry-breaking sequences, denoted S1 and S2.  For comparison, results
for the two analogous sequences $G_a$ and $G_b$, without the
$\omega^\alpha_{p,R}$ fields, are also listed.  For each entry, a
``y'' indicates that the mixing occurs, while a blank indicates that it is
absent.  For each mixing of the form $V_i^4 \leftrightarrow V^j_5$, it is
understood that one also include the corresponding transition with the
indices 4 and 5 interchanged. For mixings of the form $V_i^t \leftrightarrow
V_j^t$ with $i,j \in \{1,2,3\}$, the index $t$ takes on the values
$t=4,5$.}}
\begin{center}
\begin{tabular}{||c|c|c|c|c||}
ETC transition                  &$G_a$& S1 &$G_b$ & S2 \\ \hline\hline
$V_1^4 \leftrightarrow V^2_5$   &     &    &      &    \\ \hline $V_1^4
\leftrightarrow V^3_5$   &  y  &  y &      &    \\ \hline $V_2^4
\leftrightarrow V^3_5$   &     &    &  y   &  y \\ \hline\hline $V_1^4
\leftrightarrow V^1_5$   &     &    &  y   &  y \\ \hline $V_2^4
\leftrightarrow V^2_5$   &  y  &  y &      &  y \\ \hline $V_3^4
\leftrightarrow V^3_5$   &     &  y &      &  y \\ \hline\hline $V_1^t
\leftrightarrow V_2^t$   &     &    &      &    \\ \hline $V_1^t
\leftrightarrow V_3^t$   &     &  y &      &    \\ \hline $V_2^t
\leftrightarrow V_3^t$   &     &    &      &  y \\ \hline\hline $V^1_2
\leftrightarrow V_a$     &     &    &      &    \\ \hline $V^1_3
\leftrightarrow V_a$     &     &    &      &    \\ \hline $V^2_3
\leftrightarrow V_a$     &     &    &      &    \\ \hline\hline $V_1^2
\leftrightarrow V_2^1$   &     &    &  y   &  y \\ \hline $V_1^3
\leftrightarrow V_3^1$   &     &    &      &  y \\ \hline $V_2^3
\leftrightarrow V_3^2$   &     &    &      &  y \\ \hline $V_1^2
\leftrightarrow V_2^3$   & y   &  y &      &    \\ \hline
$V_1^2 \leftrightarrow V^1_3$   &     &    &  y   &  y \\
\end{tabular}
\end{center}
\label{etc_mixings}
\end{table}

\begin{table}
\caption{\footnotesize{Elements of fermion mass matrices for the two
symmetry-breaking sequences, denoted S1 and S2.  For comparison, results for
the two analogous sequences $G_a$ and $G_b$, without the
$\omega^\alpha_{p,R}$
fields, are also listed.  For each entry, a ``y'' and blank indicate
that the entry is nonzero and zero, respectively.  The symmetries
$M^{(f)}_{ij}=M^{(f)}_{ji}$ are implicit.}}
\begin{center}
\begin{tabular}{||c|c|c|c|c||}
matrix element               &$G_a$& S1 &$G_b$& S2 \\ \hline\hline
$M^{(u)}_{11}$               &  y  & y  &  y  & y  \\ \hline $M^{(u)}_{12}$
&     &    &     &    \\ \hline $M^{(u)}_{13}$               &  y  & y  &
&    \\ \hline $M^{(u)}_{22}$               &  y  & y  &  y  &  y \\ \hline
$M^{(u)}_{23}$               &     &    &  y  &  y \\ \hline $M^{(u)}_{33}$
&  y  & y  &  y  &  y \\ \hline\hline $M^{(d,e)}_{11}$             &     & y
&  y  &  y \\ \hline $M^{(d,e)}_{12}$, $b_{12}$   &     &    &     &    \\
\hline $M^{(d,e)}_{13}$, $b_{13}$   &  y  & y  &     &    \\ \hline
$M^{(d,e)}_{22}$, $b_{22}$   &  y  & y  &     &  y \\ \hline
$M^{(d,e)}_{23}$, $b_{23}$   &     &    &  y  &  y \\ \hline
$M^{(d,e)}_{33}$, $b_{33}$   &     & y  &     &  y \\ \hline\hline $r_{22}$
&     &    &     &    \\ \hline $r_{23}$                     &     & y  &  y
&  y \\ \hline
$r_{33}$                     &     &    &     &    \\
\end{tabular}
\end{center}
\label{matrix_elements}
\end{table}


\begin{thebibliography}{99}
\bibitem{gap} In the approximation of a single-gauge-boson exchange, the
critical coupling for chiral fermions transforming according to the
representations $R_1$ and $R_2$ to form a condensate transforming as
$R_{cond.}$ in an SU($N$) gauge theory is given by the condition
$\frac{3\alpha}{2\pi}\Delta C_2 = 1$, where $\Delta C_2 =
[C_2(R_1)+C_2(R_2)-C_2(R_c)]$, and $C_2(R)$ is the quadratic Casimir invariant.
\bibitem{global}
Recall that the number of fermion doublets of SU(2) must be even to avoid a
global anomaly.
\bibitem{qmass}
This value of $m_u$ is the current quark mass, i.e. the mass
measured at a scale well above $\Lambda_{QCD}$.  We use the value
$(m_u+m_d)/2 = 4.3$ MeV \cite{mlat} in conjunction with the current algebra
result $m_d/m_u = 1.8$ \cite{gl} to obtain $m_u = 3.1$ MeV.
\bibitem{gen1}
It is plausible that the fact that this inequality is different for the
first generation, where $m_e < m_u < m_d$, is due to the dominant effect of
off-diagonal entries in the up- and down-quark mass matrices in determining
the respective lightest up and down quark masses.
\bibitem{nuc} An alternate definition for $|\bar\nu_e\rangle$ is the
antineutrino that is emitted in a nuclear beta decay of the form $(Z,A) \to
(Z+1,A) + e^- + \bar\nu_e$, and similarly for $|\nu_e\rangle$ as the neutrino
emitted in the decay $(Z,A) \to (Z-1,A) + e^+ + \nu_e$ or the electron-capture
process $(Z,A) + e^- \to (Z-1,A) + \nu_e$.
\bibitem{imp}
Implications for measurements of weak interaction quantities such as $G_F$
and $BR(\pi^+ \to e^+ \nu_e)/BR(\pi^+ \to \mu^+ \nu_\mu)$ decays are
discussed
in \cite{ls,rs80}.
\bibitem{afactor} Parenthetically, we note the values of the exponential
coupling-dependent factor.  To obtain these we use the fact that at a given
energy scale $\mu$ the ETC coupling $\alpha_{_{ETC}}(\mu) =
g_{_{ETC}}^2(\mu)/(4\pi)$ satisfies the approximate criticality relation
\cite{gap} $3 \alpha_{_{ETC}}\Delta C_2/(2\pi) \simeq 1$.  Solving the
criticality relation for $\alpha_{_{ETC}}(\mu)$ and substituting in the
expression for $A(\mu)$, we obtain $A(\mu)={\rm exp}(-8\pi^2/g_{_{ETC}}^2(\mu))
\simeq {\rm exp}(-3 \Delta C_2)$ at each particular ETC scale.  In sequence 1
we have $\Delta C_2= 24/5$ for the condensation at $\Lambda_1$, whence
$A(\Lambda_1) \simeq {\rm exp}(-72/5) \simeq 6 \times 10^{-7}$; by similar
means we have $A(\Lambda_2) \simeq e^{-15/2} \simeq 6 \times 10^{-4}$ and
$A(\Lambda_3) \simeq e^{-4} \simeq 0.02$).  The power-law prefactor also
contributes to the dimensionless coefficient.  As indicated in the text,
because of the strong-coupling nature of the ETC theory at the relevant scales,
we will simply use an effective-field theory approach of approximating this
dimensionless coefficient as roughly unity.
\bibitem{tr}
We recall the standard definition for the Casimir invariant
$T(R)$: for a representation $R$ of a group SU($N$), $\sum_{i,j=1}^{dim(R)}
D_R(T_a)^i_j D_R(T_b)^j_i = T(R)\delta_{ab}$, where $D_R(T_a)$ is the
generator $T_a$ in this representation, with $a,b$ denoting group indices
and $i,j$ denoting the representation indices.
\end{thebibliography}
\end{document}